%% file: paper-arxiv.tex
\pdfoutput=1 
\documentclass[conference, 10pt]{IEEEtran}

\usepackage[utf8]{inputenc}
\usepackage[british]{babel}
\usepackage{amsmath, amsfonts}
\usepackage{newtx}

\usepackage[T1]{fontenc}
\usepackage{microtype}
\usepackage[dvipsnames]{xcolor}
\usepackage[style=ieee]{biblatex}
\usepackage[inline]{enumitem}
\usepackage{tikz}
\usepackage{graphicx}
\usepackage[caption=false, font=footnotesize]{subfig}
\usepackage[per-mode=symbol]{siunitx}
\usepackage{textcomp}
\usepackage{booktabs}
\usepackage{orcidlink}

\usepackage{hyperref}
\hypersetup{
    pdfauthor={Gregory Stock, Juan A. Fraire, Santiago Henn, Holger Hermanns, and Andreas Schmidt},
    pdftitle={A Stability-first Approach to Running TCP over Starlink},
}

\newcommand{\algDijkstra}{\textsc{Dijkstra}}
\newcommand{\algStubborn}{\textsc{Stubborn}}
\newcommand{\algTenacious}{\textsc{Tenacious}}
\newcommand{\algSetcover}{\textsc{SetCover}}

\newcommand{\artcp}{\textsc{artCP}}

\newcommand\copyrighttext{%
    \footnotesize \textcopyright 2024 IEEE.
    Personal use of this material is permitted.
    Permission from IEEE must be obtained for all other uses, in any current or future media, including reprinting/republishing this material for advertising or promotional purposes, creating new collective works, for resale or redistribution to servers or lists, or reuse of any copyrighted component of this work in other works.
    DOI: \href{https://doi.org/10.1109/ICCWorkshops59551.2024.10615714}{10.1109/ICCWorkshops59551.2024.10615714}
}
\newcommand\copyrightnotice{%
    \begin{tikzpicture}[remember picture, overlay]
        \node[anchor=south, yshift=10pt] at (current page.south) {\fbox{\parbox{\dimexpr\textwidth-\fboxsep-\fboxrule\relax}{\copyrighttext}}};
    \end{tikzpicture}%
}

\begin{document}
\renewcommand{\sectionautorefname}{Section} 
\renewcommand{\subsectionautorefname}{Section} 
\newcommand{\subfigureautorefname}{\figureautorefname} 

\title{\texorpdfstring{\resizebox{\linewidth}{!}{A Stability-first Approach to Running TCP over Starlink}}{A Stability-first Approach to Running TCP over Starlink}}

\author{%
    \IEEEauthorblockN{%
        Gregory Stock\IEEEauthorrefmark{1} \orcidlink{0000-0001-5170-2019},
        Juan A. Fraire\IEEEauthorrefmark{1}\IEEEauthorrefmark{2}\IEEEauthorrefmark{3} \orcidlink{0000-0001-9816-6989},
        Santiago Henn\IEEEauthorrefmark{2} \orcidlink{0000-0002-2888-961X},
        Holger Hermanns\IEEEauthorrefmark{1} \orcidlink{0000-0002-2766-9615}, and
        Andreas Schmidt\IEEEauthorrefmark{1} \orcidlink{0000-0002-7113-7376}
    }
    \IEEEauthorblockA{%
        \IEEEauthorrefmark{1}Saarland University -- Computer Science, Saarland Informatics Campus, Saarbrücken, Germany
    }
    \IEEEauthorblockA{%
        \IEEEauthorrefmark{2}CONICET -- Universidad Nacional de Córdoba, Córdoba, Argentina
    }

    \IEEEauthorblockA{%
        \IEEEauthorrefmark{3}Inria, INSA Lyon, CITI, UR3720, 69621 Villeurbanne, France
    }
}

\maketitle

\begin{abstract}
    The end-to-end connectivity patterns between two points on Earth are highly volatile if mediated via a Low-Earth orbit (LEO) satellite constellation.
    This is rooted in the enormous speeds at which satellites in LEO must travel relative to the Earth's surface.
    While changes in end-to-end routes are rare events in stationary and terrestrial applications, they are a dominating factor for connection-oriented services running over LEO constellations and mega-constellations.
    This paper discusses how TCP-over-constellations is affected by the need for rerouting and how orbital route selection algorithms impact the end-to-end performance of communication.
    In contrast to the state of the art that primarily optimizes for instantaneous shortest routes (i.e.\@ lowest delay), we propose several algorithms that have route stability and longevity in their focus.
    We show that this shift in focus comes with vastly improved end-to-end communication performance, and we discuss peculiar effects of the typical TCP-like implementations, taking inspiration from the Starlink constellation in our empirical investigations.
    The spectrum of algorithms proposed provides a basis for co-designing suitable orbital route selection algorithms and tailored transport control algorithms.
\end{abstract}

\section{Introduction}

\copyrightnotice{}
Low-Earth orbit (LEO) mega-constellations are thriving.
Over half of the orbiting satellites are part of large-scale networked fleets (e.g.\@ Starlink, Project Kuiper) aiming at global high-throughput internet connectivity and potentially other services.
This development poses new challenges in network management to achieve both \emph{low delay} and \emph{high throughput}.

Traditional routing algorithms have predominantly honed in on optimizing the instantaneous state of the network, aligning with the stochastic nature of the internet.
There is a plethora of works on the stability of routes on the internet~\cite{DBLP:conf/infocom/GovindanR97,DBLP:journals/ton/GaoR01,DBLP:conf/iwmn/CanbazBG17,DBLP:journals/access/IodiceCB19}, as well as the interaction of this stability with transport layer performance~\cite{DBLP:conf/ipccc/SrijithJA03,DBLP:conf/usenix/YanMHRWLW18}.

The dynamics in LEO however induces rapidly changing end-to-end routes, characterized by highly volatile round-trip times (RTTs) when traversing through diverse sets of inter-satellite links (ISLs)~\cite{DBLP:conf/hotnets/HauriBGS20}.
Furthermore, ground stations are forced to continually switch access satellites as visibility periods are generally short-lived~\cite{DBLP:conf/hotnets/VasishtC20}, albeit being nearly periodic.
As a consequence, cross-orbit routes exhibit a high degree of variability and are of limited stability over time.

Thus, it becomes obvious that route stability within a satellite mega-constellation is an important factor in ensuring the efficacy of end-to-end protocols.
Furthermore, unavoidable route changes need to be set up in such a way that they mitigate prevalent issues.
One important such issue is the reordering of packets caused by overtaking, which can happen if the new route is shorter in delay.
This can have a detrimental impact on communication reliability and efficiency~\cite{DBLP:conf/pam/BhosaleSBG23}, rooted in the fact that in every standard Transmission Control Protocol~(TCP) implementation such a reordering is likely to lead to \emph{duplicate acknowledgements (ACKs)}.
But also changes to considerably slower routes can trigger adverse effects due to \emph{timeouts}~\cite{rfc6298}.
While the individual handling is partly up to the TCP variant, both effects tend to lead to a substantial reduction in the in-flight packet window size and hence a reduction in achievable data rate.

On the other hand, routing solutions tailored for satellite constellations can leverage the predictably periodic nature of orbital networks to address stability issues adeptly.
This is in contrast to terrestrial mobile networks that also have dynamics, but which are less predictable as space constellations.

To the authors' knowledge, a route selection procedure that judiciously considers the instability inherent in routes within mega-constellations remains an open research topic~\cite{DBLP:conf/pam/BhosaleSBG23}.
In response, this paper introduces a family of algorithms, offering a spectrum of solutions to navigate the trade-offs between \emph{end-to-end delay} and \emph{route stability} in satellite communication.
Despite their varied computational complexities, these algorithms are unified in their objective to enhance route stability and longevity, thereby elevating overall communication performance within in-orbit networks.
The contributions of this work are as follows:
\begin{itemize}
    \item We provide a background on route stability and its consequential impact on transport-layer network functions.
    \item We introduce a new evaluation framework to compute route stability figures and apply them to TCP-like communication models within Walker Delta constellations.
    \item We propose a suite of route selection algorithms, each with distinct computational complexities, and articulate the trade-offs between end-to-end delay and route stability.
    \item We comprehensively evaluate the well-known Starlink constellation, providing insights into stability metrics and end-to-end performance in TCP-like communication.
\end{itemize}

The remainder of this paper starts off with detailing our methodological framework in \autoref{sec:framework}, followed by a description of the different route selection algorithms in \autoref{sec:algorithms}.
Empirical results are summarized in \autoref{sec:results}, and \autoref{sec:conclusion} discusses the findings and draws final conclusions.

\section{Context \& Methodological Framework}\label{sec:framework}

In this section, we present the contextual background together with the methodological details of our evaluation framework.

\subsection{Walker Delta Constellations \& Orbital Dynamics}

This study focuses on Walker Delta constellations due to the popularity of this constellation type for upcoming mega-constellations.
All satellites in a Walker Delta constellation follow circular orbits and share the same altitude~\(h\) and inclination~\(\alpha\).
They thus move at the same speed.
The satellites are distributed across \(P\)~evenly-spaced orbital planes, where each plane contains~\(Q\) evenly-spaced satellites.
Formally, we describe Walker Delta constellations by \(\alpha\colon P\cdot Q / P / F \), where~\(F\) specifies the relative phase shift between adjacent planes.
Each satellite can be distinguished at any time as either ascending or descending, depending on whether they move from South to North or from North to South, respectively.

We assume that each satellite maintains four permanent inter-satellite links: two intra-plane links (to the successor and predecessor in same orbital plane) and two inter-plane links (to the left and right neighbour on the adjacent planes)~\cite{DBLP:conf/asms-spsc/StockFH22}.
So, from the perspective of an individual satellite, the underlying connection topology resembles that of a Manhattan Street Network that gets distorted at the most polar positions, where satellites switch from ascending to descending and vice versa.

For the routing algorithmics, future satellite positions need to be predicted.
We will use a simple two-body propagator, i.e.\@ assuming ideal conditions and unperturbed central force motion.
This means that all orbital elements (except the argument of latitude) are assumed constant over time.
More sophisticated propagators, such as SGP4, could be used, but the additional precision comes with a computation time penalty without having a significant effect on the analysis results.

\subsection{Efficiency and Stability Metrics}

We work with the following metrics to measure different aspects of efficiency and stability.

\paragraph*{Route Delay}
The route delay is determined by transmission delay, propagation delay, and queuing delays.
We define the one-way delay (OWD) of a route as \(d_\mathit{ow} = d_p + d_t + d_q\), where \(d_p\) is the propagation delay (i.e.\@ the time the signal needs to travel, depending only on the physical distance), \(d_t\)~is the transmission delay (i.e.\@ the time to send the whole packet, depending only on the packet size and the data rate of the link), and \(d_q\) the queueing delay, an additional per-hop overhead for on-board processing and packet queueing.
We assume store-and-forward-switching and consider a packet size of \qty{1.5}{\kilo\byte} (standard Ethernet frame) and \qty{1}{\giga\bit\per\second} links for all satellites and ground stations.
This results in a transmission delay of \(d_t = \frac{\qty{1.5}{\kilo\byte}}{\qty{1}{\giga\bit\per\second}} = \qty{12}{\micro\second}\).
Further, we assume \(d_q = \qty{1}{\milli\second}\) for queueing and on-board processing.
Assuming that the signals travel with the speed of light~\(c\), the propagation delay for some distance~\(l\) is given by \(d_p = \frac{l}{c}\).

\paragraph*{Route Validity}
Route validity is defined as how long a particular route remains accessible.
Disregarding node failures, any route between two satellites in a Walker Delta constellation is valid ad infinitum since neighbourship remains unchanged~(while link distances oscillate across planes).
However, end-to-end routes between two ground stations are always limited by the visibility periods of the two access satellites~(i.e.\@ the first and last satellite of the route that directly communicate with the ground).
Generally, the aim is maximizing end-to-end route validity to obtain less frequent route changes.

\paragraph*{Delay Delta and Bad Changes}
The delay delta is defined as \(d_\mathit{ow}^\mathit{new} - d_\mathit{ow}^\mathit{old}\) and specifies by how much the delay changes at the points in time in which route changes occur.
Negative delay delta values (switching from high to low delay) cause out-of-order arrivals, disrupting the expected sequential delivery of packets at the transport layer, particularly impacting protocols like TCP that rely on ordered delivery.
If three duplicate acknowledgements are received, this scenario can trigger mechanisms such as TCP fast retransmission.
Conversely, positive delta values (switching from low to high delay) can provoke unnecessary retransmissions at the sender, as the sender might interpret the increased delay as packet loss if acknowledgements from the receiver are not received within the expected time frame.
Based on recent RTT measurements, TCP dynamically adjusts its retransmission timeout (RTO).
If the delay suddenly increases due to a route change, the sender might not receive acknowledgements within the expected RTO, triggering unnecessary retransmissions.
However, RFC6298 recommends (``SHOULD'') a minimum RTO of \qty{1}{\second}~\cite{rfc6298}.
In our evaluation, the delays and deltas were never close to this bound, so we ignored retransmits and the corresponding in-flight window reductions.
The RFC also suggests that this minimum value is up to further research---hence we plan to reinvestigate this in future work.

\paragraph*{TCP Data Rate}
The average data rate is a tangible value allowing for a performance-oriented comparison of different route selection algorithms.
In general, we want to maximize this quantity that is derived from the route delays and validities using our abstract TCP implementation~(\autoref{sec:artcp}).

\section{Route Selection Algorithms}\label{sec:algorithms}

This section introduces a family of route selection algorithms that cover the full spectrum between minimizing delay and minimizing the number of route changes.

Each algorithm takes as input the position of two points on Earth (ground stations) some time interval~\([t_0, t_h)\) (typically one orbital period), and a time granularity (typically \qty{1}{\second}).
The output is a mapping of each time point in the interval to a specific end-to-end route connecting the two ground stations via the constellation.

\paragraph*{Dijkstra (\algDijkstra{})}
Dijkstra's algorithm is the state-of-the-art solution.
It optimizes for the shortest route at a given time instant.
In the following, we denote a route as shortest if it is minimal with respect to one-way delay, i.e.\@ the ``fastest'' route.
(By setting \(d_t = d_q = \qty{0}{\milli\second}\), the classical shortest route w.r.t.\@ distance results.)
As a baseline, our algorithm \algDijkstra{} runs Dijkstra's algorithm for each time point in~\([t_0, t_h)\).
While this minimizes the overall delay all along the interval, it has no concept of route stability.

\paragraph*{Stubborn Dijkstra (\algStubborn{})}
As a variation of \algDijkstra{}, we introduce and analyse a ``stubborn'' variant of it.
Here, the current shortest route is selected at~\(t_0\).
This route is then used for as long as it is valid, i.e.\@ until one of the two access satellites loses visibility to its respective ground station.
Then, a new shortest route is computed, and this procedure is repeated until the end of the time interval.

\paragraph*{Tenacious Routing (\algTenacious{})}
Since visibility periods are decisive for route validity, another alternative for maximizing route validity is to identify, for both ground stations, the satellite with the longest remaining visibility period.
This is done at \(t_0\), and the resulting pair is then used as access satellites for a route that remains unchanged across the entire period in which both satellites have visibility.
Once the current route becomes invalid, two new satellites are selected with the now longest remaining visibility time and the procedure is repeated.

This algorithm yields a list of access satellite pairs together with the time interval during which they are to be used.
In a second step, a route is computed per entry.
Assume two access satellites and an interval \([t_s, t_e)\).
As the route length (and hence incurred delay) changes over time, there is no obvious optimal in-orbit route for the whole time interval.
One possibility would be to compute the route with the lowest average delay during \([t_s, t_e)\).
However, as this approach is compute-intensive, we just compute the shortest route at some point within \([t_s, t_e)\).
There is a trade-off between minimizing average route delay (typically best if we pick a point in the middle) and having a low delay at~\(t_e\) to increase the chances of avoiding reordering events when switching to another route.
Our extensive empirical studies have shown that---compared to the risk of timeouts if changing to a fast route---reordering is clearly the crucial obstruction to performance, and that \(t_s + \frac{3}{4}(t_e - t_s)\) provides overall good performance.

While experimenting with this algorithm, we noticed exorbitantly high delays in some cases.
This is due to the algorithm not distinguishing between ascending and descending satellites.
This movement direction, however, has a considerable effect on the route length.
Assume, for example, two ground stations close to each other on the Equator.
Now it may be that these stations can be connected by two ascending satellites with just one intermediate ISL hop.
On the contrary, if one of the satellites is exchanged for a nearby descending satellite, these satellites can no longer communicate directly with each other, and multiple hops are required.
Therefore, instead of selecting only two access satellites with the longest remaining visibility, we select at each station the ascending and the descending satellite with the longest remaining visibility (if it exists).
In the second step, we consider all combinations (at most four) to select the pair with the lowest delay.
This optimization significantly reduces the overall delay at the cost of a few more route changes.

\paragraph*{(Weighted) Set Cover (\algSetcover{})}

The previous algorithms are all greedy at specific time points, in one way or another.
Now, we introduce a solution that considers the entire time interval \([t_0, t_h)\) to find some global optimum.
For this, we have formulated the route selection problem as a (weighted) set cover problem.
First, all satellite visibility intervals at both ground stations are computed.
This information is used to compute all possible combinations of valid access satellite pairs, together with the maximal intervals~\([t_s, t_e)\) during which this pair can be used.
The goal is now to select as few of these combinations as possible so that the whole time interval is covered.
We solve this problem by modelling it as an integer linear program (ILP).
For each combination of access satellite pair \((\mathit{src}, \mathit{dst})\) and interval~\([t_s,t_e)\) during which it is valid, we introduce a binary variable \(x_{[t_s,t_e)}^{\mathit{src}, \mathit{dst}} \in \{0, 1\}\).
A value of \(1\) means that this element is part of the overall solution.
Let \(I\) be the set of all such variables.
The objective function \(\min \sum_{x \in I} w_x \cdot x\) now minimizes the weighted sum of variables, where~\(w_x\) is some non-negative weight for~\(x\).
In this paper, our algorithm minimizes the total number of route changes, i.e.\@ \(w_x=1\) for all \(x\in I\).
It is, however, straightforward to tweak the optimization objective by assigning different weights, such as the hop count in between the two satellites.

Let \(I\vert_t = \bigl\{x_{[t_s,t_e)}^{\mathit{src}, \mathit{dst}} \in I \bigm\vert t_s \leq t < t_e\bigr\}\) be the subset of~\(I\) that contains all binary decision variables whose interval includes time point~\(t\).
Now, we add constraints to ensure that every time point in the time interval is covered by (at least) one element:
\[
    \forall t \in \{t_0, t_1, \ldots, t_h\}. \sum\nolimits_{x \in I\vert_t} x \geq 1
\]

A solution of this ILP specifies which access satellites are best to use and at what intervals they should be used.
However, the resulting intervals usually have a small overlap.
To get disjoint intervals, we cut each interval in the middle of the overlapping part.
More precisely, two overlapping intervals \([t_s, t_e)\) and \([t'_s, t'_e)\) are mapped to \([t_s, m)\) and \([m, t'_e)\) where \(m = \frac{1}{2}(t_e + t'_s)\).
Finally, we proceed the same way as described in the \algTenacious{} algorithm to compute good routes given the access satellite pairs.

\section{Evaluation and Results}\label{sec:results}

In this section, we present our analysis results of an empirical performance evaluation using simulations.
First, we introduce a framework for transport layer simulations, our tool \artcp{}, which is used to evaluate the performance of the algorithms with respect to the transport layer.
We then report some detailed findings for an exemplary pair of ground stations using the Starlink constellation.
Finally, we provide some aggregated results obtained by placing ground stations on all possible locations on a grid, followed by an analysis of the computational costs of the algorithms.
All benchmarks were run on a Linux machine equipped with an Intel Core{\footnotesize\texttrademark{}} \mbox{i7-6700} CPU running at \qty{3.40}{\GHz} and \qty{32}{\giga\byte} of main memory.
Our toolchain is implemented in the Rust programming language, and \texttt{lp\_solve} is used to solve the ILP problems.

\begin{figure*}[p]
    \centering%
    \input{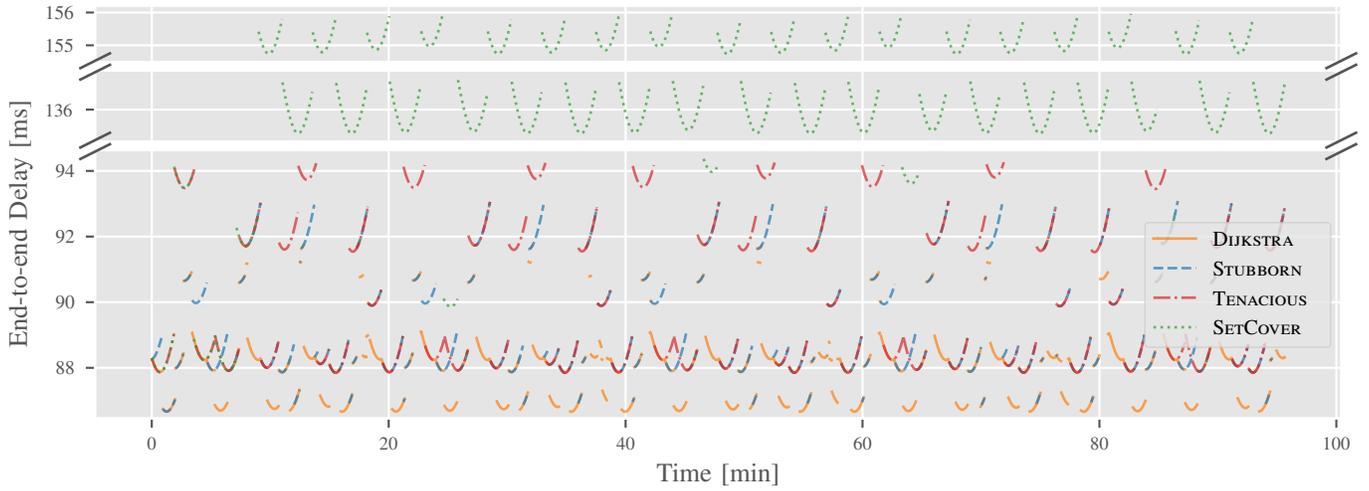}
    \caption{Comparison of end-to-end route delay and validity for the ground station pair Bariloche and Beijing on Starlink.}
    \label{fig:bariloche-beijing-delay}
\end{figure*}

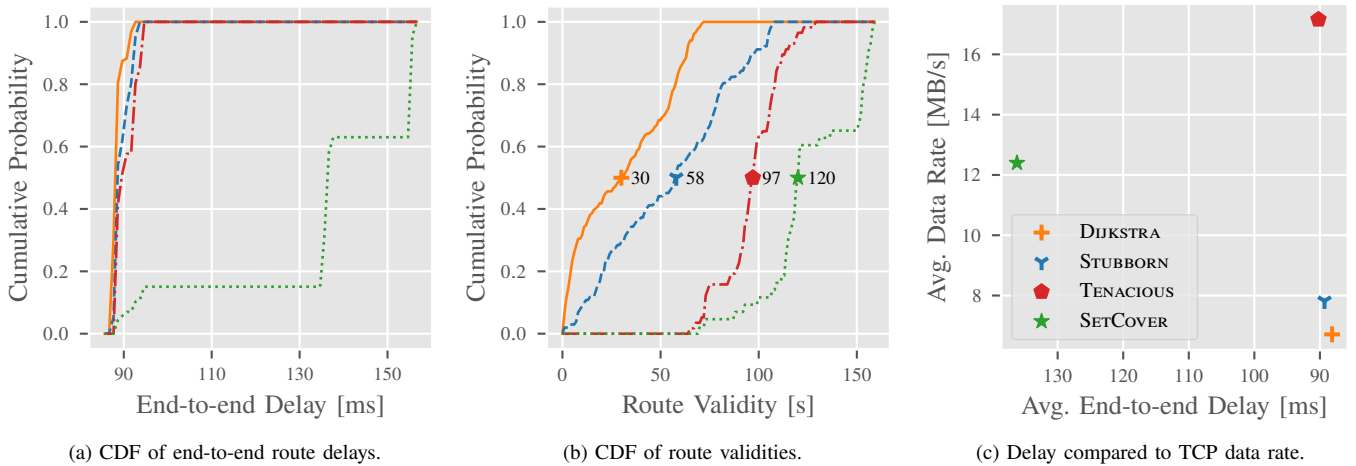
\begin{figure*}[p]
    \centering%
    \subfloat[CDF of end-to-end route delays.]{%
        \input{figures/starlink_all_bariloche_beijing_cdf_delay.pgf}
        \label{fig:bariloche-beijing-map-cdf-delay}%
    }
    \hfill%
    \subfloat[CDF of route validities.]{%
        \input{figures/starlink_all_bariloche_beijing_cdf_validity.pgf}
        \label{fig:bariloche-beijing-map-cdf-validity}%
    }
    \hfill%
    \subfloat[Delay compared to TCP data rate.]{%
        \input{figures/starlink_all_bariloche_beijing_pareto.pgf}
        \label{fig:bariloche-beijing-map-cdf-rate}%
    }
    \caption{Cumulative statistics of the four algorithms on the ground station pair Bariloche and Beijing on Starlink.}%
    \label{fig:bariloche-beijing-map-cdf}%
\end{figure*}

\begin{figure*}[p]
    \centering%
    \input{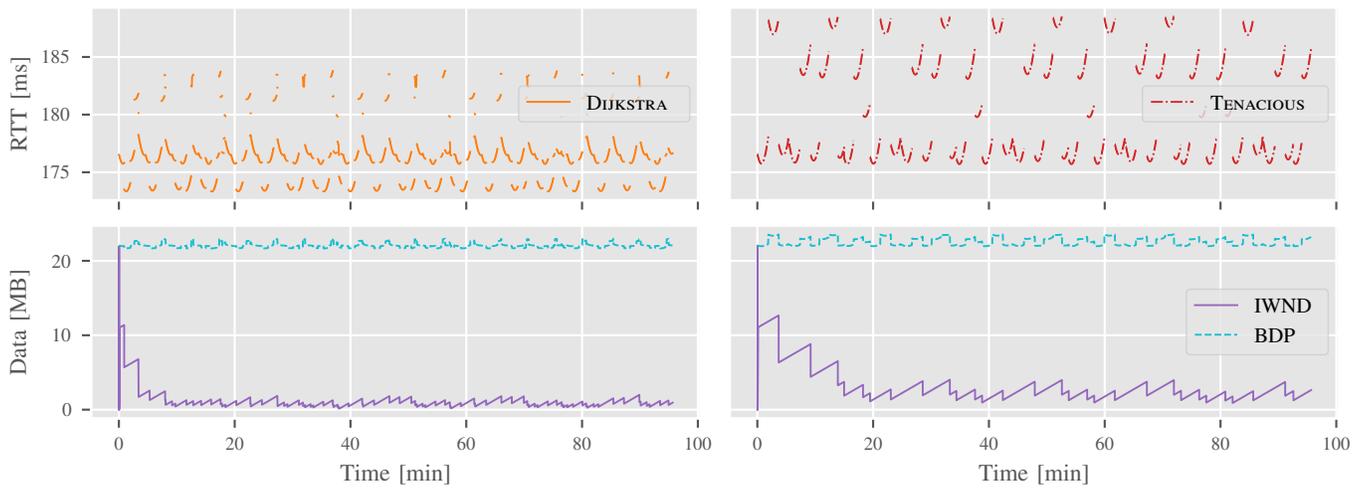}
    \caption{\artcp{} results showing the round-trip time and in-flight window (IWND) for the algorithms \algDijkstra{} and \algTenacious{}.}%
    \label{fig:bariloche-beijing-artcp}%
\end{figure*}

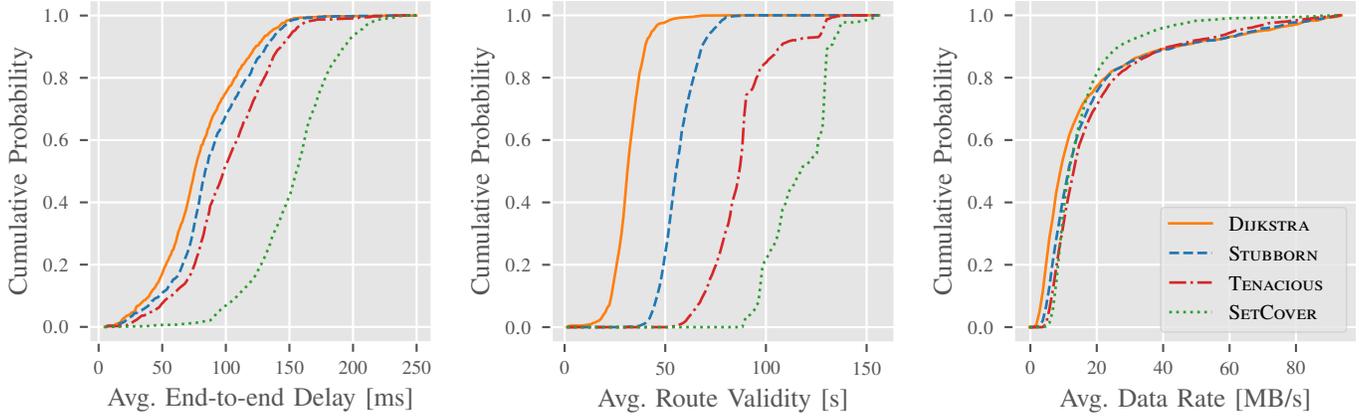
\begin{figure*}[!t]
    \centering%
    \subfloat{\input{figures/grid_avg_delay_validity_starlink_delay.pgf}}
    \hfill%
    \subfloat{\input{figures/grid_avg_delay_validity_starlink_valid.pgf}}
    \hfill%
    \subfloat{\input{figures/grid_avg_delay_validity_starlink_rate.pgf}}
    \caption{CDF plots over aggregated data for points on a latitude/longitude grid for Starlink.}%
    \label{fig:aggregated-cdf-starlink}%
\end{figure*}

\subsection{Transport Layer Performance Evaluation Framework}\label{sec:artcp}

To study the transport-layer effects of routing decisions, we have developed \artcp{}, an abstract simulation of TCP variants.
Comparing figures of delay and route validity times is a non-obvious task, as they impact the communication differently.
Instead of doing multivariate comparisons of these two parameters, we decided to use them as inputs for a transport layer simulation that works as follows:
As input, it takes both time series of RTT and reordering events~(i.e.\@ when a route change led to subsequent packets overtaking preceding ones).

During simulation, the evolution of an in-flight window is modelled, using the dynamics of well-known TCP variants (e.g.\@ Reno).
This means that we do not model individual packets via discrete event simulation, but rather assume ideal window progression under reordering events and RTT changes.
Finally, the in-flight window over time is reported, together with aggregate metrics (e.g.\@ average throughput).

In \artcp{}, we assume
\begin{enumerate*}[label=\alph*)]
    \item the sender always has packets to send, i.e.\@ infinite maximum data rate~(in practice still limited by transmission algorithms) and
    \item the receiver can always accept packets, i.e.\@ infinite receive buffer and application read rate~(no flow control effects on in-flight window are considered)
\end{enumerate*}.

\subsection{Example: Bariloche and Beijing on Starlink}

This example serves to demonstrate the different analyses we perform.
For this benchmark, we assume two ground stations in Bariloche, Argentina and Beijing, China.
For the constellation, we selected the (first) orbital shell of the initial deployment phase of SpaceX Starlink.
The reason is that Starlink is well known and one of the few already operational Walker Delta constellations.
Based on the publicly available information~\cite{FCC:SAT-MOD-20190830-00087,DBLP:conf/asms-spsc/StockFH22}, we model Starlink as a Walker Delta \(\ang{53}\colon 1584 / 72 / 39\) at \qty{550}{\km}.
We consider satellites with a minimum elevation angle of \ang{40} to be visible from the ground station.
The great-circle distance between Bariloche and Beijing is \qty{19350}{\km} and the shortest route has an average ISL hop count of~11.
Specifics aside, the results shown here are representative for the general behaviour of the algorithms, regardless of whether the ground stations are
\begin{enumerate*}[label=\alph*)]
    \item on the same or different Earth hemispheres, and
    \item rather close or rather far apart
\end{enumerate*}.

\autoref{fig:bariloche-beijing-delay} shows the dynamics of end-to-end route delays over one orbital period for the four algorithms.
It can be seen that \algDijkstra{} is indeed the algorithm with the lowest delays while the other algorithms have higher delays, \algSetcover{} almost twice as high as \algDijkstra{}.
On the other side, \algDijkstra{} induces 185 (101) route changes, \algStubborn{} 103 (61), \algTenacious{} 58 (29), and \algSetcover{} just 44 (21), where the number in parentheses indicates the number of ``bad'' route changes, i.e.\@ changes from high to low delay.
This illustrates the trade-off between minimizing delay and maximizing route validity.

The trade-off can be seen more clearly by looking at the two cumulative distribution function (CDF) plots showing the delay (\autoref{fig:bariloche-beijing-map-cdf-delay}) and validity (\autoref{fig:bariloche-beijing-map-cdf-validity}).
The median delay of \algSetcover{} is \qty{136}{\ms} while it is just below \qty{89}{\ms} for the other three algorithms.
In terms of route validity, \algSetcover{} is clearly superior to the \algDijkstra{} or \algStubborn{} approaches.

Finally, we ran the resulting routes through \artcp{}.
The average TCP data rate for all algorithms is shown in \autoref{fig:bariloche-beijing-map-cdf-rate}.
The plot shows that \algDijkstra{} and \algStubborn{} have the lowest performance.
The average data rate for \algTenacious{} is more than doubled, while still being competitive in terms of average delay.
Even that \algSetcover{} minimizes the number of route changes, we observe that the \artcp~performance falls behind that of the \algTenacious{} algorithm.
The reason is that the higher delay for \algSetcover{} prevents a good performance.

The full \artcp~comparison between \algDijkstra{} and \algTenacious{} is in \autoref{fig:bariloche-beijing-artcp}, as these provide the worst and best performance.
The plot shows that both algorithms feature rather poor performance in general, as neither of them gets close to the bandwidth-delay product (BDP).
The two key takeaways are:
\begin{enumerate*}[label=\alph*)]
    \item not just the number of route changes is important but rather how many of them are from high to low delay (i.e.\@ cause reordering events), and
    \item other, optimized TCP congestion control algorithms (e.g.\@ CUBIC) should be explored.
\end{enumerate*}

\subsection{Aggregated Grid Analysis}

The previous evaluation results were all specific to the pair of ground stations at Bariloche and Beijing.
Now, we explore how the different algorithms perform when other ground stations are chosen.
For this, we first create a grid covering the entire globe between \(-\alpha\) and \(\alpha\) degrees latitude, where \(\alpha\) is the inclination of the constellation, e.g.\@ \(\alpha = \ang{53}\) for Starlink.
In this case, we use a grid granularity of \ang{10} for longitude and \(\frac{\ang{53}}{5} = \ang{10.6}\) for latitude.
Initially, one ground station is fixed at \ang{0}\,E, \ang{0}\,N.
For the other ground station, every possible point on the grid is considered, and all algorithms are executed on that pair.
Then, the first ground station is relocated to \ang{0}\,E, \ang{10.6}\,N and again all grid points are considered for the second ground station.
This whole process is repeated until the first ground station is at \ang{0}\,E, \ang{53}\,N.
This considers all relevant combinations of two ground station positions.
Due to the symmetries of a Walker Delta constellation, it is not necessary to consider negative latitudes for the first ground station nor different longitudes than \ang{0}\,E.

In each run, the average end-to-end delay and average route validity is stored.
Each trace is also run through \artcp{} to get the average data rate.
\autoref{fig:aggregated-cdf-starlink} shows the respective CDF plots.

To evaluate the effect of the constellation topology on the results, we repeated the same analysis on two other Walker Delta constellations (\(\ang{60}\colon 779 / 41 / 5\) at \qty{500}{\km} and \(\ang{60}\colon 399 / 21 / 5\) at \qty{1000}{\km}, both with a minimum elevation angle of \ang{25} and a grid granularity of \ang{10} in each direction).
While the individual figures changed, the overall picture remained the same.

\subsection{Run Time / Computational Complexity}

\begin{table}[!t]
    \caption{Statistics of run times (in seconds) to execute the algorithms on a ground station pair for one orbital period.}%
    \label{tab:run-time-starlink}%
    \footnotesize\centering%
    \begin{tabular}{lrrrrr}
        \toprule
        Algorithm       & Min   & Q1    & Median & Q3     & Max    \\\midrule
        \algDijkstra{}  & 6.637 & 9.087 & 12.537 & 18.684 & 92.503 \\
        \algStubborn{}  & 2.104 & 2.186 & 2.243  & 2.351  & 4.829  \\
        \algTenacious{} & 2.060 & 2.110 & 2.118  & 2.127  & 2.167  \\
        \algSetcover{}  & 2.310 & 2.415 & 2.566  & 2.833  & 4.196  \\
        \bottomrule
    \end{tabular}%
\end{table}

As a final metric, we evaluate the computational cost of the four algorithms.
For every ground station pair of the Starlink grid analysis, we measured the run time of each algorithm.
The results are shown in \autoref{tab:run-time-starlink}.
The first observation is that \algStubborn{}, \algTenacious{}, and \algSetcover{} typically require about two seconds for each run while the run time of \algDijkstra{} is significantly higher.
However, this comparison is biased.
Since \algDijkstra{} should minimize the overall delay, it needs to perform a shortest-route computation every second.
This time granularity is a factor chosen by us that punishes the performance of \algDijkstra{}.
The other algorithms only sporadically have to compute shortest routes because they stick to a route longer and, in the case of \algStubborn{} and \algTenacious{}, only need to check whether the current route is still valid, i.e.\@ whether the two access satellites are still visible from the ground.
Since the median run time of \algDijkstra{} is six times higher than the other algorithms, we could increase the time granularity to six seconds to obtain a median run time that can compete with the other algorithms.
On the other side, this would increase the overall delay of \algDijkstra{} slightly.

\section{Conclusion}\label{sec:conclusion}

\enlargethispage{-1.9pt}

This paper has introduced dedicated route selection algorithms for LEO satellite networks that aim to significantly increase overall network performance while being fully compatible with standard transport layer management.

We have presented empirical studies that highlight the potential of routing algorithms optimized for stability, as opposed to the state of the art, which primarily optimizes for minimum delays.
Our evaluation results are encouraging but suggest that in addition to specialized route selection algorithms, specialized congestion control algorithms (CCAs) may also be necessary.
We are currently analysing other TCP CCAs (e.g.\@ CUBIC and BBR) and are using these findings to design a route selection algorithm that maximizes network performance.
Using our simulator \artcp{}, we are making first steps in co-designing route selection and transport control approaches for in-orbit communication.
This can be combined with performance-enhancing proxies and related approaches (e.g.\@ Yuan et~al.~\cite{DBLP:conf/hotnets/YuanZSWW22}) to further improve connection stability.

This work has so far focused on Walker Delta constellations such as Starlink, but we will extend this work to other types of constellations.
Further investigation will include packet-level simulation and/or real-world measurements to prove the validity of the results in less idealized environments.

\section*{Acknowledgements}

This project has received funding from the European Union's Horizon 2020 research and innovation programme under the Marie Skłodowska-Curie grant agreement \href{https://cordis.europa.eu/project/id/101008233}{No 101008233} -- \href{https://mission-project.eu}{MISSION}, see \url{https://mission-project.eu}, and by DFG grant 389792660 as part of TRR~248 -- CPEC, see \url{https://perspicuous-computing.science}.

\renewcommand*{\bibfont}{\footnotesize}
\printbibliography

\end{document}

%% file: figures/starlink_all_bariloche_beijing_cdf_delay.pgf
\begingroup%
\makeatletter%
\begin{pgfpicture}%
\pgfpathrectangle{\pgfpointorigin}{\pgfqpoint{2.300000in}{2.300000in}}%
\pgfusepath{use as bounding box, clip}%
\begin{pgfscope}%
\pgfsetbuttcap%
\pgfsetmiterjoin%
\definecolor{currentfill}{rgb}{1.000000,1.000000,1.000000}%
\pgfsetfillcolor{currentfill}%
\pgfsetlinewidth{0.000000pt}%
\definecolor{currentstroke}{rgb}{0.500000,0.500000,0.500000}%
\pgfsetstrokecolor{currentstroke}%
\pgfsetdash{}{0pt}%
\pgfpathmoveto{\pgfqpoint{0.000000in}{0.000000in}}%
\pgfpathlineto{\pgfqpoint{2.300000in}{0.000000in}}%
\pgfpathlineto{\pgfqpoint{2.300000in}{2.300000in}}%
\pgfpathlineto{\pgfqpoint{0.000000in}{2.300000in}}%
\pgfpathlineto{\pgfqpoint{0.000000in}{0.000000in}}%
\pgfpathclose%
\pgfusepath{fill}%
\end{pgfscope}%
\begin{pgfscope}%
\pgfsetbuttcap%
\pgfsetmiterjoin%
\definecolor{currentfill}{rgb}{0.898039,0.898039,0.898039}%
\pgfsetfillcolor{currentfill}%
\pgfsetlinewidth{0.000000pt}%
\definecolor{currentstroke}{rgb}{0.000000,0.000000,0.000000}%
\pgfsetstrokecolor{currentstroke}%
\pgfsetstrokeopacity{0.000000}%
\pgfsetdash{}{0pt}%
\pgfpathmoveto{\pgfqpoint{0.461616in}{0.431659in}}%
\pgfpathlineto{\pgfqpoint{2.258330in}{0.431659in}}%
\pgfpathlineto{\pgfqpoint{2.258330in}{2.228372in}}%
\pgfpathlineto{\pgfqpoint{0.461616in}{2.228372in}}%
\pgfpathlineto{\pgfqpoint{0.461616in}{0.431659in}}%
\pgfpathclose%
\pgfusepath{fill}%
\end{pgfscope}%
\begin{pgfscope}%
\pgfpathrectangle{\pgfqpoint{0.461616in}{0.431659in}}{\pgfqpoint{1.796714in}{1.796714in}}%
\pgfusepath{clip}%
\pgfsetrectcap%
\pgfsetroundjoin%
\pgfsetlinewidth{0.803000pt}%
\definecolor{currentstroke}{rgb}{1.000000,1.000000,1.000000}%
\pgfsetstrokecolor{currentstroke}%
\pgfsetdash{}{0pt}%
\pgfpathmoveto{\pgfqpoint{0.643107in}{0.431659in}}%
\pgfpathlineto{\pgfqpoint{0.643107in}{2.228372in}}%
\pgfusepath{stroke}%
\end{pgfscope}%
\begin{pgfscope}%
\pgfsetbuttcap%
\pgfsetroundjoin%
\definecolor{currentfill}{rgb}{0.333333,0.333333,0.333333}%
\pgfsetfillcolor{currentfill}%
\pgfsetlinewidth{0.803000pt}%
\definecolor{currentstroke}{rgb}{0.333333,0.333333,0.333333}%
\pgfsetstrokecolor{currentstroke}%
\pgfsetdash{}{0pt}%
\pgfsys@defobject{currentmarker}{\pgfqpoint{0.000000in}{-0.048611in}}{\pgfqpoint{0.000000in}{0.000000in}}{%
\pgfpathmoveto{\pgfqpoint{0.000000in}{0.000000in}}%
\pgfpathlineto{\pgfqpoint{0.000000in}{-0.048611in}}%
\pgfusepath{stroke,fill}%
}%
\begin{pgfscope}%
\pgfsys@transformshift{0.643107in}{0.431659in}%
\pgfsys@useobject{currentmarker}{}%
\end{pgfscope}%
\end{pgfscope}%
\begin{pgfscope}%
\definecolor{textcolor}{rgb}{0.333333,0.333333,0.333333}%
\pgfsetstrokecolor{textcolor}%
\pgfsetfillcolor{textcolor}%
\pgftext[x=0.643107in,y=0.334436in,,top]{\color{textcolor}\rmfamily\fontsize{7.000000}{8.400000}\selectfont 90}%
\end{pgfscope}%
\begin{pgfscope}%
\pgfpathrectangle{\pgfqpoint{0.461616in}{0.431659in}}{\pgfqpoint{1.796714in}{1.796714in}}%
\pgfusepath{clip}%
\pgfsetrectcap%
\pgfsetroundjoin%
\pgfsetlinewidth{0.803000pt}%
\definecolor{currentstroke}{rgb}{1.000000,1.000000,1.000000}%
\pgfsetstrokecolor{currentstroke}%
\pgfsetdash{}{0pt}%
\pgfpathmoveto{\pgfqpoint{1.103213in}{0.431659in}}%
\pgfpathlineto{\pgfqpoint{1.103213in}{2.228372in}}%
\pgfusepath{stroke}%
\end{pgfscope}%
\begin{pgfscope}%
\pgfsetbuttcap%
\pgfsetroundjoin%
\definecolor{currentfill}{rgb}{0.333333,0.333333,0.333333}%
\pgfsetfillcolor{currentfill}%
\pgfsetlinewidth{0.803000pt}%
\definecolor{currentstroke}{rgb}{0.333333,0.333333,0.333333}%
\pgfsetstrokecolor{currentstroke}%
\pgfsetdash{}{0pt}%
\pgfsys@defobject{currentmarker}{\pgfqpoint{0.000000in}{-0.048611in}}{\pgfqpoint{0.000000in}{0.000000in}}{%
\pgfpathmoveto{\pgfqpoint{0.000000in}{0.000000in}}%
\pgfpathlineto{\pgfqpoint{0.000000in}{-0.048611in}}%
\pgfusepath{stroke,fill}%
}%
\begin{pgfscope}%
\pgfsys@transformshift{1.103213in}{0.431659in}%
\pgfsys@useobject{currentmarker}{}%
\end{pgfscope}%
\end{pgfscope}%
\begin{pgfscope}%
\definecolor{textcolor}{rgb}{0.333333,0.333333,0.333333}%
\pgfsetstrokecolor{textcolor}%
\pgfsetfillcolor{textcolor}%
\pgftext[x=1.103213in,y=0.334436in,,top]{\color{textcolor}\rmfamily\fontsize{7.000000}{8.400000}\selectfont 110}%
\end{pgfscope}%
\begin{pgfscope}%
\pgfpathrectangle{\pgfqpoint{0.461616in}{0.431659in}}{\pgfqpoint{1.796714in}{1.796714in}}%
\pgfusepath{clip}%
\pgfsetrectcap%
\pgfsetroundjoin%
\pgfsetlinewidth{0.803000pt}%
\definecolor{currentstroke}{rgb}{1.000000,1.000000,1.000000}%
\pgfsetstrokecolor{currentstroke}%
\pgfsetdash{}{0pt}%
\pgfpathmoveto{\pgfqpoint{1.563319in}{0.431659in}}%
\pgfpathlineto{\pgfqpoint{1.563319in}{2.228372in}}%
\pgfusepath{stroke}%
\end{pgfscope}%
\begin{pgfscope}%
\pgfsetbuttcap%
\pgfsetroundjoin%
\definecolor{currentfill}{rgb}{0.333333,0.333333,0.333333}%
\pgfsetfillcolor{currentfill}%
\pgfsetlinewidth{0.803000pt}%
\definecolor{currentstroke}{rgb}{0.333333,0.333333,0.333333}%
\pgfsetstrokecolor{currentstroke}%
\pgfsetdash{}{0pt}%
\pgfsys@defobject{currentmarker}{\pgfqpoint{0.000000in}{-0.048611in}}{\pgfqpoint{0.000000in}{0.000000in}}{%
\pgfpathmoveto{\pgfqpoint{0.000000in}{0.000000in}}%
\pgfpathlineto{\pgfqpoint{0.000000in}{-0.048611in}}%
\pgfusepath{stroke,fill}%
}%
\begin{pgfscope}%
\pgfsys@transformshift{1.563319in}{0.431659in}%
\pgfsys@useobject{currentmarker}{}%
\end{pgfscope}%
\end{pgfscope}%
\begin{pgfscope}%
\definecolor{textcolor}{rgb}{0.333333,0.333333,0.333333}%
\pgfsetstrokecolor{textcolor}%
\pgfsetfillcolor{textcolor}%
\pgftext[x=1.563319in,y=0.334436in,,top]{\color{textcolor}\rmfamily\fontsize{7.000000}{8.400000}\selectfont 130}%
\end{pgfscope}%
\begin{pgfscope}%
\pgfpathrectangle{\pgfqpoint{0.461616in}{0.431659in}}{\pgfqpoint{1.796714in}{1.796714in}}%
\pgfusepath{clip}%
\pgfsetrectcap%
\pgfsetroundjoin%
\pgfsetlinewidth{0.803000pt}%
\definecolor{currentstroke}{rgb}{1.000000,1.000000,1.000000}%
\pgfsetstrokecolor{currentstroke}%
\pgfsetdash{}{0pt}%
\pgfpathmoveto{\pgfqpoint{2.023425in}{0.431659in}}%
\pgfpathlineto{\pgfqpoint{2.023425in}{2.228372in}}%
\pgfusepath{stroke}%
\end{pgfscope}%
\begin{pgfscope}%
\pgfsetbuttcap%
\pgfsetroundjoin%
\definecolor{currentfill}{rgb}{0.333333,0.333333,0.333333}%
\pgfsetfillcolor{currentfill}%
\pgfsetlinewidth{0.803000pt}%
\definecolor{currentstroke}{rgb}{0.333333,0.333333,0.333333}%
\pgfsetstrokecolor{currentstroke}%
\pgfsetdash{}{0pt}%
\pgfsys@defobject{currentmarker}{\pgfqpoint{0.000000in}{-0.048611in}}{\pgfqpoint{0.000000in}{0.000000in}}{%
\pgfpathmoveto{\pgfqpoint{0.000000in}{0.000000in}}%
\pgfpathlineto{\pgfqpoint{0.000000in}{-0.048611in}}%
\pgfusepath{stroke,fill}%
}%
\begin{pgfscope}%
\pgfsys@transformshift{2.023425in}{0.431659in}%
\pgfsys@useobject{currentmarker}{}%
\end{pgfscope}%
\end{pgfscope}%
\begin{pgfscope}%
\definecolor{textcolor}{rgb}{0.333333,0.333333,0.333333}%
\pgfsetstrokecolor{textcolor}%
\pgfsetfillcolor{textcolor}%
\pgftext[x=2.023425in,y=0.334436in,,top]{\color{textcolor}\rmfamily\fontsize{7.000000}{8.400000}\selectfont 150}%
\end{pgfscope}%
\begin{pgfscope}%
\definecolor{textcolor}{rgb}{0.333333,0.333333,0.333333}%
\pgfsetstrokecolor{textcolor}%
\pgfsetfillcolor{textcolor}%
\pgftext[x=1.359973in,y=0.192461in,,top]{\color{textcolor}\rmfamily\fontsize{10.000000}{12.000000}\selectfont End-to-end Delay [ms]}%
\end{pgfscope}%
\begin{pgfscope}%
\pgfpathrectangle{\pgfqpoint{0.461616in}{0.431659in}}{\pgfqpoint{1.796714in}{1.796714in}}%
\pgfusepath{clip}%
\pgfsetrectcap%
\pgfsetroundjoin%
\pgfsetlinewidth{0.803000pt}%
\definecolor{currentstroke}{rgb}{1.000000,1.000000,1.000000}%
\pgfsetstrokecolor{currentstroke}%
\pgfsetdash{}{0pt}%
\pgfpathmoveto{\pgfqpoint{0.461616in}{0.513327in}}%
\pgfpathlineto{\pgfqpoint{2.258330in}{0.513327in}}%
\pgfusepath{stroke}%
\end{pgfscope}%
\begin{pgfscope}%
\pgfsetbuttcap%
\pgfsetroundjoin%
\definecolor{currentfill}{rgb}{0.333333,0.333333,0.333333}%
\pgfsetfillcolor{currentfill}%
\pgfsetlinewidth{0.803000pt}%
\definecolor{currentstroke}{rgb}{0.333333,0.333333,0.333333}%
\pgfsetstrokecolor{currentstroke}%
\pgfsetdash{}{0pt}%
\pgfsys@defobject{currentmarker}{\pgfqpoint{-0.048611in}{0.000000in}}{\pgfqpoint{-0.000000in}{0.000000in}}{%
\pgfpathmoveto{\pgfqpoint{-0.000000in}{0.000000in}}%
\pgfpathlineto{\pgfqpoint{-0.048611in}{0.000000in}}%
\pgfusepath{stroke,fill}%
}%
\begin{pgfscope}%
\pgfsys@transformshift{0.461616in}{0.513327in}%
\pgfsys@useobject{currentmarker}{}%
\end{pgfscope}%
\end{pgfscope}%
\begin{pgfscope}%
\definecolor{textcolor}{rgb}{0.333333,0.333333,0.333333}%
\pgfsetstrokecolor{textcolor}%
\pgfsetfillcolor{textcolor}%
\pgftext[x=0.220682in, y=0.479570in, left, base]{\color{textcolor}\rmfamily\fontsize{7.000000}{8.400000}\selectfont \(\displaystyle {0.0}\)}%
\end{pgfscope}%
\begin{pgfscope}%
\pgfpathrectangle{\pgfqpoint{0.461616in}{0.431659in}}{\pgfqpoint{1.796714in}{1.796714in}}%
\pgfusepath{clip}%
\pgfsetrectcap%
\pgfsetroundjoin%
\pgfsetlinewidth{0.803000pt}%
\definecolor{currentstroke}{rgb}{1.000000,1.000000,1.000000}%
\pgfsetstrokecolor{currentstroke}%
\pgfsetdash{}{0pt}%
\pgfpathmoveto{\pgfqpoint{0.461616in}{0.840003in}}%
\pgfpathlineto{\pgfqpoint{2.258330in}{0.840003in}}%
\pgfusepath{stroke}%
\end{pgfscope}%
\begin{pgfscope}%
\pgfsetbuttcap%
\pgfsetroundjoin%
\definecolor{currentfill}{rgb}{0.333333,0.333333,0.333333}%
\pgfsetfillcolor{currentfill}%
\pgfsetlinewidth{0.803000pt}%
\definecolor{currentstroke}{rgb}{0.333333,0.333333,0.333333}%
\pgfsetstrokecolor{currentstroke}%
\pgfsetdash{}{0pt}%
\pgfsys@defobject{currentmarker}{\pgfqpoint{-0.048611in}{0.000000in}}{\pgfqpoint{-0.000000in}{0.000000in}}{%
\pgfpathmoveto{\pgfqpoint{-0.000000in}{0.000000in}}%
\pgfpathlineto{\pgfqpoint{-0.048611in}{0.000000in}}%
\pgfusepath{stroke,fill}%
}%
\begin{pgfscope}%
\pgfsys@transformshift{0.461616in}{0.840003in}%
\pgfsys@useobject{currentmarker}{}%
\end{pgfscope}%
\end{pgfscope}%
\begin{pgfscope}%
\definecolor{textcolor}{rgb}{0.333333,0.333333,0.333333}%
\pgfsetstrokecolor{textcolor}%
\pgfsetfillcolor{textcolor}%
\pgftext[x=0.220682in, y=0.806245in, left, base]{\color{textcolor}\rmfamily\fontsize{7.000000}{8.400000}\selectfont \(\displaystyle {0.2}\)}%
\end{pgfscope}%
\begin{pgfscope}%
\pgfpathrectangle{\pgfqpoint{0.461616in}{0.431659in}}{\pgfqpoint{1.796714in}{1.796714in}}%
\pgfusepath{clip}%
\pgfsetrectcap%
\pgfsetroundjoin%
\pgfsetlinewidth{0.803000pt}%
\definecolor{currentstroke}{rgb}{1.000000,1.000000,1.000000}%
\pgfsetstrokecolor{currentstroke}%
\pgfsetdash{}{0pt}%
\pgfpathmoveto{\pgfqpoint{0.461616in}{1.166678in}}%
\pgfpathlineto{\pgfqpoint{2.258330in}{1.166678in}}%
\pgfusepath{stroke}%
\end{pgfscope}%
\begin{pgfscope}%
\pgfsetbuttcap%
\pgfsetroundjoin%
\definecolor{currentfill}{rgb}{0.333333,0.333333,0.333333}%
\pgfsetfillcolor{currentfill}%
\pgfsetlinewidth{0.803000pt}%
\definecolor{currentstroke}{rgb}{0.333333,0.333333,0.333333}%
\pgfsetstrokecolor{currentstroke}%
\pgfsetdash{}{0pt}%
\pgfsys@defobject{currentmarker}{\pgfqpoint{-0.048611in}{0.000000in}}{\pgfqpoint{-0.000000in}{0.000000in}}{%
\pgfpathmoveto{\pgfqpoint{-0.000000in}{0.000000in}}%
\pgfpathlineto{\pgfqpoint{-0.048611in}{0.000000in}}%
\pgfusepath{stroke,fill}%
}%
\begin{pgfscope}%
\pgfsys@transformshift{0.461616in}{1.166678in}%
\pgfsys@useobject{currentmarker}{}%
\end{pgfscope}%
\end{pgfscope}%
\begin{pgfscope}%
\definecolor{textcolor}{rgb}{0.333333,0.333333,0.333333}%
\pgfsetstrokecolor{textcolor}%
\pgfsetfillcolor{textcolor}%
\pgftext[x=0.220682in, y=1.132920in, left, base]{\color{textcolor}\rmfamily\fontsize{7.000000}{8.400000}\selectfont \(\displaystyle {0.4}\)}%
\end{pgfscope}%
\begin{pgfscope}%
\pgfpathrectangle{\pgfqpoint{0.461616in}{0.431659in}}{\pgfqpoint{1.796714in}{1.796714in}}%
\pgfusepath{clip}%
\pgfsetrectcap%
\pgfsetroundjoin%
\pgfsetlinewidth{0.803000pt}%
\definecolor{currentstroke}{rgb}{1.000000,1.000000,1.000000}%
\pgfsetstrokecolor{currentstroke}%
\pgfsetdash{}{0pt}%
\pgfpathmoveto{\pgfqpoint{0.461616in}{1.493353in}}%
\pgfpathlineto{\pgfqpoint{2.258330in}{1.493353in}}%
\pgfusepath{stroke}%
\end{pgfscope}%
\begin{pgfscope}%
\pgfsetbuttcap%
\pgfsetroundjoin%
\definecolor{currentfill}{rgb}{0.333333,0.333333,0.333333}%
\pgfsetfillcolor{currentfill}%
\pgfsetlinewidth{0.803000pt}%
\definecolor{currentstroke}{rgb}{0.333333,0.333333,0.333333}%
\pgfsetstrokecolor{currentstroke}%
\pgfsetdash{}{0pt}%
\pgfsys@defobject{currentmarker}{\pgfqpoint{-0.048611in}{0.000000in}}{\pgfqpoint{-0.000000in}{0.000000in}}{%
\pgfpathmoveto{\pgfqpoint{-0.000000in}{0.000000in}}%
\pgfpathlineto{\pgfqpoint{-0.048611in}{0.000000in}}%
\pgfusepath{stroke,fill}%
}%
\begin{pgfscope}%
\pgfsys@transformshift{0.461616in}{1.493353in}%
\pgfsys@useobject{currentmarker}{}%
\end{pgfscope}%
\end{pgfscope}%
\begin{pgfscope}%
\definecolor{textcolor}{rgb}{0.333333,0.333333,0.333333}%
\pgfsetstrokecolor{textcolor}%
\pgfsetfillcolor{textcolor}%
\pgftext[x=0.220682in, y=1.459595in, left, base]{\color{textcolor}\rmfamily\fontsize{7.000000}{8.400000}\selectfont \(\displaystyle {0.6}\)}%
\end{pgfscope}%
\begin{pgfscope}%
\pgfpathrectangle{\pgfqpoint{0.461616in}{0.431659in}}{\pgfqpoint{1.796714in}{1.796714in}}%
\pgfusepath{clip}%
\pgfsetrectcap%
\pgfsetroundjoin%
\pgfsetlinewidth{0.803000pt}%
\definecolor{currentstroke}{rgb}{1.000000,1.000000,1.000000}%
\pgfsetstrokecolor{currentstroke}%
\pgfsetdash{}{0pt}%
\pgfpathmoveto{\pgfqpoint{0.461616in}{1.820028in}}%
\pgfpathlineto{\pgfqpoint{2.258330in}{1.820028in}}%
\pgfusepath{stroke}%
\end{pgfscope}%
\begin{pgfscope}%
\pgfsetbuttcap%
\pgfsetroundjoin%
\definecolor{currentfill}{rgb}{0.333333,0.333333,0.333333}%
\pgfsetfillcolor{currentfill}%
\pgfsetlinewidth{0.803000pt}%
\definecolor{currentstroke}{rgb}{0.333333,0.333333,0.333333}%
\pgfsetstrokecolor{currentstroke}%
\pgfsetdash{}{0pt}%
\pgfsys@defobject{currentmarker}{\pgfqpoint{-0.048611in}{0.000000in}}{\pgfqpoint{-0.000000in}{0.000000in}}{%
\pgfpathmoveto{\pgfqpoint{-0.000000in}{0.000000in}}%
\pgfpathlineto{\pgfqpoint{-0.048611in}{0.000000in}}%
\pgfusepath{stroke,fill}%
}%
\begin{pgfscope}%
\pgfsys@transformshift{0.461616in}{1.820028in}%
\pgfsys@useobject{currentmarker}{}%
\end{pgfscope}%
\end{pgfscope}%
\begin{pgfscope}%
\definecolor{textcolor}{rgb}{0.333333,0.333333,0.333333}%
\pgfsetstrokecolor{textcolor}%
\pgfsetfillcolor{textcolor}%
\pgftext[x=0.220682in, y=1.786271in, left, base]{\color{textcolor}\rmfamily\fontsize{7.000000}{8.400000}\selectfont \(\displaystyle {0.8}\)}%
\end{pgfscope}%
\begin{pgfscope}%
\pgfpathrectangle{\pgfqpoint{0.461616in}{0.431659in}}{\pgfqpoint{1.796714in}{1.796714in}}%
\pgfusepath{clip}%
\pgfsetrectcap%
\pgfsetroundjoin%
\pgfsetlinewidth{0.803000pt}%
\definecolor{currentstroke}{rgb}{1.000000,1.000000,1.000000}%
\pgfsetstrokecolor{currentstroke}%
\pgfsetdash{}{0pt}%
\pgfpathmoveto{\pgfqpoint{0.461616in}{2.146703in}}%
\pgfpathlineto{\pgfqpoint{2.258330in}{2.146703in}}%
\pgfusepath{stroke}%
\end{pgfscope}%
\begin{pgfscope}%
\pgfsetbuttcap%
\pgfsetroundjoin%
\definecolor{currentfill}{rgb}{0.333333,0.333333,0.333333}%
\pgfsetfillcolor{currentfill}%
\pgfsetlinewidth{0.803000pt}%
\definecolor{currentstroke}{rgb}{0.333333,0.333333,0.333333}%
\pgfsetstrokecolor{currentstroke}%
\pgfsetdash{}{0pt}%
\pgfsys@defobject{currentmarker}{\pgfqpoint{-0.048611in}{0.000000in}}{\pgfqpoint{-0.000000in}{0.000000in}}{%
\pgfpathmoveto{\pgfqpoint{-0.000000in}{0.000000in}}%
\pgfpathlineto{\pgfqpoint{-0.048611in}{0.000000in}}%
\pgfusepath{stroke,fill}%
}%
\begin{pgfscope}%
\pgfsys@transformshift{0.461616in}{2.146703in}%
\pgfsys@useobject{currentmarker}{}%
\end{pgfscope}%
\end{pgfscope}%
\begin{pgfscope}%
\definecolor{textcolor}{rgb}{0.333333,0.333333,0.333333}%
\pgfsetstrokecolor{textcolor}%
\pgfsetfillcolor{textcolor}%
\pgftext[x=0.220682in, y=2.112946in, left, base]{\color{textcolor}\rmfamily\fontsize{7.000000}{8.400000}\selectfont \(\displaystyle {1.0}\)}%
\end{pgfscope}%
\begin{pgfscope}%
\definecolor{textcolor}{rgb}{0.333333,0.333333,0.333333}%
\pgfsetstrokecolor{textcolor}%
\pgfsetfillcolor{textcolor}%
\pgftext[x=0.165127in,y=1.330015in,,bottom,rotate=90.000000]{\color{textcolor}\rmfamily\fontsize{10.000000}{12.000000}\selectfont Cumulative Probability}%
\end{pgfscope}%
\begin{pgfscope}%
\pgfpathrectangle{\pgfqpoint{0.461616in}{0.431659in}}{\pgfqpoint{1.796714in}{1.796714in}}%
\pgfusepath{clip}%
\pgfsetrectcap%
\pgfsetroundjoin%
\pgfsetlinewidth{1.003750pt}%
\definecolor{currentstroke}{rgb}{1.000000,0.498039,0.054902}%
\pgfsetstrokecolor{currentstroke}%
\pgfsetdash{}{0pt}%
\pgfpathmoveto{\pgfqpoint{0.543285in}{0.513327in}}%
\pgfpathlineto{\pgfqpoint{0.566290in}{0.513327in}}%
\pgfpathlineto{\pgfqpoint{0.589296in}{1.003426in}}%
\pgfpathlineto{\pgfqpoint{0.612301in}{1.826233in}}%
\pgfpathlineto{\pgfqpoint{0.635306in}{1.942354in}}%
\pgfpathlineto{\pgfqpoint{0.658312in}{1.954592in}}%
\pgfpathlineto{\pgfqpoint{0.681317in}{2.094051in}}%
\pgfpathlineto{\pgfqpoint{0.704322in}{2.146703in}}%
\pgfpathlineto{\pgfqpoint{0.727328in}{2.146703in}}%
\pgfpathlineto{\pgfqpoint{0.750333in}{2.146703in}}%
\pgfpathlineto{\pgfqpoint{0.773338in}{2.146703in}}%
\pgfpathlineto{\pgfqpoint{0.796343in}{2.146703in}}%
\pgfpathlineto{\pgfqpoint{0.819349in}{2.146703in}}%
\pgfpathlineto{\pgfqpoint{0.842354in}{2.146703in}}%
\pgfpathlineto{\pgfqpoint{0.865359in}{2.146703in}}%
\pgfpathlineto{\pgfqpoint{0.888365in}{2.146703in}}%
\pgfpathlineto{\pgfqpoint{0.911370in}{2.146703in}}%
\pgfpathlineto{\pgfqpoint{0.934375in}{2.146703in}}%
\pgfpathlineto{\pgfqpoint{0.957381in}{2.146703in}}%
\pgfpathlineto{\pgfqpoint{0.980386in}{2.146703in}}%
\pgfpathlineto{\pgfqpoint{1.003391in}{2.146703in}}%
\pgfpathlineto{\pgfqpoint{1.026396in}{2.146703in}}%
\pgfpathlineto{\pgfqpoint{1.049402in}{2.146703in}}%
\pgfpathlineto{\pgfqpoint{1.072407in}{2.146703in}}%
\pgfpathlineto{\pgfqpoint{1.095412in}{2.146703in}}%
\pgfpathlineto{\pgfqpoint{1.118418in}{2.146703in}}%
\pgfpathlineto{\pgfqpoint{1.141423in}{2.146703in}}%
\pgfpathlineto{\pgfqpoint{1.164428in}{2.146703in}}%
\pgfpathlineto{\pgfqpoint{1.187433in}{2.146703in}}%
\pgfpathlineto{\pgfqpoint{1.210439in}{2.146703in}}%
\pgfpathlineto{\pgfqpoint{1.233444in}{2.146703in}}%
\pgfpathlineto{\pgfqpoint{1.256449in}{2.146703in}}%
\pgfpathlineto{\pgfqpoint{1.279455in}{2.146703in}}%
\pgfpathlineto{\pgfqpoint{1.302460in}{2.146703in}}%
\pgfpathlineto{\pgfqpoint{1.325465in}{2.146703in}}%
\pgfpathlineto{\pgfqpoint{1.348471in}{2.146703in}}%
\pgfpathlineto{\pgfqpoint{1.371476in}{2.146703in}}%
\pgfpathlineto{\pgfqpoint{1.394481in}{2.146703in}}%
\pgfpathlineto{\pgfqpoint{1.417486in}{2.146703in}}%
\pgfpathlineto{\pgfqpoint{1.440492in}{2.146703in}}%
\pgfpathlineto{\pgfqpoint{1.463497in}{2.146703in}}%
\pgfpathlineto{\pgfqpoint{1.486502in}{2.146703in}}%
\pgfpathlineto{\pgfqpoint{1.509508in}{2.146703in}}%
\pgfpathlineto{\pgfqpoint{1.532513in}{2.146703in}}%
\pgfpathlineto{\pgfqpoint{1.555518in}{2.146703in}}%
\pgfpathlineto{\pgfqpoint{1.578524in}{2.146703in}}%
\pgfpathlineto{\pgfqpoint{1.601529in}{2.146703in}}%
\pgfpathlineto{\pgfqpoint{1.624534in}{2.146703in}}%
\pgfpathlineto{\pgfqpoint{1.647539in}{2.146703in}}%
\pgfpathlineto{\pgfqpoint{1.670545in}{2.146703in}}%
\pgfpathlineto{\pgfqpoint{1.693550in}{2.146703in}}%
\pgfpathlineto{\pgfqpoint{1.716555in}{2.146703in}}%
\pgfpathlineto{\pgfqpoint{1.739561in}{2.146703in}}%
\pgfpathlineto{\pgfqpoint{1.762566in}{2.146703in}}%
\pgfpathlineto{\pgfqpoint{1.785571in}{2.146703in}}%
\pgfpathlineto{\pgfqpoint{1.808576in}{2.146703in}}%
\pgfpathlineto{\pgfqpoint{1.831582in}{2.146703in}}%
\pgfpathlineto{\pgfqpoint{1.854587in}{2.146703in}}%
\pgfpathlineto{\pgfqpoint{1.877592in}{2.146703in}}%
\pgfpathlineto{\pgfqpoint{1.900598in}{2.146703in}}%
\pgfpathlineto{\pgfqpoint{1.923603in}{2.146703in}}%
\pgfpathlineto{\pgfqpoint{1.946608in}{2.146703in}}%
\pgfpathlineto{\pgfqpoint{1.969614in}{2.146703in}}%
\pgfpathlineto{\pgfqpoint{1.992619in}{2.146703in}}%
\pgfpathlineto{\pgfqpoint{2.015624in}{2.146703in}}%
\pgfpathlineto{\pgfqpoint{2.038629in}{2.146703in}}%
\pgfpathlineto{\pgfqpoint{2.061635in}{2.146703in}}%
\pgfpathlineto{\pgfqpoint{2.084640in}{2.146703in}}%
\pgfpathlineto{\pgfqpoint{2.107645in}{2.146703in}}%
\pgfpathlineto{\pgfqpoint{2.130651in}{2.146703in}}%
\pgfpathlineto{\pgfqpoint{2.153656in}{2.146703in}}%
\pgfpathlineto{\pgfqpoint{2.176661in}{2.146703in}}%
\pgfusepath{stroke}%
\end{pgfscope}%
\begin{pgfscope}%
\pgfpathrectangle{\pgfqpoint{0.461616in}{0.431659in}}{\pgfqpoint{1.796714in}{1.796714in}}%
\pgfusepath{clip}%
\pgfsetbuttcap%
\pgfsetroundjoin%
\pgfsetlinewidth{1.003750pt}%
\definecolor{currentstroke}{rgb}{0.121569,0.466667,0.705882}%
\pgfsetstrokecolor{currentstroke}%
\pgfsetdash{{3.700000pt}{1.600000pt}}{0.000000pt}%
\pgfpathmoveto{\pgfqpoint{0.543285in}{0.513327in}}%
\pgfpathlineto{\pgfqpoint{0.566290in}{0.513327in}}%
\pgfpathlineto{\pgfqpoint{0.589296in}{0.598710in}}%
\pgfpathlineto{\pgfqpoint{0.612301in}{1.396472in}}%
\pgfpathlineto{\pgfqpoint{0.635306in}{1.524831in}}%
\pgfpathlineto{\pgfqpoint{0.658312in}{1.707835in}}%
\pgfpathlineto{\pgfqpoint{0.681317in}{1.822533in}}%
\pgfpathlineto{\pgfqpoint{0.704322in}{2.069574in}}%
\pgfpathlineto{\pgfqpoint{0.727328in}{2.146703in}}%
\pgfpathlineto{\pgfqpoint{0.750333in}{2.146703in}}%
\pgfpathlineto{\pgfqpoint{0.773338in}{2.146703in}}%
\pgfpathlineto{\pgfqpoint{0.796343in}{2.146703in}}%
\pgfpathlineto{\pgfqpoint{0.819349in}{2.146703in}}%
\pgfpathlineto{\pgfqpoint{0.842354in}{2.146703in}}%
\pgfpathlineto{\pgfqpoint{0.865359in}{2.146703in}}%
\pgfpathlineto{\pgfqpoint{0.888365in}{2.146703in}}%
\pgfpathlineto{\pgfqpoint{0.911370in}{2.146703in}}%
\pgfpathlineto{\pgfqpoint{0.934375in}{2.146703in}}%
\pgfpathlineto{\pgfqpoint{0.957381in}{2.146703in}}%
\pgfpathlineto{\pgfqpoint{0.980386in}{2.146703in}}%
\pgfpathlineto{\pgfqpoint{1.003391in}{2.146703in}}%
\pgfpathlineto{\pgfqpoint{1.026396in}{2.146703in}}%
\pgfpathlineto{\pgfqpoint{1.049402in}{2.146703in}}%
\pgfpathlineto{\pgfqpoint{1.072407in}{2.146703in}}%
\pgfpathlineto{\pgfqpoint{1.095412in}{2.146703in}}%
\pgfpathlineto{\pgfqpoint{1.118418in}{2.146703in}}%
\pgfpathlineto{\pgfqpoint{1.141423in}{2.146703in}}%
\pgfpathlineto{\pgfqpoint{1.164428in}{2.146703in}}%
\pgfpathlineto{\pgfqpoint{1.187433in}{2.146703in}}%
\pgfpathlineto{\pgfqpoint{1.210439in}{2.146703in}}%
\pgfpathlineto{\pgfqpoint{1.233444in}{2.146703in}}%
\pgfpathlineto{\pgfqpoint{1.256449in}{2.146703in}}%
\pgfpathlineto{\pgfqpoint{1.279455in}{2.146703in}}%
\pgfpathlineto{\pgfqpoint{1.302460in}{2.146703in}}%
\pgfpathlineto{\pgfqpoint{1.325465in}{2.146703in}}%
\pgfpathlineto{\pgfqpoint{1.348471in}{2.146703in}}%
\pgfpathlineto{\pgfqpoint{1.371476in}{2.146703in}}%
\pgfpathlineto{\pgfqpoint{1.394481in}{2.146703in}}%
\pgfpathlineto{\pgfqpoint{1.417486in}{2.146703in}}%
\pgfpathlineto{\pgfqpoint{1.440492in}{2.146703in}}%
\pgfpathlineto{\pgfqpoint{1.463497in}{2.146703in}}%
\pgfpathlineto{\pgfqpoint{1.486502in}{2.146703in}}%
\pgfpathlineto{\pgfqpoint{1.509508in}{2.146703in}}%
\pgfpathlineto{\pgfqpoint{1.532513in}{2.146703in}}%
\pgfpathlineto{\pgfqpoint{1.555518in}{2.146703in}}%
\pgfpathlineto{\pgfqpoint{1.578524in}{2.146703in}}%
\pgfpathlineto{\pgfqpoint{1.601529in}{2.146703in}}%
\pgfpathlineto{\pgfqpoint{1.624534in}{2.146703in}}%
\pgfpathlineto{\pgfqpoint{1.647539in}{2.146703in}}%
\pgfpathlineto{\pgfqpoint{1.670545in}{2.146703in}}%
\pgfpathlineto{\pgfqpoint{1.693550in}{2.146703in}}%
\pgfpathlineto{\pgfqpoint{1.716555in}{2.146703in}}%
\pgfpathlineto{\pgfqpoint{1.739561in}{2.146703in}}%
\pgfpathlineto{\pgfqpoint{1.762566in}{2.146703in}}%
\pgfpathlineto{\pgfqpoint{1.785571in}{2.146703in}}%
\pgfpathlineto{\pgfqpoint{1.808576in}{2.146703in}}%
\pgfpathlineto{\pgfqpoint{1.831582in}{2.146703in}}%
\pgfpathlineto{\pgfqpoint{1.854587in}{2.146703in}}%
\pgfpathlineto{\pgfqpoint{1.877592in}{2.146703in}}%
\pgfpathlineto{\pgfqpoint{1.900598in}{2.146703in}}%
\pgfpathlineto{\pgfqpoint{1.923603in}{2.146703in}}%
\pgfpathlineto{\pgfqpoint{1.946608in}{2.146703in}}%
\pgfpathlineto{\pgfqpoint{1.969614in}{2.146703in}}%
\pgfpathlineto{\pgfqpoint{1.992619in}{2.146703in}}%
\pgfpathlineto{\pgfqpoint{2.015624in}{2.146703in}}%
\pgfpathlineto{\pgfqpoint{2.038629in}{2.146703in}}%
\pgfpathlineto{\pgfqpoint{2.061635in}{2.146703in}}%
\pgfpathlineto{\pgfqpoint{2.084640in}{2.146703in}}%
\pgfpathlineto{\pgfqpoint{2.107645in}{2.146703in}}%
\pgfpathlineto{\pgfqpoint{2.130651in}{2.146703in}}%
\pgfpathlineto{\pgfqpoint{2.153656in}{2.146703in}}%
\pgfpathlineto{\pgfqpoint{2.176661in}{2.146703in}}%
\pgfusepath{stroke}%
\end{pgfscope}%
\begin{pgfscope}%
\pgfpathrectangle{\pgfqpoint{0.461616in}{0.431659in}}{\pgfqpoint{1.796714in}{1.796714in}}%
\pgfusepath{clip}%
\pgfsetbuttcap%
\pgfsetroundjoin%
\pgfsetlinewidth{1.003750pt}%
\definecolor{currentstroke}{rgb}{0.839216,0.152941,0.156863}%
\pgfsetstrokecolor{currentstroke}%
\pgfsetdash{{6.400000pt}{1.600000pt}{1.000000pt}{1.600000pt}}{0.000000pt}%
\pgfpathmoveto{\pgfqpoint{0.543285in}{0.513327in}}%
\pgfpathlineto{\pgfqpoint{0.566290in}{0.513327in}}%
\pgfpathlineto{\pgfqpoint{0.589296in}{0.513327in}}%
\pgfpathlineto{\pgfqpoint{0.612301in}{1.192976in}}%
\pgfpathlineto{\pgfqpoint{0.635306in}{1.351788in}}%
\pgfpathlineto{\pgfqpoint{0.658312in}{1.455671in}}%
\pgfpathlineto{\pgfqpoint{0.681317in}{1.456240in}}%
\pgfpathlineto{\pgfqpoint{0.704322in}{1.826802in}}%
\pgfpathlineto{\pgfqpoint{0.727328in}{1.905639in}}%
\pgfpathlineto{\pgfqpoint{0.750333in}{2.146703in}}%
\pgfpathlineto{\pgfqpoint{0.773338in}{2.146703in}}%
\pgfpathlineto{\pgfqpoint{0.796343in}{2.146703in}}%
\pgfpathlineto{\pgfqpoint{0.819349in}{2.146703in}}%
\pgfpathlineto{\pgfqpoint{0.842354in}{2.146703in}}%
\pgfpathlineto{\pgfqpoint{0.865359in}{2.146703in}}%
\pgfpathlineto{\pgfqpoint{0.888365in}{2.146703in}}%
\pgfpathlineto{\pgfqpoint{0.911370in}{2.146703in}}%
\pgfpathlineto{\pgfqpoint{0.934375in}{2.146703in}}%
\pgfpathlineto{\pgfqpoint{0.957381in}{2.146703in}}%
\pgfpathlineto{\pgfqpoint{0.980386in}{2.146703in}}%
\pgfpathlineto{\pgfqpoint{1.003391in}{2.146703in}}%
\pgfpathlineto{\pgfqpoint{1.026396in}{2.146703in}}%
\pgfpathlineto{\pgfqpoint{1.049402in}{2.146703in}}%
\pgfpathlineto{\pgfqpoint{1.072407in}{2.146703in}}%
\pgfpathlineto{\pgfqpoint{1.095412in}{2.146703in}}%
\pgfpathlineto{\pgfqpoint{1.118418in}{2.146703in}}%
\pgfpathlineto{\pgfqpoint{1.141423in}{2.146703in}}%
\pgfpathlineto{\pgfqpoint{1.164428in}{2.146703in}}%
\pgfpathlineto{\pgfqpoint{1.187433in}{2.146703in}}%
\pgfpathlineto{\pgfqpoint{1.210439in}{2.146703in}}%
\pgfpathlineto{\pgfqpoint{1.233444in}{2.146703in}}%
\pgfpathlineto{\pgfqpoint{1.256449in}{2.146703in}}%
\pgfpathlineto{\pgfqpoint{1.279455in}{2.146703in}}%
\pgfpathlineto{\pgfqpoint{1.302460in}{2.146703in}}%
\pgfpathlineto{\pgfqpoint{1.325465in}{2.146703in}}%
\pgfpathlineto{\pgfqpoint{1.348471in}{2.146703in}}%
\pgfpathlineto{\pgfqpoint{1.371476in}{2.146703in}}%
\pgfpathlineto{\pgfqpoint{1.394481in}{2.146703in}}%
\pgfpathlineto{\pgfqpoint{1.417486in}{2.146703in}}%
\pgfpathlineto{\pgfqpoint{1.440492in}{2.146703in}}%
\pgfpathlineto{\pgfqpoint{1.463497in}{2.146703in}}%
\pgfpathlineto{\pgfqpoint{1.486502in}{2.146703in}}%
\pgfpathlineto{\pgfqpoint{1.509508in}{2.146703in}}%
\pgfpathlineto{\pgfqpoint{1.532513in}{2.146703in}}%
\pgfpathlineto{\pgfqpoint{1.555518in}{2.146703in}}%
\pgfpathlineto{\pgfqpoint{1.578524in}{2.146703in}}%
\pgfpathlineto{\pgfqpoint{1.601529in}{2.146703in}}%
\pgfpathlineto{\pgfqpoint{1.624534in}{2.146703in}}%
\pgfpathlineto{\pgfqpoint{1.647539in}{2.146703in}}%
\pgfpathlineto{\pgfqpoint{1.670545in}{2.146703in}}%
\pgfpathlineto{\pgfqpoint{1.693550in}{2.146703in}}%
\pgfpathlineto{\pgfqpoint{1.716555in}{2.146703in}}%
\pgfpathlineto{\pgfqpoint{1.739561in}{2.146703in}}%
\pgfpathlineto{\pgfqpoint{1.762566in}{2.146703in}}%
\pgfpathlineto{\pgfqpoint{1.785571in}{2.146703in}}%
\pgfpathlineto{\pgfqpoint{1.808576in}{2.146703in}}%
\pgfpathlineto{\pgfqpoint{1.831582in}{2.146703in}}%
\pgfpathlineto{\pgfqpoint{1.854587in}{2.146703in}}%
\pgfpathlineto{\pgfqpoint{1.877592in}{2.146703in}}%
\pgfpathlineto{\pgfqpoint{1.900598in}{2.146703in}}%
\pgfpathlineto{\pgfqpoint{1.923603in}{2.146703in}}%
\pgfpathlineto{\pgfqpoint{1.946608in}{2.146703in}}%
\pgfpathlineto{\pgfqpoint{1.969614in}{2.146703in}}%
\pgfpathlineto{\pgfqpoint{1.992619in}{2.146703in}}%
\pgfpathlineto{\pgfqpoint{2.015624in}{2.146703in}}%
\pgfpathlineto{\pgfqpoint{2.038629in}{2.146703in}}%
\pgfpathlineto{\pgfqpoint{2.061635in}{2.146703in}}%
\pgfpathlineto{\pgfqpoint{2.084640in}{2.146703in}}%
\pgfpathlineto{\pgfqpoint{2.107645in}{2.146703in}}%
\pgfpathlineto{\pgfqpoint{2.130651in}{2.146703in}}%
\pgfpathlineto{\pgfqpoint{2.153656in}{2.146703in}}%
\pgfpathlineto{\pgfqpoint{2.176661in}{2.146703in}}%
\pgfusepath{stroke}%
\end{pgfscope}%
\begin{pgfscope}%
\pgfpathrectangle{\pgfqpoint{0.461616in}{0.431659in}}{\pgfqpoint{1.796714in}{1.796714in}}%
\pgfusepath{clip}%
\pgfsetbuttcap%
\pgfsetroundjoin%
\pgfsetlinewidth{1.003750pt}%
\definecolor{currentstroke}{rgb}{0.172549,0.627451,0.172549}%
\pgfsetstrokecolor{currentstroke}%
\pgfsetdash{{1.000000pt}{1.650000pt}}{0.000000pt}%
\pgfpathmoveto{\pgfqpoint{0.543285in}{0.513327in}}%
\pgfpathlineto{\pgfqpoint{0.566290in}{0.513327in}}%
\pgfpathlineto{\pgfqpoint{0.589296in}{0.513327in}}%
\pgfpathlineto{\pgfqpoint{0.612301in}{0.584195in}}%
\pgfpathlineto{\pgfqpoint{0.635306in}{0.606680in}}%
\pgfpathlineto{\pgfqpoint{0.658312in}{0.627456in}}%
\pgfpathlineto{\pgfqpoint{0.681317in}{0.627456in}}%
\pgfpathlineto{\pgfqpoint{0.704322in}{0.681247in}}%
\pgfpathlineto{\pgfqpoint{0.727328in}{0.691209in}}%
\pgfpathlineto{\pgfqpoint{0.750333in}{0.759230in}}%
\pgfpathlineto{\pgfqpoint{0.773338in}{0.759230in}}%
\pgfpathlineto{\pgfqpoint{0.796343in}{0.759230in}}%
\pgfpathlineto{\pgfqpoint{0.819349in}{0.759230in}}%
\pgfpathlineto{\pgfqpoint{0.842354in}{0.759230in}}%
\pgfpathlineto{\pgfqpoint{0.865359in}{0.759230in}}%
\pgfpathlineto{\pgfqpoint{0.888365in}{0.759230in}}%
\pgfpathlineto{\pgfqpoint{0.911370in}{0.759230in}}%
\pgfpathlineto{\pgfqpoint{0.934375in}{0.759230in}}%
\pgfpathlineto{\pgfqpoint{0.957381in}{0.759230in}}%
\pgfpathlineto{\pgfqpoint{0.980386in}{0.759230in}}%
\pgfpathlineto{\pgfqpoint{1.003391in}{0.759230in}}%
\pgfpathlineto{\pgfqpoint{1.026396in}{0.759230in}}%
\pgfpathlineto{\pgfqpoint{1.049402in}{0.759230in}}%
\pgfpathlineto{\pgfqpoint{1.072407in}{0.759230in}}%
\pgfpathlineto{\pgfqpoint{1.095412in}{0.759230in}}%
\pgfpathlineto{\pgfqpoint{1.118418in}{0.759230in}}%
\pgfpathlineto{\pgfqpoint{1.141423in}{0.759230in}}%
\pgfpathlineto{\pgfqpoint{1.164428in}{0.759230in}}%
\pgfpathlineto{\pgfqpoint{1.187433in}{0.759230in}}%
\pgfpathlineto{\pgfqpoint{1.210439in}{0.759230in}}%
\pgfpathlineto{\pgfqpoint{1.233444in}{0.759230in}}%
\pgfpathlineto{\pgfqpoint{1.256449in}{0.759230in}}%
\pgfpathlineto{\pgfqpoint{1.279455in}{0.759230in}}%
\pgfpathlineto{\pgfqpoint{1.302460in}{0.759230in}}%
\pgfpathlineto{\pgfqpoint{1.325465in}{0.759230in}}%
\pgfpathlineto{\pgfqpoint{1.348471in}{0.759230in}}%
\pgfpathlineto{\pgfqpoint{1.371476in}{0.759230in}}%
\pgfpathlineto{\pgfqpoint{1.394481in}{0.759230in}}%
\pgfpathlineto{\pgfqpoint{1.417486in}{0.759230in}}%
\pgfpathlineto{\pgfqpoint{1.440492in}{0.759230in}}%
\pgfpathlineto{\pgfqpoint{1.463497in}{0.759230in}}%
\pgfpathlineto{\pgfqpoint{1.486502in}{0.759230in}}%
\pgfpathlineto{\pgfqpoint{1.509508in}{0.759230in}}%
\pgfpathlineto{\pgfqpoint{1.532513in}{0.759230in}}%
\pgfpathlineto{\pgfqpoint{1.555518in}{0.759230in}}%
\pgfpathlineto{\pgfqpoint{1.578524in}{0.759230in}}%
\pgfpathlineto{\pgfqpoint{1.601529in}{0.759230in}}%
\pgfpathlineto{\pgfqpoint{1.624534in}{0.759230in}}%
\pgfpathlineto{\pgfqpoint{1.647539in}{0.759230in}}%
\pgfpathlineto{\pgfqpoint{1.670545in}{0.759230in}}%
\pgfpathlineto{\pgfqpoint{1.693550in}{1.031317in}}%
\pgfpathlineto{\pgfqpoint{1.716555in}{1.463355in}}%
\pgfpathlineto{\pgfqpoint{1.739561in}{1.542477in}}%
\pgfpathlineto{\pgfqpoint{1.762566in}{1.542477in}}%
\pgfpathlineto{\pgfqpoint{1.785571in}{1.542477in}}%
\pgfpathlineto{\pgfqpoint{1.808576in}{1.542477in}}%
\pgfpathlineto{\pgfqpoint{1.831582in}{1.542477in}}%
\pgfpathlineto{\pgfqpoint{1.854587in}{1.542477in}}%
\pgfpathlineto{\pgfqpoint{1.877592in}{1.542477in}}%
\pgfpathlineto{\pgfqpoint{1.900598in}{1.542477in}}%
\pgfpathlineto{\pgfqpoint{1.923603in}{1.542477in}}%
\pgfpathlineto{\pgfqpoint{1.946608in}{1.542477in}}%
\pgfpathlineto{\pgfqpoint{1.969614in}{1.542477in}}%
\pgfpathlineto{\pgfqpoint{1.992619in}{1.542477in}}%
\pgfpathlineto{\pgfqpoint{2.015624in}{1.542477in}}%
\pgfpathlineto{\pgfqpoint{2.038629in}{1.542477in}}%
\pgfpathlineto{\pgfqpoint{2.061635in}{1.542477in}}%
\pgfpathlineto{\pgfqpoint{2.084640in}{1.542477in}}%
\pgfpathlineto{\pgfqpoint{2.107645in}{1.542477in}}%
\pgfpathlineto{\pgfqpoint{2.130651in}{1.542477in}}%
\pgfpathlineto{\pgfqpoint{2.153656in}{2.072705in}}%
\pgfpathlineto{\pgfqpoint{2.176661in}{2.146703in}}%
\pgfusepath{stroke}%
\end{pgfscope}%
\begin{pgfscope}%
\pgfsetrectcap%
\pgfsetmiterjoin%
\pgfsetlinewidth{1.003750pt}%
\definecolor{currentstroke}{rgb}{1.000000,1.000000,1.000000}%
\pgfsetstrokecolor{currentstroke}%
\pgfsetdash{}{0pt}%
\pgfpathmoveto{\pgfqpoint{0.461616in}{0.431659in}}%
\pgfpathlineto{\pgfqpoint{0.461616in}{2.228372in}}%
\pgfusepath{stroke}%
\end{pgfscope}%
\begin{pgfscope}%
\pgfsetrectcap%
\pgfsetmiterjoin%
\pgfsetlinewidth{1.003750pt}%
\definecolor{currentstroke}{rgb}{1.000000,1.000000,1.000000}%
\pgfsetstrokecolor{currentstroke}%
\pgfsetdash{}{0pt}%
\pgfpathmoveto{\pgfqpoint{2.258330in}{0.431659in}}%
\pgfpathlineto{\pgfqpoint{2.258330in}{2.228372in}}%
\pgfusepath{stroke}%
\end{pgfscope}%
\begin{pgfscope}%
\pgfsetrectcap%
\pgfsetmiterjoin%
\pgfsetlinewidth{1.003750pt}%
\definecolor{currentstroke}{rgb}{1.000000,1.000000,1.000000}%
\pgfsetstrokecolor{currentstroke}%
\pgfsetdash{}{0pt}%
\pgfpathmoveto{\pgfqpoint{0.461616in}{0.431659in}}%
\pgfpathlineto{\pgfqpoint{2.258330in}{0.431659in}}%
\pgfusepath{stroke}%
\end{pgfscope}%
\begin{pgfscope}%
\pgfsetrectcap%
\pgfsetmiterjoin%
\pgfsetlinewidth{1.003750pt}%
\definecolor{currentstroke}{rgb}{1.000000,1.000000,1.000000}%
\pgfsetstrokecolor{currentstroke}%
\pgfsetdash{}{0pt}%
\pgfpathmoveto{\pgfqpoint{0.461616in}{2.228372in}}%
\pgfpathlineto{\pgfqpoint{2.258330in}{2.228372in}}%
\pgfusepath{stroke}%
\end{pgfscope}%
\end{pgfpicture}%
\makeatother%
\endgroup%

%% file: figures/starlink_all_bariloche_beijing_cdf_validity.pgf
\begingroup%
\makeatletter%
\begin{pgfpicture}%
\pgfpathrectangle{\pgfpointorigin}{\pgfqpoint{2.300000in}{2.300000in}}%
\pgfusepath{use as bounding box, clip}%
\begin{pgfscope}%
\pgfsetbuttcap%
\pgfsetmiterjoin%
\definecolor{currentfill}{rgb}{1.000000,1.000000,1.000000}%
\pgfsetfillcolor{currentfill}%
\pgfsetlinewidth{0.000000pt}%
\definecolor{currentstroke}{rgb}{0.500000,0.500000,0.500000}%
\pgfsetstrokecolor{currentstroke}%
\pgfsetdash{}{0pt}%
\pgfpathmoveto{\pgfqpoint{0.000000in}{0.000000in}}%
\pgfpathlineto{\pgfqpoint{2.300000in}{0.000000in}}%
\pgfpathlineto{\pgfqpoint{2.300000in}{2.300000in}}%
\pgfpathlineto{\pgfqpoint{0.000000in}{2.300000in}}%
\pgfpathlineto{\pgfqpoint{0.000000in}{0.000000in}}%
\pgfpathclose%
\pgfusepath{fill}%
\end{pgfscope}%
\begin{pgfscope}%
\pgfsetbuttcap%
\pgfsetmiterjoin%
\definecolor{currentfill}{rgb}{0.898039,0.898039,0.898039}%
\pgfsetfillcolor{currentfill}%
\pgfsetlinewidth{0.000000pt}%
\definecolor{currentstroke}{rgb}{0.000000,0.000000,0.000000}%
\pgfsetstrokecolor{currentstroke}%
\pgfsetstrokeopacity{0.000000}%
\pgfsetdash{}{0pt}%
\pgfpathmoveto{\pgfqpoint{0.461616in}{0.431659in}}%
\pgfpathlineto{\pgfqpoint{2.258330in}{0.431659in}}%
\pgfpathlineto{\pgfqpoint{2.258330in}{2.228372in}}%
\pgfpathlineto{\pgfqpoint{0.461616in}{2.228372in}}%
\pgfpathlineto{\pgfqpoint{0.461616in}{0.431659in}}%
\pgfpathclose%
\pgfusepath{fill}%
\end{pgfscope}%
\begin{pgfscope}%
\pgfpathrectangle{\pgfqpoint{0.461616in}{0.431659in}}{\pgfqpoint{1.796714in}{1.796714in}}%
\pgfusepath{clip}%
\pgfsetrectcap%
\pgfsetroundjoin%
\pgfsetlinewidth{0.803000pt}%
\definecolor{currentstroke}{rgb}{1.000000,1.000000,1.000000}%
\pgfsetstrokecolor{currentstroke}%
\pgfsetdash{}{0pt}%
\pgfpathmoveto{\pgfqpoint{0.543285in}{0.431659in}}%
\pgfpathlineto{\pgfqpoint{0.543285in}{2.228372in}}%
\pgfusepath{stroke}%
\end{pgfscope}%
\begin{pgfscope}%
\pgfsetbuttcap%
\pgfsetroundjoin%
\definecolor{currentfill}{rgb}{0.333333,0.333333,0.333333}%
\pgfsetfillcolor{currentfill}%
\pgfsetlinewidth{0.803000pt}%
\definecolor{currentstroke}{rgb}{0.333333,0.333333,0.333333}%
\pgfsetstrokecolor{currentstroke}%
\pgfsetdash{}{0pt}%
\pgfsys@defobject{currentmarker}{\pgfqpoint{0.000000in}{-0.048611in}}{\pgfqpoint{0.000000in}{0.000000in}}{%
\pgfpathmoveto{\pgfqpoint{0.000000in}{0.000000in}}%
\pgfpathlineto{\pgfqpoint{0.000000in}{-0.048611in}}%
\pgfusepath{stroke,fill}%
}%
\begin{pgfscope}%
\pgfsys@transformshift{0.543285in}{0.431659in}%
\pgfsys@useobject{currentmarker}{}%
\end{pgfscope}%
\end{pgfscope}%
\begin{pgfscope}%
\definecolor{textcolor}{rgb}{0.333333,0.333333,0.333333}%
\pgfsetstrokecolor{textcolor}%
\pgfsetfillcolor{textcolor}%
\pgftext[x=0.543285in,y=0.334436in,,top]{\color{textcolor}\rmfamily\fontsize{7.000000}{8.400000}\selectfont \(\displaystyle {0}\)}%
\end{pgfscope}%
\begin{pgfscope}%
\pgfpathrectangle{\pgfqpoint{0.461616in}{0.431659in}}{\pgfqpoint{1.796714in}{1.796714in}}%
\pgfusepath{clip}%
\pgfsetrectcap%
\pgfsetroundjoin%
\pgfsetlinewidth{0.803000pt}%
\definecolor{currentstroke}{rgb}{1.000000,1.000000,1.000000}%
\pgfsetstrokecolor{currentstroke}%
\pgfsetdash{}{0pt}%
\pgfpathmoveto{\pgfqpoint{1.056925in}{0.431659in}}%
\pgfpathlineto{\pgfqpoint{1.056925in}{2.228372in}}%
\pgfusepath{stroke}%
\end{pgfscope}%
\begin{pgfscope}%
\pgfsetbuttcap%
\pgfsetroundjoin%
\definecolor{currentfill}{rgb}{0.333333,0.333333,0.333333}%
\pgfsetfillcolor{currentfill}%
\pgfsetlinewidth{0.803000pt}%
\definecolor{currentstroke}{rgb}{0.333333,0.333333,0.333333}%
\pgfsetstrokecolor{currentstroke}%
\pgfsetdash{}{0pt}%
\pgfsys@defobject{currentmarker}{\pgfqpoint{0.000000in}{-0.048611in}}{\pgfqpoint{0.000000in}{0.000000in}}{%
\pgfpathmoveto{\pgfqpoint{0.000000in}{0.000000in}}%
\pgfpathlineto{\pgfqpoint{0.000000in}{-0.048611in}}%
\pgfusepath{stroke,fill}%
}%
\begin{pgfscope}%
\pgfsys@transformshift{1.056925in}{0.431659in}%
\pgfsys@useobject{currentmarker}{}%
\end{pgfscope}%
\end{pgfscope}%
\begin{pgfscope}%
\definecolor{textcolor}{rgb}{0.333333,0.333333,0.333333}%
\pgfsetstrokecolor{textcolor}%
\pgfsetfillcolor{textcolor}%
\pgftext[x=1.056925in,y=0.334436in,,top]{\color{textcolor}\rmfamily\fontsize{7.000000}{8.400000}\selectfont \(\displaystyle {50}\)}%
\end{pgfscope}%
\begin{pgfscope}%
\pgfpathrectangle{\pgfqpoint{0.461616in}{0.431659in}}{\pgfqpoint{1.796714in}{1.796714in}}%
\pgfusepath{clip}%
\pgfsetrectcap%
\pgfsetroundjoin%
\pgfsetlinewidth{0.803000pt}%
\definecolor{currentstroke}{rgb}{1.000000,1.000000,1.000000}%
\pgfsetstrokecolor{currentstroke}%
\pgfsetdash{}{0pt}%
\pgfpathmoveto{\pgfqpoint{1.570566in}{0.431659in}}%
\pgfpathlineto{\pgfqpoint{1.570566in}{2.228372in}}%
\pgfusepath{stroke}%
\end{pgfscope}%
\begin{pgfscope}%
\pgfsetbuttcap%
\pgfsetroundjoin%
\definecolor{currentfill}{rgb}{0.333333,0.333333,0.333333}%
\pgfsetfillcolor{currentfill}%
\pgfsetlinewidth{0.803000pt}%
\definecolor{currentstroke}{rgb}{0.333333,0.333333,0.333333}%
\pgfsetstrokecolor{currentstroke}%
\pgfsetdash{}{0pt}%
\pgfsys@defobject{currentmarker}{\pgfqpoint{0.000000in}{-0.048611in}}{\pgfqpoint{0.000000in}{0.000000in}}{%
\pgfpathmoveto{\pgfqpoint{0.000000in}{0.000000in}}%
\pgfpathlineto{\pgfqpoint{0.000000in}{-0.048611in}}%
\pgfusepath{stroke,fill}%
}%
\begin{pgfscope}%
\pgfsys@transformshift{1.570566in}{0.431659in}%
\pgfsys@useobject{currentmarker}{}%
\end{pgfscope}%
\end{pgfscope}%
\begin{pgfscope}%
\definecolor{textcolor}{rgb}{0.333333,0.333333,0.333333}%
\pgfsetstrokecolor{textcolor}%
\pgfsetfillcolor{textcolor}%
\pgftext[x=1.570566in,y=0.334436in,,top]{\color{textcolor}\rmfamily\fontsize{7.000000}{8.400000}\selectfont \(\displaystyle {100}\)}%
\end{pgfscope}%
\begin{pgfscope}%
\pgfpathrectangle{\pgfqpoint{0.461616in}{0.431659in}}{\pgfqpoint{1.796714in}{1.796714in}}%
\pgfusepath{clip}%
\pgfsetrectcap%
\pgfsetroundjoin%
\pgfsetlinewidth{0.803000pt}%
\definecolor{currentstroke}{rgb}{1.000000,1.000000,1.000000}%
\pgfsetstrokecolor{currentstroke}%
\pgfsetdash{}{0pt}%
\pgfpathmoveto{\pgfqpoint{2.084206in}{0.431659in}}%
\pgfpathlineto{\pgfqpoint{2.084206in}{2.228372in}}%
\pgfusepath{stroke}%
\end{pgfscope}%
\begin{pgfscope}%
\pgfsetbuttcap%
\pgfsetroundjoin%
\definecolor{currentfill}{rgb}{0.333333,0.333333,0.333333}%
\pgfsetfillcolor{currentfill}%
\pgfsetlinewidth{0.803000pt}%
\definecolor{currentstroke}{rgb}{0.333333,0.333333,0.333333}%
\pgfsetstrokecolor{currentstroke}%
\pgfsetdash{}{0pt}%
\pgfsys@defobject{currentmarker}{\pgfqpoint{0.000000in}{-0.048611in}}{\pgfqpoint{0.000000in}{0.000000in}}{%
\pgfpathmoveto{\pgfqpoint{0.000000in}{0.000000in}}%
\pgfpathlineto{\pgfqpoint{0.000000in}{-0.048611in}}%
\pgfusepath{stroke,fill}%
}%
\begin{pgfscope}%
\pgfsys@transformshift{2.084206in}{0.431659in}%
\pgfsys@useobject{currentmarker}{}%
\end{pgfscope}%
\end{pgfscope}%
\begin{pgfscope}%
\definecolor{textcolor}{rgb}{0.333333,0.333333,0.333333}%
\pgfsetstrokecolor{textcolor}%
\pgfsetfillcolor{textcolor}%
\pgftext[x=2.084206in,y=0.334436in,,top]{\color{textcolor}\rmfamily\fontsize{7.000000}{8.400000}\selectfont \(\displaystyle {150}\)}%
\end{pgfscope}%
\begin{pgfscope}%
\definecolor{textcolor}{rgb}{0.333333,0.333333,0.333333}%
\pgfsetstrokecolor{textcolor}%
\pgfsetfillcolor{textcolor}%
\pgftext[x=1.359973in,y=0.192461in,,top]{\color{textcolor}\rmfamily\fontsize{10.000000}{12.000000}\selectfont Route Validity [s]}%
\end{pgfscope}%
\begin{pgfscope}%
\pgfpathrectangle{\pgfqpoint{0.461616in}{0.431659in}}{\pgfqpoint{1.796714in}{1.796714in}}%
\pgfusepath{clip}%
\pgfsetrectcap%
\pgfsetroundjoin%
\pgfsetlinewidth{0.803000pt}%
\definecolor{currentstroke}{rgb}{1.000000,1.000000,1.000000}%
\pgfsetstrokecolor{currentstroke}%
\pgfsetdash{}{0pt}%
\pgfpathmoveto{\pgfqpoint{0.461616in}{0.513327in}}%
\pgfpathlineto{\pgfqpoint{2.258330in}{0.513327in}}%
\pgfusepath{stroke}%
\end{pgfscope}%
\begin{pgfscope}%
\pgfsetbuttcap%
\pgfsetroundjoin%
\definecolor{currentfill}{rgb}{0.333333,0.333333,0.333333}%
\pgfsetfillcolor{currentfill}%
\pgfsetlinewidth{0.803000pt}%
\definecolor{currentstroke}{rgb}{0.333333,0.333333,0.333333}%
\pgfsetstrokecolor{currentstroke}%
\pgfsetdash{}{0pt}%
\pgfsys@defobject{currentmarker}{\pgfqpoint{-0.048611in}{0.000000in}}{\pgfqpoint{-0.000000in}{0.000000in}}{%
\pgfpathmoveto{\pgfqpoint{-0.000000in}{0.000000in}}%
\pgfpathlineto{\pgfqpoint{-0.048611in}{0.000000in}}%
\pgfusepath{stroke,fill}%
}%
\begin{pgfscope}%
\pgfsys@transformshift{0.461616in}{0.513327in}%
\pgfsys@useobject{currentmarker}{}%
\end{pgfscope}%
\end{pgfscope}%
\begin{pgfscope}%
\definecolor{textcolor}{rgb}{0.333333,0.333333,0.333333}%
\pgfsetstrokecolor{textcolor}%
\pgfsetfillcolor{textcolor}%
\pgftext[x=0.220682in, y=0.479570in, left, base]{\color{textcolor}\rmfamily\fontsize{7.000000}{8.400000}\selectfont \(\displaystyle {0.0}\)}%
\end{pgfscope}%
\begin{pgfscope}%
\pgfpathrectangle{\pgfqpoint{0.461616in}{0.431659in}}{\pgfqpoint{1.796714in}{1.796714in}}%
\pgfusepath{clip}%
\pgfsetrectcap%
\pgfsetroundjoin%
\pgfsetlinewidth{0.803000pt}%
\definecolor{currentstroke}{rgb}{1.000000,1.000000,1.000000}%
\pgfsetstrokecolor{currentstroke}%
\pgfsetdash{}{0pt}%
\pgfpathmoveto{\pgfqpoint{0.461616in}{0.840003in}}%
\pgfpathlineto{\pgfqpoint{2.258330in}{0.840003in}}%
\pgfusepath{stroke}%
\end{pgfscope}%
\begin{pgfscope}%
\pgfsetbuttcap%
\pgfsetroundjoin%
\definecolor{currentfill}{rgb}{0.333333,0.333333,0.333333}%
\pgfsetfillcolor{currentfill}%
\pgfsetlinewidth{0.803000pt}%
\definecolor{currentstroke}{rgb}{0.333333,0.333333,0.333333}%
\pgfsetstrokecolor{currentstroke}%
\pgfsetdash{}{0pt}%
\pgfsys@defobject{currentmarker}{\pgfqpoint{-0.048611in}{0.000000in}}{\pgfqpoint{-0.000000in}{0.000000in}}{%
\pgfpathmoveto{\pgfqpoint{-0.000000in}{0.000000in}}%
\pgfpathlineto{\pgfqpoint{-0.048611in}{0.000000in}}%
\pgfusepath{stroke,fill}%
}%
\begin{pgfscope}%
\pgfsys@transformshift{0.461616in}{0.840003in}%
\pgfsys@useobject{currentmarker}{}%
\end{pgfscope}%
\end{pgfscope}%
\begin{pgfscope}%
\definecolor{textcolor}{rgb}{0.333333,0.333333,0.333333}%
\pgfsetstrokecolor{textcolor}%
\pgfsetfillcolor{textcolor}%
\pgftext[x=0.220682in, y=0.806245in, left, base]{\color{textcolor}\rmfamily\fontsize{7.000000}{8.400000}\selectfont \(\displaystyle {0.2}\)}%
\end{pgfscope}%
\begin{pgfscope}%
\pgfpathrectangle{\pgfqpoint{0.461616in}{0.431659in}}{\pgfqpoint{1.796714in}{1.796714in}}%
\pgfusepath{clip}%
\pgfsetrectcap%
\pgfsetroundjoin%
\pgfsetlinewidth{0.803000pt}%
\definecolor{currentstroke}{rgb}{1.000000,1.000000,1.000000}%
\pgfsetstrokecolor{currentstroke}%
\pgfsetdash{}{0pt}%
\pgfpathmoveto{\pgfqpoint{0.461616in}{1.166678in}}%
\pgfpathlineto{\pgfqpoint{2.258330in}{1.166678in}}%
\pgfusepath{stroke}%
\end{pgfscope}%
\begin{pgfscope}%
\pgfsetbuttcap%
\pgfsetroundjoin%
\definecolor{currentfill}{rgb}{0.333333,0.333333,0.333333}%
\pgfsetfillcolor{currentfill}%
\pgfsetlinewidth{0.803000pt}%
\definecolor{currentstroke}{rgb}{0.333333,0.333333,0.333333}%
\pgfsetstrokecolor{currentstroke}%
\pgfsetdash{}{0pt}%
\pgfsys@defobject{currentmarker}{\pgfqpoint{-0.048611in}{0.000000in}}{\pgfqpoint{-0.000000in}{0.000000in}}{%
\pgfpathmoveto{\pgfqpoint{-0.000000in}{0.000000in}}%
\pgfpathlineto{\pgfqpoint{-0.048611in}{0.000000in}}%
\pgfusepath{stroke,fill}%
}%
\begin{pgfscope}%
\pgfsys@transformshift{0.461616in}{1.166678in}%
\pgfsys@useobject{currentmarker}{}%
\end{pgfscope}%
\end{pgfscope}%
\begin{pgfscope}%
\definecolor{textcolor}{rgb}{0.333333,0.333333,0.333333}%
\pgfsetstrokecolor{textcolor}%
\pgfsetfillcolor{textcolor}%
\pgftext[x=0.220682in, y=1.132920in, left, base]{\color{textcolor}\rmfamily\fontsize{7.000000}{8.400000}\selectfont \(\displaystyle {0.4}\)}%
\end{pgfscope}%
\begin{pgfscope}%
\pgfpathrectangle{\pgfqpoint{0.461616in}{0.431659in}}{\pgfqpoint{1.796714in}{1.796714in}}%
\pgfusepath{clip}%
\pgfsetrectcap%
\pgfsetroundjoin%
\pgfsetlinewidth{0.803000pt}%
\definecolor{currentstroke}{rgb}{1.000000,1.000000,1.000000}%
\pgfsetstrokecolor{currentstroke}%
\pgfsetdash{}{0pt}%
\pgfpathmoveto{\pgfqpoint{0.461616in}{1.493353in}}%
\pgfpathlineto{\pgfqpoint{2.258330in}{1.493353in}}%
\pgfusepath{stroke}%
\end{pgfscope}%
\begin{pgfscope}%
\pgfsetbuttcap%
\pgfsetroundjoin%
\definecolor{currentfill}{rgb}{0.333333,0.333333,0.333333}%
\pgfsetfillcolor{currentfill}%
\pgfsetlinewidth{0.803000pt}%
\definecolor{currentstroke}{rgb}{0.333333,0.333333,0.333333}%
\pgfsetstrokecolor{currentstroke}%
\pgfsetdash{}{0pt}%
\pgfsys@defobject{currentmarker}{\pgfqpoint{-0.048611in}{0.000000in}}{\pgfqpoint{-0.000000in}{0.000000in}}{%
\pgfpathmoveto{\pgfqpoint{-0.000000in}{0.000000in}}%
\pgfpathlineto{\pgfqpoint{-0.048611in}{0.000000in}}%
\pgfusepath{stroke,fill}%
}%
\begin{pgfscope}%
\pgfsys@transformshift{0.461616in}{1.493353in}%
\pgfsys@useobject{currentmarker}{}%
\end{pgfscope}%
\end{pgfscope}%
\begin{pgfscope}%
\definecolor{textcolor}{rgb}{0.333333,0.333333,0.333333}%
\pgfsetstrokecolor{textcolor}%
\pgfsetfillcolor{textcolor}%
\pgftext[x=0.220682in, y=1.459595in, left, base]{\color{textcolor}\rmfamily\fontsize{7.000000}{8.400000}\selectfont \(\displaystyle {0.6}\)}%
\end{pgfscope}%
\begin{pgfscope}%
\pgfpathrectangle{\pgfqpoint{0.461616in}{0.431659in}}{\pgfqpoint{1.796714in}{1.796714in}}%
\pgfusepath{clip}%
\pgfsetrectcap%
\pgfsetroundjoin%
\pgfsetlinewidth{0.803000pt}%
\definecolor{currentstroke}{rgb}{1.000000,1.000000,1.000000}%
\pgfsetstrokecolor{currentstroke}%
\pgfsetdash{}{0pt}%
\pgfpathmoveto{\pgfqpoint{0.461616in}{1.820028in}}%
\pgfpathlineto{\pgfqpoint{2.258330in}{1.820028in}}%
\pgfusepath{stroke}%
\end{pgfscope}%
\begin{pgfscope}%
\pgfsetbuttcap%
\pgfsetroundjoin%
\definecolor{currentfill}{rgb}{0.333333,0.333333,0.333333}%
\pgfsetfillcolor{currentfill}%
\pgfsetlinewidth{0.803000pt}%
\definecolor{currentstroke}{rgb}{0.333333,0.333333,0.333333}%
\pgfsetstrokecolor{currentstroke}%
\pgfsetdash{}{0pt}%
\pgfsys@defobject{currentmarker}{\pgfqpoint{-0.048611in}{0.000000in}}{\pgfqpoint{-0.000000in}{0.000000in}}{%
\pgfpathmoveto{\pgfqpoint{-0.000000in}{0.000000in}}%
\pgfpathlineto{\pgfqpoint{-0.048611in}{0.000000in}}%
\pgfusepath{stroke,fill}%
}%
\begin{pgfscope}%
\pgfsys@transformshift{0.461616in}{1.820028in}%
\pgfsys@useobject{currentmarker}{}%
\end{pgfscope}%
\end{pgfscope}%
\begin{pgfscope}%
\definecolor{textcolor}{rgb}{0.333333,0.333333,0.333333}%
\pgfsetstrokecolor{textcolor}%
\pgfsetfillcolor{textcolor}%
\pgftext[x=0.220682in, y=1.786271in, left, base]{\color{textcolor}\rmfamily\fontsize{7.000000}{8.400000}\selectfont \(\displaystyle {0.8}\)}%
\end{pgfscope}%
\begin{pgfscope}%
\pgfpathrectangle{\pgfqpoint{0.461616in}{0.431659in}}{\pgfqpoint{1.796714in}{1.796714in}}%
\pgfusepath{clip}%
\pgfsetrectcap%
\pgfsetroundjoin%
\pgfsetlinewidth{0.803000pt}%
\definecolor{currentstroke}{rgb}{1.000000,1.000000,1.000000}%
\pgfsetstrokecolor{currentstroke}%
\pgfsetdash{}{0pt}%
\pgfpathmoveto{\pgfqpoint{0.461616in}{2.146703in}}%
\pgfpathlineto{\pgfqpoint{2.258330in}{2.146703in}}%
\pgfusepath{stroke}%
\end{pgfscope}%
\begin{pgfscope}%
\pgfsetbuttcap%
\pgfsetroundjoin%
\definecolor{currentfill}{rgb}{0.333333,0.333333,0.333333}%
\pgfsetfillcolor{currentfill}%
\pgfsetlinewidth{0.803000pt}%
\definecolor{currentstroke}{rgb}{0.333333,0.333333,0.333333}%
\pgfsetstrokecolor{currentstroke}%
\pgfsetdash{}{0pt}%
\pgfsys@defobject{currentmarker}{\pgfqpoint{-0.048611in}{0.000000in}}{\pgfqpoint{-0.000000in}{0.000000in}}{%
\pgfpathmoveto{\pgfqpoint{-0.000000in}{0.000000in}}%
\pgfpathlineto{\pgfqpoint{-0.048611in}{0.000000in}}%
\pgfusepath{stroke,fill}%
}%
\begin{pgfscope}%
\pgfsys@transformshift{0.461616in}{2.146703in}%
\pgfsys@useobject{currentmarker}{}%
\end{pgfscope}%
\end{pgfscope}%
\begin{pgfscope}%
\definecolor{textcolor}{rgb}{0.333333,0.333333,0.333333}%
\pgfsetstrokecolor{textcolor}%
\pgfsetfillcolor{textcolor}%
\pgftext[x=0.220682in, y=2.112946in, left, base]{\color{textcolor}\rmfamily\fontsize{7.000000}{8.400000}\selectfont \(\displaystyle {1.0}\)}%
\end{pgfscope}%
\begin{pgfscope}%
\definecolor{textcolor}{rgb}{0.333333,0.333333,0.333333}%
\pgfsetstrokecolor{textcolor}%
\pgfsetfillcolor{textcolor}%
\pgftext[x=0.165127in,y=1.330015in,,bottom,rotate=90.000000]{\color{textcolor}\rmfamily\fontsize{10.000000}{12.000000}\selectfont Cumulative Probability}%
\end{pgfscope}%
\begin{pgfscope}%
\pgfpathrectangle{\pgfqpoint{0.461616in}{0.431659in}}{\pgfqpoint{1.796714in}{1.796714in}}%
\pgfusepath{clip}%
\pgfsetbuttcap%
\pgfsetroundjoin%
\definecolor{currentfill}{rgb}{1.000000,0.498039,0.054902}%
\pgfsetfillcolor{currentfill}%
\pgfsetlinewidth{1.505625pt}%
\definecolor{currentstroke}{rgb}{1.000000,0.498039,0.054902}%
\pgfsetstrokecolor{currentstroke}%
\pgfsetdash{}{0pt}%
\pgfsys@defobject{currentmarker}{\pgfqpoint{-0.041667in}{-0.041667in}}{\pgfqpoint{0.041667in}{0.041667in}}{%
\pgfpathmoveto{\pgfqpoint{-0.041667in}{0.000000in}}%
\pgfpathlineto{\pgfqpoint{0.041667in}{0.000000in}}%
\pgfpathmoveto{\pgfqpoint{0.000000in}{-0.041667in}}%
\pgfpathlineto{\pgfqpoint{0.000000in}{0.041667in}}%
\pgfusepath{stroke,fill}%
}%
\begin{pgfscope}%
\pgfsys@transformshift{0.851469in}{1.330015in}%
\pgfsys@useobject{currentmarker}{}%
\end{pgfscope}%
\end{pgfscope}%
\begin{pgfscope}%
\pgfpathrectangle{\pgfqpoint{0.461616in}{0.431659in}}{\pgfqpoint{1.796714in}{1.796714in}}%
\pgfusepath{clip}%
\pgfsetbuttcap%
\pgfsetroundjoin%
\definecolor{currentfill}{rgb}{0.121569,0.466667,0.705882}%
\pgfsetfillcolor{currentfill}%
\pgfsetlinewidth{1.505625pt}%
\definecolor{currentstroke}{rgb}{0.121569,0.466667,0.705882}%
\pgfsetstrokecolor{currentstroke}%
\pgfsetdash{}{0pt}%
\pgfsys@defobject{currentmarker}{\pgfqpoint{-0.033333in}{-0.041667in}}{\pgfqpoint{0.033333in}{0.020833in}}{%
\pgfpathmoveto{\pgfqpoint{0.000000in}{0.000000in}}%
\pgfpathlineto{\pgfqpoint{0.000000in}{-0.041667in}}%
\pgfpathmoveto{\pgfqpoint{0.000000in}{0.000000in}}%
\pgfpathlineto{\pgfqpoint{0.033333in}{0.020833in}}%
\pgfpathmoveto{\pgfqpoint{0.000000in}{0.000000in}}%
\pgfpathlineto{\pgfqpoint{-0.033333in}{0.020833in}}%
\pgfusepath{stroke,fill}%
}%
\begin{pgfscope}%
\pgfsys@transformshift{1.139108in}{1.330015in}%
\pgfsys@useobject{currentmarker}{}%
\end{pgfscope}%
\end{pgfscope}%
\begin{pgfscope}%
\pgfpathrectangle{\pgfqpoint{0.461616in}{0.431659in}}{\pgfqpoint{1.796714in}{1.796714in}}%
\pgfusepath{clip}%
\pgfsetbuttcap%
\pgfsetroundjoin%
\definecolor{currentfill}{rgb}{0.839216,0.152941,0.156863}%
\pgfsetfillcolor{currentfill}%
\pgfsetlinewidth{0.501875pt}%
\definecolor{currentstroke}{rgb}{0.839216,0.152941,0.156863}%
\pgfsetstrokecolor{currentstroke}%
\pgfsetdash{}{0pt}%
\pgfsys@defobject{currentmarker}{\pgfqpoint{-0.039627in}{-0.033709in}}{\pgfqpoint{0.039627in}{0.041667in}}{%
\pgfpathmoveto{\pgfqpoint{0.000000in}{0.041667in}}%
\pgfpathlineto{\pgfqpoint{-0.039627in}{0.012876in}}%
\pgfpathlineto{\pgfqpoint{-0.024491in}{-0.033709in}}%
\pgfpathlineto{\pgfqpoint{0.024491in}{-0.033709in}}%
\pgfpathlineto{\pgfqpoint{0.039627in}{0.012876in}}%
\pgfpathlineto{\pgfqpoint{0.000000in}{0.041667in}}%
\pgfpathclose%
\pgfusepath{stroke,fill}%
}%
\begin{pgfscope}%
\pgfsys@transformshift{1.539747in}{1.330015in}%
\pgfsys@useobject{currentmarker}{}%
\end{pgfscope}%
\end{pgfscope}%
\begin{pgfscope}%
\pgfpathrectangle{\pgfqpoint{0.461616in}{0.431659in}}{\pgfqpoint{1.796714in}{1.796714in}}%
\pgfusepath{clip}%
\pgfsetbuttcap%
\pgfsetroundjoin%
\definecolor{currentfill}{rgb}{0.172549,0.627451,0.172549}%
\pgfsetfillcolor{currentfill}%
\pgfsetlinewidth{0.501875pt}%
\definecolor{currentstroke}{rgb}{0.172549,0.627451,0.172549}%
\pgfsetstrokecolor{currentstroke}%
\pgfsetdash{}{0pt}%
\pgfsys@defobject{currentmarker}{\pgfqpoint{-0.039627in}{-0.033709in}}{\pgfqpoint{0.039627in}{0.041667in}}{%
\pgfpathmoveto{\pgfqpoint{0.000000in}{0.041667in}}%
\pgfpathlineto{\pgfqpoint{-0.009355in}{0.012876in}}%
\pgfpathlineto{\pgfqpoint{-0.039627in}{0.012876in}}%
\pgfpathlineto{\pgfqpoint{-0.015136in}{-0.004918in}}%
\pgfpathlineto{\pgfqpoint{-0.024491in}{-0.033709in}}%
\pgfpathlineto{\pgfqpoint{-0.000000in}{-0.015915in}}%
\pgfpathlineto{\pgfqpoint{0.024491in}{-0.033709in}}%
\pgfpathlineto{\pgfqpoint{0.015136in}{-0.004918in}}%
\pgfpathlineto{\pgfqpoint{0.039627in}{0.012876in}}%
\pgfpathlineto{\pgfqpoint{0.009355in}{0.012876in}}%
\pgfpathlineto{\pgfqpoint{0.000000in}{0.041667in}}%
\pgfpathclose%
\pgfusepath{stroke,fill}%
}%
\begin{pgfscope}%
\pgfsys@transformshift{1.776022in}{1.330015in}%
\pgfsys@useobject{currentmarker}{}%
\end{pgfscope}%
\end{pgfscope}%
\begin{pgfscope}%
\pgfpathrectangle{\pgfqpoint{0.461616in}{0.431659in}}{\pgfqpoint{1.796714in}{1.796714in}}%
\pgfusepath{clip}%
\pgfsetrectcap%
\pgfsetroundjoin%
\pgfsetlinewidth{1.003750pt}%
\definecolor{currentstroke}{rgb}{1.000000,0.498039,0.054902}%
\pgfsetstrokecolor{currentstroke}%
\pgfsetdash{}{0pt}%
\pgfpathmoveto{\pgfqpoint{0.543285in}{0.513327in}}%
\pgfpathlineto{\pgfqpoint{0.553558in}{0.619852in}}%
\pgfpathlineto{\pgfqpoint{0.563831in}{0.699745in}}%
\pgfpathlineto{\pgfqpoint{0.574104in}{0.753008in}}%
\pgfpathlineto{\pgfqpoint{0.594649in}{0.895040in}}%
\pgfpathlineto{\pgfqpoint{0.604922in}{0.939426in}}%
\pgfpathlineto{\pgfqpoint{0.625468in}{1.010442in}}%
\pgfpathlineto{\pgfqpoint{0.635740in}{1.010442in}}%
\pgfpathlineto{\pgfqpoint{0.646013in}{1.028196in}}%
\pgfpathlineto{\pgfqpoint{0.656286in}{1.072581in}}%
\pgfpathlineto{\pgfqpoint{0.666559in}{1.081458in}}%
\pgfpathlineto{\pgfqpoint{0.687104in}{1.134721in}}%
\pgfpathlineto{\pgfqpoint{0.697377in}{1.143598in}}%
\pgfpathlineto{\pgfqpoint{0.707650in}{1.161352in}}%
\pgfpathlineto{\pgfqpoint{0.728196in}{1.179106in}}%
\pgfpathlineto{\pgfqpoint{0.738468in}{1.196860in}}%
\pgfpathlineto{\pgfqpoint{0.748741in}{1.196860in}}%
\pgfpathlineto{\pgfqpoint{0.759014in}{1.232368in}}%
\pgfpathlineto{\pgfqpoint{0.769287in}{1.250122in}}%
\pgfpathlineto{\pgfqpoint{0.779560in}{1.258999in}}%
\pgfpathlineto{\pgfqpoint{0.800105in}{1.258999in}}%
\pgfpathlineto{\pgfqpoint{0.810378in}{1.276753in}}%
\pgfpathlineto{\pgfqpoint{0.820651in}{1.285630in}}%
\pgfpathlineto{\pgfqpoint{0.830924in}{1.303384in}}%
\pgfpathlineto{\pgfqpoint{0.841197in}{1.312261in}}%
\pgfpathlineto{\pgfqpoint{0.851469in}{1.338893in}}%
\pgfpathlineto{\pgfqpoint{0.861742in}{1.338893in}}%
\pgfpathlineto{\pgfqpoint{0.882288in}{1.374401in}}%
\pgfpathlineto{\pgfqpoint{0.902833in}{1.427663in}}%
\pgfpathlineto{\pgfqpoint{0.923379in}{1.445417in}}%
\pgfpathlineto{\pgfqpoint{0.933652in}{1.472048in}}%
\pgfpathlineto{\pgfqpoint{0.943925in}{1.480925in}}%
\pgfpathlineto{\pgfqpoint{0.954197in}{1.516433in}}%
\pgfpathlineto{\pgfqpoint{0.974743in}{1.551942in}}%
\pgfpathlineto{\pgfqpoint{0.985016in}{1.560819in}}%
\pgfpathlineto{\pgfqpoint{0.995289in}{1.560819in}}%
\pgfpathlineto{\pgfqpoint{1.015834in}{1.578573in}}%
\pgfpathlineto{\pgfqpoint{1.026107in}{1.596327in}}%
\pgfpathlineto{\pgfqpoint{1.036380in}{1.605204in}}%
\pgfpathlineto{\pgfqpoint{1.046653in}{1.631835in}}%
\pgfpathlineto{\pgfqpoint{1.056925in}{1.631835in}}%
\pgfpathlineto{\pgfqpoint{1.067198in}{1.649589in}}%
\pgfpathlineto{\pgfqpoint{1.077471in}{1.658466in}}%
\pgfpathlineto{\pgfqpoint{1.098017in}{1.693974in}}%
\pgfpathlineto{\pgfqpoint{1.118562in}{1.782745in}}%
\pgfpathlineto{\pgfqpoint{1.128835in}{1.800499in}}%
\pgfpathlineto{\pgfqpoint{1.139108in}{1.836007in}}%
\pgfpathlineto{\pgfqpoint{1.149381in}{1.880392in}}%
\pgfpathlineto{\pgfqpoint{1.159653in}{1.907023in}}%
\pgfpathlineto{\pgfqpoint{1.190472in}{1.960286in}}%
\pgfpathlineto{\pgfqpoint{1.200745in}{2.022425in}}%
\pgfpathlineto{\pgfqpoint{1.211018in}{2.049056in}}%
\pgfpathlineto{\pgfqpoint{1.221290in}{2.049056in}}%
\pgfpathlineto{\pgfqpoint{1.231563in}{2.084564in}}%
\pgfpathlineto{\pgfqpoint{1.241836in}{2.093441in}}%
\pgfpathlineto{\pgfqpoint{1.252109in}{2.111195in}}%
\pgfpathlineto{\pgfqpoint{1.262382in}{2.120072in}}%
\pgfpathlineto{\pgfqpoint{1.272654in}{2.137826in}}%
\pgfpathlineto{\pgfqpoint{1.282927in}{2.146703in}}%
\pgfpathlineto{\pgfqpoint{2.176661in}{2.146703in}}%
\pgfpathlineto{\pgfqpoint{2.176661in}{2.146703in}}%
\pgfusepath{stroke}%
\end{pgfscope}%
\begin{pgfscope}%
\pgfpathrectangle{\pgfqpoint{0.461616in}{0.431659in}}{\pgfqpoint{1.796714in}{1.796714in}}%
\pgfusepath{clip}%
\pgfsetbuttcap%
\pgfsetroundjoin%
\pgfsetlinewidth{1.003750pt}%
\definecolor{currentstroke}{rgb}{0.121569,0.466667,0.705882}%
\pgfsetstrokecolor{currentstroke}%
\pgfsetdash{{3.700000pt}{1.600000pt}}{0.000000pt}%
\pgfpathmoveto{\pgfqpoint{0.543285in}{0.513327in}}%
\pgfpathlineto{\pgfqpoint{0.553558in}{0.545354in}}%
\pgfpathlineto{\pgfqpoint{0.574104in}{0.545354in}}%
\pgfpathlineto{\pgfqpoint{0.584376in}{0.561368in}}%
\pgfpathlineto{\pgfqpoint{0.604922in}{0.561368in}}%
\pgfpathlineto{\pgfqpoint{0.615195in}{0.577381in}}%
\pgfpathlineto{\pgfqpoint{0.625468in}{0.625422in}}%
\pgfpathlineto{\pgfqpoint{0.666559in}{0.689476in}}%
\pgfpathlineto{\pgfqpoint{0.676832in}{0.689476in}}%
\pgfpathlineto{\pgfqpoint{0.687104in}{0.705489in}}%
\pgfpathlineto{\pgfqpoint{0.707650in}{0.705489in}}%
\pgfpathlineto{\pgfqpoint{0.728196in}{0.769543in}}%
\pgfpathlineto{\pgfqpoint{0.738468in}{0.785557in}}%
\pgfpathlineto{\pgfqpoint{0.748741in}{0.833597in}}%
\pgfpathlineto{\pgfqpoint{0.759014in}{0.833597in}}%
\pgfpathlineto{\pgfqpoint{0.769287in}{0.897651in}}%
\pgfpathlineto{\pgfqpoint{0.779560in}{0.897651in}}%
\pgfpathlineto{\pgfqpoint{0.789833in}{0.929678in}}%
\pgfpathlineto{\pgfqpoint{0.810378in}{0.961705in}}%
\pgfpathlineto{\pgfqpoint{0.820651in}{0.961705in}}%
\pgfpathlineto{\pgfqpoint{0.830924in}{0.977719in}}%
\pgfpathlineto{\pgfqpoint{0.841197in}{0.977719in}}%
\pgfpathlineto{\pgfqpoint{0.851469in}{0.993732in}}%
\pgfpathlineto{\pgfqpoint{0.861742in}{0.993732in}}%
\pgfpathlineto{\pgfqpoint{0.872015in}{1.025759in}}%
\pgfpathlineto{\pgfqpoint{0.902833in}{1.073800in}}%
\pgfpathlineto{\pgfqpoint{0.913106in}{1.073800in}}%
\pgfpathlineto{\pgfqpoint{0.933652in}{1.105827in}}%
\pgfpathlineto{\pgfqpoint{0.943925in}{1.105827in}}%
\pgfpathlineto{\pgfqpoint{0.954197in}{1.121840in}}%
\pgfpathlineto{\pgfqpoint{0.964470in}{1.121840in}}%
\pgfpathlineto{\pgfqpoint{0.974743in}{1.153867in}}%
\pgfpathlineto{\pgfqpoint{0.995289in}{1.185894in}}%
\pgfpathlineto{\pgfqpoint{1.015834in}{1.185894in}}%
\pgfpathlineto{\pgfqpoint{1.026107in}{1.201908in}}%
\pgfpathlineto{\pgfqpoint{1.036380in}{1.201908in}}%
\pgfpathlineto{\pgfqpoint{1.046653in}{1.233935in}}%
\pgfpathlineto{\pgfqpoint{1.067198in}{1.233935in}}%
\pgfpathlineto{\pgfqpoint{1.077471in}{1.249948in}}%
\pgfpathlineto{\pgfqpoint{1.098017in}{1.249948in}}%
\pgfpathlineto{\pgfqpoint{1.118562in}{1.281975in}}%
\pgfpathlineto{\pgfqpoint{1.128835in}{1.314002in}}%
\pgfpathlineto{\pgfqpoint{1.139108in}{1.362042in}}%
\pgfpathlineto{\pgfqpoint{1.149381in}{1.394069in}}%
\pgfpathlineto{\pgfqpoint{1.159653in}{1.394069in}}%
\pgfpathlineto{\pgfqpoint{1.169926in}{1.410083in}}%
\pgfpathlineto{\pgfqpoint{1.180199in}{1.410083in}}%
\pgfpathlineto{\pgfqpoint{1.211018in}{1.458123in}}%
\pgfpathlineto{\pgfqpoint{1.221290in}{1.458123in}}%
\pgfpathlineto{\pgfqpoint{1.241836in}{1.490150in}}%
\pgfpathlineto{\pgfqpoint{1.252109in}{1.522177in}}%
\pgfpathlineto{\pgfqpoint{1.272654in}{1.522177in}}%
\pgfpathlineto{\pgfqpoint{1.282927in}{1.538191in}}%
\pgfpathlineto{\pgfqpoint{1.293200in}{1.570218in}}%
\pgfpathlineto{\pgfqpoint{1.313746in}{1.602245in}}%
\pgfpathlineto{\pgfqpoint{1.344564in}{1.698326in}}%
\pgfpathlineto{\pgfqpoint{1.354837in}{1.746366in}}%
\pgfpathlineto{\pgfqpoint{1.375382in}{1.810420in}}%
\pgfpathlineto{\pgfqpoint{1.385655in}{1.826434in}}%
\pgfpathlineto{\pgfqpoint{1.406201in}{1.826434in}}%
\pgfpathlineto{\pgfqpoint{1.426746in}{1.858461in}}%
\pgfpathlineto{\pgfqpoint{1.447292in}{1.858461in}}%
\pgfpathlineto{\pgfqpoint{1.457565in}{1.874474in}}%
\pgfpathlineto{\pgfqpoint{1.467838in}{1.874474in}}%
\pgfpathlineto{\pgfqpoint{1.478110in}{1.890488in}}%
\pgfpathlineto{\pgfqpoint{1.488383in}{1.890488in}}%
\pgfpathlineto{\pgfqpoint{1.498656in}{1.922515in}}%
\pgfpathlineto{\pgfqpoint{1.508929in}{1.922515in}}%
\pgfpathlineto{\pgfqpoint{1.519202in}{1.938528in}}%
\pgfpathlineto{\pgfqpoint{1.529474in}{1.970555in}}%
\pgfpathlineto{\pgfqpoint{1.539747in}{1.970555in}}%
\pgfpathlineto{\pgfqpoint{1.560293in}{2.002582in}}%
\pgfpathlineto{\pgfqpoint{1.591111in}{2.002582in}}%
\pgfpathlineto{\pgfqpoint{1.601384in}{2.018596in}}%
\pgfpathlineto{\pgfqpoint{1.611657in}{2.018596in}}%
\pgfpathlineto{\pgfqpoint{1.621930in}{2.082650in}}%
\pgfpathlineto{\pgfqpoint{1.632203in}{2.114677in}}%
\pgfpathlineto{\pgfqpoint{1.652748in}{2.146703in}}%
\pgfpathlineto{\pgfqpoint{2.176661in}{2.146703in}}%
\pgfpathlineto{\pgfqpoint{2.176661in}{2.146703in}}%
\pgfusepath{stroke}%
\end{pgfscope}%
\begin{pgfscope}%
\pgfpathrectangle{\pgfqpoint{0.461616in}{0.431659in}}{\pgfqpoint{1.796714in}{1.796714in}}%
\pgfusepath{clip}%
\pgfsetbuttcap%
\pgfsetroundjoin%
\pgfsetlinewidth{1.003750pt}%
\definecolor{currentstroke}{rgb}{0.839216,0.152941,0.156863}%
\pgfsetstrokecolor{currentstroke}%
\pgfsetdash{{6.400000pt}{1.600000pt}{1.000000pt}{1.600000pt}}{0.000000pt}%
\pgfpathmoveto{\pgfqpoint{0.543285in}{0.513327in}}%
\pgfpathlineto{\pgfqpoint{1.200745in}{0.513327in}}%
\pgfpathlineto{\pgfqpoint{1.211018in}{0.541983in}}%
\pgfpathlineto{\pgfqpoint{1.231563in}{0.541983in}}%
\pgfpathlineto{\pgfqpoint{1.241836in}{0.570639in}}%
\pgfpathlineto{\pgfqpoint{1.262382in}{0.570639in}}%
\pgfpathlineto{\pgfqpoint{1.272654in}{0.599295in}}%
\pgfpathlineto{\pgfqpoint{1.282927in}{0.599295in}}%
\pgfpathlineto{\pgfqpoint{1.293200in}{0.713918in}}%
\pgfpathlineto{\pgfqpoint{1.313746in}{0.771229in}}%
\pgfpathlineto{\pgfqpoint{1.406201in}{0.771229in}}%
\pgfpathlineto{\pgfqpoint{1.426746in}{0.828540in}}%
\pgfpathlineto{\pgfqpoint{1.447292in}{0.828540in}}%
\pgfpathlineto{\pgfqpoint{1.467838in}{0.885852in}}%
\pgfpathlineto{\pgfqpoint{1.478110in}{0.943163in}}%
\pgfpathlineto{\pgfqpoint{1.488383in}{0.971819in}}%
\pgfpathlineto{\pgfqpoint{1.498656in}{1.086442in}}%
\pgfpathlineto{\pgfqpoint{1.508929in}{1.115098in}}%
\pgfpathlineto{\pgfqpoint{1.529474in}{1.287032in}}%
\pgfpathlineto{\pgfqpoint{1.539747in}{1.344343in}}%
\pgfpathlineto{\pgfqpoint{1.560293in}{1.516278in}}%
\pgfpathlineto{\pgfqpoint{1.570566in}{1.544933in}}%
\pgfpathlineto{\pgfqpoint{1.580839in}{1.544933in}}%
\pgfpathlineto{\pgfqpoint{1.591111in}{1.573589in}}%
\pgfpathlineto{\pgfqpoint{1.611657in}{1.573589in}}%
\pgfpathlineto{\pgfqpoint{1.621930in}{1.630901in}}%
\pgfpathlineto{\pgfqpoint{1.632203in}{1.716868in}}%
\pgfpathlineto{\pgfqpoint{1.642475in}{1.774179in}}%
\pgfpathlineto{\pgfqpoint{1.652748in}{1.802835in}}%
\pgfpathlineto{\pgfqpoint{1.663021in}{1.888802in}}%
\pgfpathlineto{\pgfqpoint{1.673294in}{1.917458in}}%
\pgfpathlineto{\pgfqpoint{1.683567in}{1.917458in}}%
\pgfpathlineto{\pgfqpoint{1.693839in}{1.974769in}}%
\pgfpathlineto{\pgfqpoint{1.704112in}{1.974769in}}%
\pgfpathlineto{\pgfqpoint{1.714385in}{2.003425in}}%
\pgfpathlineto{\pgfqpoint{1.724658in}{2.003425in}}%
\pgfpathlineto{\pgfqpoint{1.734931in}{2.032081in}}%
\pgfpathlineto{\pgfqpoint{1.755476in}{2.032081in}}%
\pgfpathlineto{\pgfqpoint{1.776022in}{2.089392in}}%
\pgfpathlineto{\pgfqpoint{1.796567in}{2.089392in}}%
\pgfpathlineto{\pgfqpoint{1.806840in}{2.118048in}}%
\pgfpathlineto{\pgfqpoint{1.858204in}{2.118048in}}%
\pgfpathlineto{\pgfqpoint{1.868477in}{2.146703in}}%
\pgfpathlineto{\pgfqpoint{2.176661in}{2.146703in}}%
\pgfpathlineto{\pgfqpoint{2.176661in}{2.146703in}}%
\pgfusepath{stroke}%
\end{pgfscope}%
\begin{pgfscope}%
\pgfpathrectangle{\pgfqpoint{0.461616in}{0.431659in}}{\pgfqpoint{1.796714in}{1.796714in}}%
\pgfusepath{clip}%
\pgfsetbuttcap%
\pgfsetroundjoin%
\pgfsetlinewidth{1.003750pt}%
\definecolor{currentstroke}{rgb}{0.172549,0.627451,0.172549}%
\pgfsetstrokecolor{currentstroke}%
\pgfsetdash{{1.000000pt}{1.650000pt}}{0.000000pt}%
\pgfpathmoveto{\pgfqpoint{0.543285in}{0.513327in}}%
\pgfpathlineto{\pgfqpoint{1.252109in}{0.513327in}}%
\pgfpathlineto{\pgfqpoint{1.262382in}{0.551313in}}%
\pgfpathlineto{\pgfqpoint{1.282927in}{0.551313in}}%
\pgfpathlineto{\pgfqpoint{1.293200in}{0.589298in}}%
\pgfpathlineto{\pgfqpoint{1.437019in}{0.589298in}}%
\pgfpathlineto{\pgfqpoint{1.447292in}{0.627284in}}%
\pgfpathlineto{\pgfqpoint{1.488383in}{0.627284in}}%
\pgfpathlineto{\pgfqpoint{1.498656in}{0.665269in}}%
\pgfpathlineto{\pgfqpoint{1.560293in}{0.665269in}}%
\pgfpathlineto{\pgfqpoint{1.570566in}{0.703255in}}%
\pgfpathlineto{\pgfqpoint{1.632203in}{0.703255in}}%
\pgfpathlineto{\pgfqpoint{1.642475in}{0.741240in}}%
\pgfpathlineto{\pgfqpoint{1.652748in}{0.741240in}}%
\pgfpathlineto{\pgfqpoint{1.663021in}{0.779226in}}%
\pgfpathlineto{\pgfqpoint{1.683567in}{0.779226in}}%
\pgfpathlineto{\pgfqpoint{1.693839in}{0.817211in}}%
\pgfpathlineto{\pgfqpoint{1.704112in}{0.817211in}}%
\pgfpathlineto{\pgfqpoint{1.714385in}{0.931168in}}%
\pgfpathlineto{\pgfqpoint{1.724658in}{1.007139in}}%
\pgfpathlineto{\pgfqpoint{1.755476in}{1.121095in}}%
\pgfpathlineto{\pgfqpoint{1.765749in}{1.273037in}}%
\pgfpathlineto{\pgfqpoint{1.776022in}{1.349008in}}%
\pgfpathlineto{\pgfqpoint{1.786295in}{1.500950in}}%
\pgfpathlineto{\pgfqpoint{1.878750in}{1.500950in}}%
\pgfpathlineto{\pgfqpoint{1.889023in}{1.538936in}}%
\pgfpathlineto{\pgfqpoint{1.940387in}{1.538936in}}%
\pgfpathlineto{\pgfqpoint{1.950659in}{1.576921in}}%
\pgfpathlineto{\pgfqpoint{2.084206in}{1.576921in}}%
\pgfpathlineto{\pgfqpoint{2.104752in}{1.652892in}}%
\pgfpathlineto{\pgfqpoint{2.115024in}{1.804834in}}%
\pgfpathlineto{\pgfqpoint{2.125297in}{1.842820in}}%
\pgfpathlineto{\pgfqpoint{2.145843in}{1.994762in}}%
\pgfpathlineto{\pgfqpoint{2.156116in}{2.032747in}}%
\pgfpathlineto{\pgfqpoint{2.166388in}{2.108718in}}%
\pgfpathlineto{\pgfqpoint{2.176661in}{2.146703in}}%
\pgfpathlineto{\pgfqpoint{2.176661in}{2.146703in}}%
\pgfusepath{stroke}%
\end{pgfscope}%
\begin{pgfscope}%
\pgfsetrectcap%
\pgfsetmiterjoin%
\pgfsetlinewidth{1.003750pt}%
\definecolor{currentstroke}{rgb}{1.000000,1.000000,1.000000}%
\pgfsetstrokecolor{currentstroke}%
\pgfsetdash{}{0pt}%
\pgfpathmoveto{\pgfqpoint{0.461616in}{0.431659in}}%
\pgfpathlineto{\pgfqpoint{0.461616in}{2.228372in}}%
\pgfusepath{stroke}%
\end{pgfscope}%
\begin{pgfscope}%
\pgfsetrectcap%
\pgfsetmiterjoin%
\pgfsetlinewidth{1.003750pt}%
\definecolor{currentstroke}{rgb}{1.000000,1.000000,1.000000}%
\pgfsetstrokecolor{currentstroke}%
\pgfsetdash{}{0pt}%
\pgfpathmoveto{\pgfqpoint{2.258330in}{0.431659in}}%
\pgfpathlineto{\pgfqpoint{2.258330in}{2.228372in}}%
\pgfusepath{stroke}%
\end{pgfscope}%
\begin{pgfscope}%
\pgfsetrectcap%
\pgfsetmiterjoin%
\pgfsetlinewidth{1.003750pt}%
\definecolor{currentstroke}{rgb}{1.000000,1.000000,1.000000}%
\pgfsetstrokecolor{currentstroke}%
\pgfsetdash{}{0pt}%
\pgfpathmoveto{\pgfqpoint{0.461616in}{0.431659in}}%
\pgfpathlineto{\pgfqpoint{2.258330in}{0.431659in}}%
\pgfusepath{stroke}%
\end{pgfscope}%
\begin{pgfscope}%
\pgfsetrectcap%
\pgfsetmiterjoin%
\pgfsetlinewidth{1.003750pt}%
\definecolor{currentstroke}{rgb}{1.000000,1.000000,1.000000}%
\pgfsetstrokecolor{currentstroke}%
\pgfsetdash{}{0pt}%
\pgfpathmoveto{\pgfqpoint{0.461616in}{2.228372in}}%
\pgfpathlineto{\pgfqpoint{2.258330in}{2.228372in}}%
\pgfusepath{stroke}%
\end{pgfscope}%
\begin{pgfscope}%
\definecolor{textcolor}{rgb}{0.000000,0.000000,0.000000}%
\pgfsetstrokecolor{textcolor}%
\pgfsetfillcolor{textcolor}%
\pgftext[x=0.902833in,y=1.330015in,left,]{\color{textcolor}\rmfamily\fontsize{7.000000}{8.400000}\selectfont 30}%
\end{pgfscope}%
\begin{pgfscope}%
\definecolor{textcolor}{rgb}{0.000000,0.000000,0.000000}%
\pgfsetstrokecolor{textcolor}%
\pgfsetfillcolor{textcolor}%
\pgftext[x=1.190472in,y=1.330015in,left,]{\color{textcolor}\rmfamily\fontsize{7.000000}{8.400000}\selectfont 58}%
\end{pgfscope}%
\begin{pgfscope}%
\definecolor{textcolor}{rgb}{0.000000,0.000000,0.000000}%
\pgfsetstrokecolor{textcolor}%
\pgfsetfillcolor{textcolor}%
\pgftext[x=1.591111in,y=1.330015in,left,]{\color{textcolor}\rmfamily\fontsize{7.000000}{8.400000}\selectfont 97}%
\end{pgfscope}%
\begin{pgfscope}%
\definecolor{textcolor}{rgb}{0.000000,0.000000,0.000000}%
\pgfsetstrokecolor{textcolor}%
\pgfsetfillcolor{textcolor}%
\pgftext[x=1.827386in,y=1.330015in,left,]{\color{textcolor}\rmfamily\fontsize{7.000000}{8.400000}\selectfont 120}%
\end{pgfscope}%
\end{pgfpicture}%
\makeatother%
\endgroup%

%% file: figures/starlink_all_bariloche_beijing_pareto.pgf
\begingroup%
\makeatletter%
\begin{pgfpicture}%
\pgfpathrectangle{\pgfpointorigin}{\pgfqpoint{2.300000in}{2.300000in}}%
\pgfusepath{use as bounding box, clip}%
\begin{pgfscope}%
\pgfsetbuttcap%
\pgfsetmiterjoin%
\definecolor{currentfill}{rgb}{1.000000,1.000000,1.000000}%
\pgfsetfillcolor{currentfill}%
\pgfsetlinewidth{0.000000pt}%
\definecolor{currentstroke}{rgb}{0.500000,0.500000,0.500000}%
\pgfsetstrokecolor{currentstroke}%
\pgfsetdash{}{0pt}%
\pgfpathmoveto{\pgfqpoint{0.000000in}{0.000000in}}%
\pgfpathlineto{\pgfqpoint{2.300000in}{0.000000in}}%
\pgfpathlineto{\pgfqpoint{2.300000in}{2.300000in}}%
\pgfpathlineto{\pgfqpoint{0.000000in}{2.300000in}}%
\pgfpathlineto{\pgfqpoint{0.000000in}{0.000000in}}%
\pgfpathclose%
\pgfusepath{fill}%
\end{pgfscope}%
\begin{pgfscope}%
\pgfsetbuttcap%
\pgfsetmiterjoin%
\definecolor{currentfill}{rgb}{0.898039,0.898039,0.898039}%
\pgfsetfillcolor{currentfill}%
\pgfsetlinewidth{0.000000pt}%
\definecolor{currentstroke}{rgb}{0.000000,0.000000,0.000000}%
\pgfsetstrokecolor{currentstroke}%
\pgfsetstrokeopacity{0.000000}%
\pgfsetdash{}{0pt}%
\pgfpathmoveto{\pgfqpoint{0.444063in}{0.427270in}}%
\pgfpathlineto{\pgfqpoint{2.258330in}{0.427270in}}%
\pgfpathlineto{\pgfqpoint{2.258330in}{2.241538in}}%
\pgfpathlineto{\pgfqpoint{0.444063in}{2.241538in}}%
\pgfpathlineto{\pgfqpoint{0.444063in}{0.427270in}}%
\pgfpathclose%
\pgfusepath{fill}%
\end{pgfscope}%
\begin{pgfscope}%
\pgfpathrectangle{\pgfqpoint{0.444063in}{0.427270in}}{\pgfqpoint{1.814267in}{1.814267in}}%
\pgfusepath{clip}%
\pgfsetrectcap%
\pgfsetroundjoin%
\pgfsetlinewidth{0.803000pt}%
\definecolor{currentstroke}{rgb}{1.000000,1.000000,1.000000}%
\pgfsetstrokecolor{currentstroke}%
\pgfsetdash{}{0pt}%
\pgfpathmoveto{\pgfqpoint{2.112115in}{0.427270in}}%
\pgfpathlineto{\pgfqpoint{2.112115in}{2.241538in}}%
\pgfusepath{stroke}%
\end{pgfscope}%
\begin{pgfscope}%
\pgfsetbuttcap%
\pgfsetroundjoin%
\definecolor{currentfill}{rgb}{0.333333,0.333333,0.333333}%
\pgfsetfillcolor{currentfill}%
\pgfsetlinewidth{0.803000pt}%
\definecolor{currentstroke}{rgb}{0.333333,0.333333,0.333333}%
\pgfsetstrokecolor{currentstroke}%
\pgfsetdash{}{0pt}%
\pgfsys@defobject{currentmarker}{\pgfqpoint{0.000000in}{-0.048611in}}{\pgfqpoint{0.000000in}{0.000000in}}{%
\pgfpathmoveto{\pgfqpoint{0.000000in}{0.000000in}}%
\pgfpathlineto{\pgfqpoint{0.000000in}{-0.048611in}}%
\pgfusepath{stroke,fill}%
}%
\begin{pgfscope}%
\pgfsys@transformshift{2.112115in}{0.427270in}%
\pgfsys@useobject{currentmarker}{}%
\end{pgfscope}%
\end{pgfscope}%
\begin{pgfscope}%
\definecolor{textcolor}{rgb}{0.333333,0.333333,0.333333}%
\pgfsetstrokecolor{textcolor}%
\pgfsetfillcolor{textcolor}%
\pgftext[x=2.112115in,y=0.330048in,,top]{\color{textcolor}\rmfamily\fontsize{7.000000}{8.400000}\selectfont \(\displaystyle {90}\)}%
\end{pgfscope}%
\begin{pgfscope}%
\pgfpathrectangle{\pgfqpoint{0.444063in}{0.427270in}}{\pgfqpoint{1.814267in}{1.814267in}}%
\pgfusepath{clip}%
\pgfsetrectcap%
\pgfsetroundjoin%
\pgfsetlinewidth{0.803000pt}%
\definecolor{currentstroke}{rgb}{1.000000,1.000000,1.000000}%
\pgfsetstrokecolor{currentstroke}%
\pgfsetdash{}{0pt}%
\pgfpathmoveto{\pgfqpoint{1.768933in}{0.427270in}}%
\pgfpathlineto{\pgfqpoint{1.768933in}{2.241538in}}%
\pgfusepath{stroke}%
\end{pgfscope}%
\begin{pgfscope}%
\pgfsetbuttcap%
\pgfsetroundjoin%
\definecolor{currentfill}{rgb}{0.333333,0.333333,0.333333}%
\pgfsetfillcolor{currentfill}%
\pgfsetlinewidth{0.803000pt}%
\definecolor{currentstroke}{rgb}{0.333333,0.333333,0.333333}%
\pgfsetstrokecolor{currentstroke}%
\pgfsetdash{}{0pt}%
\pgfsys@defobject{currentmarker}{\pgfqpoint{0.000000in}{-0.048611in}}{\pgfqpoint{0.000000in}{0.000000in}}{%
\pgfpathmoveto{\pgfqpoint{0.000000in}{0.000000in}}%
\pgfpathlineto{\pgfqpoint{0.000000in}{-0.048611in}}%
\pgfusepath{stroke,fill}%
}%
\begin{pgfscope}%
\pgfsys@transformshift{1.768933in}{0.427270in}%
\pgfsys@useobject{currentmarker}{}%
\end{pgfscope}%
\end{pgfscope}%
\begin{pgfscope}%
\definecolor{textcolor}{rgb}{0.333333,0.333333,0.333333}%
\pgfsetstrokecolor{textcolor}%
\pgfsetfillcolor{textcolor}%
\pgftext[x=1.768933in,y=0.330048in,,top]{\color{textcolor}\rmfamily\fontsize{7.000000}{8.400000}\selectfont \(\displaystyle {100}\)}%
\end{pgfscope}%
\begin{pgfscope}%
\pgfpathrectangle{\pgfqpoint{0.444063in}{0.427270in}}{\pgfqpoint{1.814267in}{1.814267in}}%
\pgfusepath{clip}%
\pgfsetrectcap%
\pgfsetroundjoin%
\pgfsetlinewidth{0.803000pt}%
\definecolor{currentstroke}{rgb}{1.000000,1.000000,1.000000}%
\pgfsetstrokecolor{currentstroke}%
\pgfsetdash{}{0pt}%
\pgfpathmoveto{\pgfqpoint{1.425751in}{0.427270in}}%
\pgfpathlineto{\pgfqpoint{1.425751in}{2.241538in}}%
\pgfusepath{stroke}%
\end{pgfscope}%
\begin{pgfscope}%
\pgfsetbuttcap%
\pgfsetroundjoin%
\definecolor{currentfill}{rgb}{0.333333,0.333333,0.333333}%
\pgfsetfillcolor{currentfill}%
\pgfsetlinewidth{0.803000pt}%
\definecolor{currentstroke}{rgb}{0.333333,0.333333,0.333333}%
\pgfsetstrokecolor{currentstroke}%
\pgfsetdash{}{0pt}%
\pgfsys@defobject{currentmarker}{\pgfqpoint{0.000000in}{-0.048611in}}{\pgfqpoint{0.000000in}{0.000000in}}{%
\pgfpathmoveto{\pgfqpoint{0.000000in}{0.000000in}}%
\pgfpathlineto{\pgfqpoint{0.000000in}{-0.048611in}}%
\pgfusepath{stroke,fill}%
}%
\begin{pgfscope}%
\pgfsys@transformshift{1.425751in}{0.427270in}%
\pgfsys@useobject{currentmarker}{}%
\end{pgfscope}%
\end{pgfscope}%
\begin{pgfscope}%
\definecolor{textcolor}{rgb}{0.333333,0.333333,0.333333}%
\pgfsetstrokecolor{textcolor}%
\pgfsetfillcolor{textcolor}%
\pgftext[x=1.425751in,y=0.330048in,,top]{\color{textcolor}\rmfamily\fontsize{7.000000}{8.400000}\selectfont \(\displaystyle {110}\)}%
\end{pgfscope}%
\begin{pgfscope}%
\pgfpathrectangle{\pgfqpoint{0.444063in}{0.427270in}}{\pgfqpoint{1.814267in}{1.814267in}}%
\pgfusepath{clip}%
\pgfsetrectcap%
\pgfsetroundjoin%
\pgfsetlinewidth{0.803000pt}%
\definecolor{currentstroke}{rgb}{1.000000,1.000000,1.000000}%
\pgfsetstrokecolor{currentstroke}%
\pgfsetdash{}{0pt}%
\pgfpathmoveto{\pgfqpoint{1.082569in}{0.427270in}}%
\pgfpathlineto{\pgfqpoint{1.082569in}{2.241538in}}%
\pgfusepath{stroke}%
\end{pgfscope}%
\begin{pgfscope}%
\pgfsetbuttcap%
\pgfsetroundjoin%
\definecolor{currentfill}{rgb}{0.333333,0.333333,0.333333}%
\pgfsetfillcolor{currentfill}%
\pgfsetlinewidth{0.803000pt}%
\definecolor{currentstroke}{rgb}{0.333333,0.333333,0.333333}%
\pgfsetstrokecolor{currentstroke}%
\pgfsetdash{}{0pt}%
\pgfsys@defobject{currentmarker}{\pgfqpoint{0.000000in}{-0.048611in}}{\pgfqpoint{0.000000in}{0.000000in}}{%
\pgfpathmoveto{\pgfqpoint{0.000000in}{0.000000in}}%
\pgfpathlineto{\pgfqpoint{0.000000in}{-0.048611in}}%
\pgfusepath{stroke,fill}%
}%
\begin{pgfscope}%
\pgfsys@transformshift{1.082569in}{0.427270in}%
\pgfsys@useobject{currentmarker}{}%
\end{pgfscope}%
\end{pgfscope}%
\begin{pgfscope}%
\definecolor{textcolor}{rgb}{0.333333,0.333333,0.333333}%
\pgfsetstrokecolor{textcolor}%
\pgfsetfillcolor{textcolor}%
\pgftext[x=1.082569in,y=0.330048in,,top]{\color{textcolor}\rmfamily\fontsize{7.000000}{8.400000}\selectfont \(\displaystyle {120}\)}%
\end{pgfscope}%
\begin{pgfscope}%
\pgfpathrectangle{\pgfqpoint{0.444063in}{0.427270in}}{\pgfqpoint{1.814267in}{1.814267in}}%
\pgfusepath{clip}%
\pgfsetrectcap%
\pgfsetroundjoin%
\pgfsetlinewidth{0.803000pt}%
\definecolor{currentstroke}{rgb}{1.000000,1.000000,1.000000}%
\pgfsetstrokecolor{currentstroke}%
\pgfsetdash{}{0pt}%
\pgfpathmoveto{\pgfqpoint{0.739387in}{0.427270in}}%
\pgfpathlineto{\pgfqpoint{0.739387in}{2.241538in}}%
\pgfusepath{stroke}%
\end{pgfscope}%
\begin{pgfscope}%
\pgfsetbuttcap%
\pgfsetroundjoin%
\definecolor{currentfill}{rgb}{0.333333,0.333333,0.333333}%
\pgfsetfillcolor{currentfill}%
\pgfsetlinewidth{0.803000pt}%
\definecolor{currentstroke}{rgb}{0.333333,0.333333,0.333333}%
\pgfsetstrokecolor{currentstroke}%
\pgfsetdash{}{0pt}%
\pgfsys@defobject{currentmarker}{\pgfqpoint{0.000000in}{-0.048611in}}{\pgfqpoint{0.000000in}{0.000000in}}{%
\pgfpathmoveto{\pgfqpoint{0.000000in}{0.000000in}}%
\pgfpathlineto{\pgfqpoint{0.000000in}{-0.048611in}}%
\pgfusepath{stroke,fill}%
}%
\begin{pgfscope}%
\pgfsys@transformshift{0.739387in}{0.427270in}%
\pgfsys@useobject{currentmarker}{}%
\end{pgfscope}%
\end{pgfscope}%
\begin{pgfscope}%
\definecolor{textcolor}{rgb}{0.333333,0.333333,0.333333}%
\pgfsetstrokecolor{textcolor}%
\pgfsetfillcolor{textcolor}%
\pgftext[x=0.739387in,y=0.330048in,,top]{\color{textcolor}\rmfamily\fontsize{7.000000}{8.400000}\selectfont \(\displaystyle {130}\)}%
\end{pgfscope}%
\begin{pgfscope}%
\definecolor{textcolor}{rgb}{0.333333,0.333333,0.333333}%
\pgfsetstrokecolor{textcolor}%
\pgfsetfillcolor{textcolor}%
\pgftext[x=1.351196in,y=0.188073in,,top]{\color{textcolor}\rmfamily\fontsize{10.000000}{12.000000}\selectfont Avg. End-to-end Delay [ms]}%
\end{pgfscope}%
\begin{pgfscope}%
\pgfpathrectangle{\pgfqpoint{0.444063in}{0.427270in}}{\pgfqpoint{1.814267in}{1.814267in}}%
\pgfusepath{clip}%
\pgfsetrectcap%
\pgfsetroundjoin%
\pgfsetlinewidth{0.803000pt}%
\definecolor{currentstroke}{rgb}{1.000000,1.000000,1.000000}%
\pgfsetstrokecolor{currentstroke}%
\pgfsetdash{}{0pt}%
\pgfpathmoveto{\pgfqpoint{0.444063in}{0.713207in}}%
\pgfpathlineto{\pgfqpoint{2.258330in}{0.713207in}}%
\pgfusepath{stroke}%
\end{pgfscope}%
\begin{pgfscope}%
\pgfsetbuttcap%
\pgfsetroundjoin%
\definecolor{currentfill}{rgb}{0.333333,0.333333,0.333333}%
\pgfsetfillcolor{currentfill}%
\pgfsetlinewidth{0.803000pt}%
\definecolor{currentstroke}{rgb}{0.333333,0.333333,0.333333}%
\pgfsetstrokecolor{currentstroke}%
\pgfsetdash{}{0pt}%
\pgfsys@defobject{currentmarker}{\pgfqpoint{-0.048611in}{0.000000in}}{\pgfqpoint{-0.000000in}{0.000000in}}{%
\pgfpathmoveto{\pgfqpoint{-0.000000in}{0.000000in}}%
\pgfpathlineto{\pgfqpoint{-0.048611in}{0.000000in}}%
\pgfusepath{stroke,fill}%
}%
\begin{pgfscope}%
\pgfsys@transformshift{0.444063in}{0.713207in}%
\pgfsys@useobject{currentmarker}{}%
\end{pgfscope}%
\end{pgfscope}%
\begin{pgfscope}%
\definecolor{textcolor}{rgb}{0.333333,0.333333,0.333333}%
\pgfsetstrokecolor{textcolor}%
\pgfsetfillcolor{textcolor}%
\pgftext[x=0.291477in, y=0.679450in, left, base]{\color{textcolor}\rmfamily\fontsize{7.000000}{8.400000}\selectfont \(\displaystyle {8}\)}%
\end{pgfscope}%
\begin{pgfscope}%
\pgfpathrectangle{\pgfqpoint{0.444063in}{0.427270in}}{\pgfqpoint{1.814267in}{1.814267in}}%
\pgfusepath{clip}%
\pgfsetrectcap%
\pgfsetroundjoin%
\pgfsetlinewidth{0.803000pt}%
\definecolor{currentstroke}{rgb}{1.000000,1.000000,1.000000}%
\pgfsetstrokecolor{currentstroke}%
\pgfsetdash{}{0pt}%
\pgfpathmoveto{\pgfqpoint{0.444063in}{1.029008in}}%
\pgfpathlineto{\pgfqpoint{2.258330in}{1.029008in}}%
\pgfusepath{stroke}%
\end{pgfscope}%
\begin{pgfscope}%
\pgfsetbuttcap%
\pgfsetroundjoin%
\definecolor{currentfill}{rgb}{0.333333,0.333333,0.333333}%
\pgfsetfillcolor{currentfill}%
\pgfsetlinewidth{0.803000pt}%
\definecolor{currentstroke}{rgb}{0.333333,0.333333,0.333333}%
\pgfsetstrokecolor{currentstroke}%
\pgfsetdash{}{0pt}%
\pgfsys@defobject{currentmarker}{\pgfqpoint{-0.048611in}{0.000000in}}{\pgfqpoint{-0.000000in}{0.000000in}}{%
\pgfpathmoveto{\pgfqpoint{-0.000000in}{0.000000in}}%
\pgfpathlineto{\pgfqpoint{-0.048611in}{0.000000in}}%
\pgfusepath{stroke,fill}%
}%
\begin{pgfscope}%
\pgfsys@transformshift{0.444063in}{1.029008in}%
\pgfsys@useobject{currentmarker}{}%
\end{pgfscope}%
\end{pgfscope}%
\begin{pgfscope}%
\definecolor{textcolor}{rgb}{0.333333,0.333333,0.333333}%
\pgfsetstrokecolor{textcolor}%
\pgfsetfillcolor{textcolor}%
\pgftext[x=0.236114in, y=0.995250in, left, base]{\color{textcolor}\rmfamily\fontsize{7.000000}{8.400000}\selectfont \(\displaystyle {10}\)}%
\end{pgfscope}%
\begin{pgfscope}%
\pgfpathrectangle{\pgfqpoint{0.444063in}{0.427270in}}{\pgfqpoint{1.814267in}{1.814267in}}%
\pgfusepath{clip}%
\pgfsetrectcap%
\pgfsetroundjoin%
\pgfsetlinewidth{0.803000pt}%
\definecolor{currentstroke}{rgb}{1.000000,1.000000,1.000000}%
\pgfsetstrokecolor{currentstroke}%
\pgfsetdash{}{0pt}%
\pgfpathmoveto{\pgfqpoint{0.444063in}{1.344808in}}%
\pgfpathlineto{\pgfqpoint{2.258330in}{1.344808in}}%
\pgfusepath{stroke}%
\end{pgfscope}%
\begin{pgfscope}%
\pgfsetbuttcap%
\pgfsetroundjoin%
\definecolor{currentfill}{rgb}{0.333333,0.333333,0.333333}%
\pgfsetfillcolor{currentfill}%
\pgfsetlinewidth{0.803000pt}%
\definecolor{currentstroke}{rgb}{0.333333,0.333333,0.333333}%
\pgfsetstrokecolor{currentstroke}%
\pgfsetdash{}{0pt}%
\pgfsys@defobject{currentmarker}{\pgfqpoint{-0.048611in}{0.000000in}}{\pgfqpoint{-0.000000in}{0.000000in}}{%
\pgfpathmoveto{\pgfqpoint{-0.000000in}{0.000000in}}%
\pgfpathlineto{\pgfqpoint{-0.048611in}{0.000000in}}%
\pgfusepath{stroke,fill}%
}%
\begin{pgfscope}%
\pgfsys@transformshift{0.444063in}{1.344808in}%
\pgfsys@useobject{currentmarker}{}%
\end{pgfscope}%
\end{pgfscope}%
\begin{pgfscope}%
\definecolor{textcolor}{rgb}{0.333333,0.333333,0.333333}%
\pgfsetstrokecolor{textcolor}%
\pgfsetfillcolor{textcolor}%
\pgftext[x=0.236114in, y=1.311050in, left, base]{\color{textcolor}\rmfamily\fontsize{7.000000}{8.400000}\selectfont \(\displaystyle {12}\)}%
\end{pgfscope}%
\begin{pgfscope}%
\pgfpathrectangle{\pgfqpoint{0.444063in}{0.427270in}}{\pgfqpoint{1.814267in}{1.814267in}}%
\pgfusepath{clip}%
\pgfsetrectcap%
\pgfsetroundjoin%
\pgfsetlinewidth{0.803000pt}%
\definecolor{currentstroke}{rgb}{1.000000,1.000000,1.000000}%
\pgfsetstrokecolor{currentstroke}%
\pgfsetdash{}{0pt}%
\pgfpathmoveto{\pgfqpoint{0.444063in}{1.660609in}}%
\pgfpathlineto{\pgfqpoint{2.258330in}{1.660609in}}%
\pgfusepath{stroke}%
\end{pgfscope}%
\begin{pgfscope}%
\pgfsetbuttcap%
\pgfsetroundjoin%
\definecolor{currentfill}{rgb}{0.333333,0.333333,0.333333}%
\pgfsetfillcolor{currentfill}%
\pgfsetlinewidth{0.803000pt}%
\definecolor{currentstroke}{rgb}{0.333333,0.333333,0.333333}%
\pgfsetstrokecolor{currentstroke}%
\pgfsetdash{}{0pt}%
\pgfsys@defobject{currentmarker}{\pgfqpoint{-0.048611in}{0.000000in}}{\pgfqpoint{-0.000000in}{0.000000in}}{%
\pgfpathmoveto{\pgfqpoint{-0.000000in}{0.000000in}}%
\pgfpathlineto{\pgfqpoint{-0.048611in}{0.000000in}}%
\pgfusepath{stroke,fill}%
}%
\begin{pgfscope}%
\pgfsys@transformshift{0.444063in}{1.660609in}%
\pgfsys@useobject{currentmarker}{}%
\end{pgfscope}%
\end{pgfscope}%
\begin{pgfscope}%
\definecolor{textcolor}{rgb}{0.333333,0.333333,0.333333}%
\pgfsetstrokecolor{textcolor}%
\pgfsetfillcolor{textcolor}%
\pgftext[x=0.236114in, y=1.626851in, left, base]{\color{textcolor}\rmfamily\fontsize{7.000000}{8.400000}\selectfont \(\displaystyle {14}\)}%
\end{pgfscope}%
\begin{pgfscope}%
\pgfpathrectangle{\pgfqpoint{0.444063in}{0.427270in}}{\pgfqpoint{1.814267in}{1.814267in}}%
\pgfusepath{clip}%
\pgfsetrectcap%
\pgfsetroundjoin%
\pgfsetlinewidth{0.803000pt}%
\definecolor{currentstroke}{rgb}{1.000000,1.000000,1.000000}%
\pgfsetstrokecolor{currentstroke}%
\pgfsetdash{}{0pt}%
\pgfpathmoveto{\pgfqpoint{0.444063in}{1.976409in}}%
\pgfpathlineto{\pgfqpoint{2.258330in}{1.976409in}}%
\pgfusepath{stroke}%
\end{pgfscope}%
\begin{pgfscope}%
\pgfsetbuttcap%
\pgfsetroundjoin%
\definecolor{currentfill}{rgb}{0.333333,0.333333,0.333333}%
\pgfsetfillcolor{currentfill}%
\pgfsetlinewidth{0.803000pt}%
\definecolor{currentstroke}{rgb}{0.333333,0.333333,0.333333}%
\pgfsetstrokecolor{currentstroke}%
\pgfsetdash{}{0pt}%
\pgfsys@defobject{currentmarker}{\pgfqpoint{-0.048611in}{0.000000in}}{\pgfqpoint{-0.000000in}{0.000000in}}{%
\pgfpathmoveto{\pgfqpoint{-0.000000in}{0.000000in}}%
\pgfpathlineto{\pgfqpoint{-0.048611in}{0.000000in}}%
\pgfusepath{stroke,fill}%
}%
\begin{pgfscope}%
\pgfsys@transformshift{0.444063in}{1.976409in}%
\pgfsys@useobject{currentmarker}{}%
\end{pgfscope}%
\end{pgfscope}%
\begin{pgfscope}%
\definecolor{textcolor}{rgb}{0.333333,0.333333,0.333333}%
\pgfsetstrokecolor{textcolor}%
\pgfsetfillcolor{textcolor}%
\pgftext[x=0.236114in, y=1.942651in, left, base]{\color{textcolor}\rmfamily\fontsize{7.000000}{8.400000}\selectfont \(\displaystyle {16}\)}%
\end{pgfscope}%
\begin{pgfscope}%
\definecolor{textcolor}{rgb}{0.333333,0.333333,0.333333}%
\pgfsetstrokecolor{textcolor}%
\pgfsetfillcolor{textcolor}%
\pgftext[x=0.180559in,y=1.334404in,,bottom,rotate=90.000000]{\color{textcolor}\rmfamily\fontsize{10.000000}{12.000000}\selectfont Avg. Data Rate [MB/s]}%
\end{pgfscope}%
\begin{pgfscope}%
\pgfpathrectangle{\pgfqpoint{0.444063in}{0.427270in}}{\pgfqpoint{1.814267in}{1.814267in}}%
\pgfusepath{clip}%
\pgfsetbuttcap%
\pgfsetroundjoin%
\definecolor{currentfill}{rgb}{1.000000,0.498039,0.054902}%
\pgfsetfillcolor{currentfill}%
\pgfsetlinewidth{1.505625pt}%
\definecolor{currentstroke}{rgb}{1.000000,0.498039,0.054902}%
\pgfsetstrokecolor{currentstroke}%
\pgfsetdash{}{0pt}%
\pgfsys@defobject{currentmarker}{\pgfqpoint{-0.041667in}{-0.041667in}}{\pgfqpoint{0.041667in}{0.041667in}}{%
\pgfpathmoveto{\pgfqpoint{-0.041667in}{0.000000in}}%
\pgfpathlineto{\pgfqpoint{0.041667in}{0.000000in}}%
\pgfpathmoveto{\pgfqpoint{0.000000in}{-0.041667in}}%
\pgfpathlineto{\pgfqpoint{0.000000in}{0.041667in}}%
\pgfusepath{stroke,fill}%
}%
\begin{pgfscope}%
\pgfsys@transformshift{2.175863in}{0.509737in}%
\pgfsys@useobject{currentmarker}{}%
\end{pgfscope}%
\end{pgfscope}%
\begin{pgfscope}%
\pgfpathrectangle{\pgfqpoint{0.444063in}{0.427270in}}{\pgfqpoint{1.814267in}{1.814267in}}%
\pgfusepath{clip}%
\pgfsetbuttcap%
\pgfsetroundjoin%
\definecolor{currentfill}{rgb}{0.121569,0.466667,0.705882}%
\pgfsetfillcolor{currentfill}%
\pgfsetlinewidth{1.505625pt}%
\definecolor{currentstroke}{rgb}{0.121569,0.466667,0.705882}%
\pgfsetstrokecolor{currentstroke}%
\pgfsetdash{}{0pt}%
\pgfsys@defobject{currentmarker}{\pgfqpoint{-0.033333in}{-0.041667in}}{\pgfqpoint{0.033333in}{0.020833in}}{%
\pgfpathmoveto{\pgfqpoint{0.000000in}{0.000000in}}%
\pgfpathlineto{\pgfqpoint{0.000000in}{-0.041667in}}%
\pgfpathmoveto{\pgfqpoint{0.000000in}{0.000000in}}%
\pgfpathlineto{\pgfqpoint{0.033333in}{0.020833in}}%
\pgfpathmoveto{\pgfqpoint{0.000000in}{0.000000in}}%
\pgfpathlineto{\pgfqpoint{-0.033333in}{0.020833in}}%
\pgfusepath{stroke,fill}%
}%
\begin{pgfscope}%
\pgfsys@transformshift{2.136202in}{0.684536in}%
\pgfsys@useobject{currentmarker}{}%
\end{pgfscope}%
\end{pgfscope}%
\begin{pgfscope}%
\pgfpathrectangle{\pgfqpoint{0.444063in}{0.427270in}}{\pgfqpoint{1.814267in}{1.814267in}}%
\pgfusepath{clip}%
\pgfsetbuttcap%
\pgfsetroundjoin%
\definecolor{currentfill}{rgb}{0.839216,0.152941,0.156863}%
\pgfsetfillcolor{currentfill}%
\pgfsetlinewidth{0.501875pt}%
\definecolor{currentstroke}{rgb}{0.839216,0.152941,0.156863}%
\pgfsetstrokecolor{currentstroke}%
\pgfsetdash{}{0pt}%
\pgfsys@defobject{currentmarker}{\pgfqpoint{-0.039627in}{-0.033709in}}{\pgfqpoint{0.039627in}{0.041667in}}{%
\pgfpathmoveto{\pgfqpoint{0.000000in}{0.041667in}}%
\pgfpathlineto{\pgfqpoint{-0.039627in}{0.012876in}}%
\pgfpathlineto{\pgfqpoint{-0.024491in}{-0.033709in}}%
\pgfpathlineto{\pgfqpoint{0.024491in}{-0.033709in}}%
\pgfpathlineto{\pgfqpoint{0.039627in}{0.012876in}}%
\pgfpathlineto{\pgfqpoint{0.000000in}{0.041667in}}%
\pgfpathclose%
\pgfusepath{stroke,fill}%
}%
\begin{pgfscope}%
\pgfsys@transformshift{2.104591in}{2.159071in}%
\pgfsys@useobject{currentmarker}{}%
\end{pgfscope}%
\end{pgfscope}%
\begin{pgfscope}%
\pgfpathrectangle{\pgfqpoint{0.444063in}{0.427270in}}{\pgfqpoint{1.814267in}{1.814267in}}%
\pgfusepath{clip}%
\pgfsetbuttcap%
\pgfsetroundjoin%
\definecolor{currentfill}{rgb}{0.172549,0.627451,0.172549}%
\pgfsetfillcolor{currentfill}%
\pgfsetlinewidth{0.501875pt}%
\definecolor{currentstroke}{rgb}{0.172549,0.627451,0.172549}%
\pgfsetstrokecolor{currentstroke}%
\pgfsetdash{}{0pt}%
\pgfsys@defobject{currentmarker}{\pgfqpoint{-0.039627in}{-0.033709in}}{\pgfqpoint{0.039627in}{0.041667in}}{%
\pgfpathmoveto{\pgfqpoint{0.000000in}{0.041667in}}%
\pgfpathlineto{\pgfqpoint{-0.009355in}{0.012876in}}%
\pgfpathlineto{\pgfqpoint{-0.039627in}{0.012876in}}%
\pgfpathlineto{\pgfqpoint{-0.015136in}{-0.004918in}}%
\pgfpathlineto{\pgfqpoint{-0.024491in}{-0.033709in}}%
\pgfpathlineto{\pgfqpoint{-0.000000in}{-0.015915in}}%
\pgfpathlineto{\pgfqpoint{0.024491in}{-0.033709in}}%
\pgfpathlineto{\pgfqpoint{0.015136in}{-0.004918in}}%
\pgfpathlineto{\pgfqpoint{0.039627in}{0.012876in}}%
\pgfpathlineto{\pgfqpoint{0.009355in}{0.012876in}}%
\pgfpathlineto{\pgfqpoint{0.000000in}{0.041667in}}%
\pgfpathclose%
\pgfusepath{stroke,fill}%
}%
\begin{pgfscope}%
\pgfsys@transformshift{0.526529in}{1.407537in}%
\pgfsys@useobject{currentmarker}{}%
\end{pgfscope}%
\end{pgfscope}%
\begin{pgfscope}%
\pgfsetrectcap%
\pgfsetmiterjoin%
\pgfsetlinewidth{1.003750pt}%
\definecolor{currentstroke}{rgb}{1.000000,1.000000,1.000000}%
\pgfsetstrokecolor{currentstroke}%
\pgfsetdash{}{0pt}%
\pgfpathmoveto{\pgfqpoint{0.444063in}{0.427270in}}%
\pgfpathlineto{\pgfqpoint{0.444063in}{2.241538in}}%
\pgfusepath{stroke}%
\end{pgfscope}%
\begin{pgfscope}%
\pgfsetrectcap%
\pgfsetmiterjoin%
\pgfsetlinewidth{1.003750pt}%
\definecolor{currentstroke}{rgb}{1.000000,1.000000,1.000000}%
\pgfsetstrokecolor{currentstroke}%
\pgfsetdash{}{0pt}%
\pgfpathmoveto{\pgfqpoint{2.258330in}{0.427270in}}%
\pgfpathlineto{\pgfqpoint{2.258330in}{2.241538in}}%
\pgfusepath{stroke}%
\end{pgfscope}%
\begin{pgfscope}%
\pgfsetrectcap%
\pgfsetmiterjoin%
\pgfsetlinewidth{1.003750pt}%
\definecolor{currentstroke}{rgb}{1.000000,1.000000,1.000000}%
\pgfsetstrokecolor{currentstroke}%
\pgfsetdash{}{0pt}%
\pgfpathmoveto{\pgfqpoint{0.444063in}{0.427270in}}%
\pgfpathlineto{\pgfqpoint{2.258330in}{0.427270in}}%
\pgfusepath{stroke}%
\end{pgfscope}%
\begin{pgfscope}%
\pgfsetrectcap%
\pgfsetmiterjoin%
\pgfsetlinewidth{1.003750pt}%
\definecolor{currentstroke}{rgb}{1.000000,1.000000,1.000000}%
\pgfsetstrokecolor{currentstroke}%
\pgfsetdash{}{0pt}%
\pgfpathmoveto{\pgfqpoint{0.444063in}{2.241538in}}%
\pgfpathlineto{\pgfqpoint{2.258330in}{2.241538in}}%
\pgfusepath{stroke}%
\end{pgfscope}%
\begin{pgfscope}%
\pgfsetbuttcap%
\pgfsetmiterjoin%
\definecolor{currentfill}{rgb}{0.898039,0.898039,0.898039}%
\pgfsetfillcolor{currentfill}%
\pgfsetfillopacity{0.800000}%
\pgfsetlinewidth{0.501875pt}%
\definecolor{currentstroke}{rgb}{0.800000,0.800000,0.800000}%
\pgfsetstrokecolor{currentstroke}%
\pgfsetstrokeopacity{0.800000}%
\pgfsetdash{}{0pt}%
\pgfpathmoveto{\pgfqpoint{0.521840in}{0.482826in}}%
\pgfpathlineto{\pgfqpoint{1.449430in}{0.482826in}}%
\pgfpathquadraticcurveto{\pgfqpoint{1.471652in}{0.482826in}}{\pgfqpoint{1.471652in}{0.505048in}}%
\pgfpathlineto{\pgfqpoint{1.471652in}{1.113690in}}%
\pgfpathquadraticcurveto{\pgfqpoint{1.471652in}{1.135912in}}{\pgfqpoint{1.449430in}{1.135912in}}%
\pgfpathlineto{\pgfqpoint{0.521840in}{1.135912in}}%
\pgfpathquadraticcurveto{\pgfqpoint{0.499618in}{1.135912in}}{\pgfqpoint{0.499618in}{1.113690in}}%
\pgfpathlineto{\pgfqpoint{0.499618in}{0.505048in}}%
\pgfpathquadraticcurveto{\pgfqpoint{0.499618in}{0.482826in}}{\pgfqpoint{0.521840in}{0.482826in}}%
\pgfpathlineto{\pgfqpoint{0.521840in}{0.482826in}}%
\pgfpathclose%
\pgfusepath{stroke,fill}%
\end{pgfscope}%
\begin{pgfscope}%
\pgfsetbuttcap%
\pgfsetroundjoin%
\definecolor{currentfill}{rgb}{1.000000,0.498039,0.054902}%
\pgfsetfillcolor{currentfill}%
\pgfsetlinewidth{1.505625pt}%
\definecolor{currentstroke}{rgb}{1.000000,0.498039,0.054902}%
\pgfsetstrokecolor{currentstroke}%
\pgfsetdash{}{0pt}%
\pgfsys@defobject{currentmarker}{\pgfqpoint{-0.041667in}{-0.041667in}}{\pgfqpoint{0.041667in}{0.041667in}}{%
\pgfpathmoveto{\pgfqpoint{-0.041667in}{0.000000in}}%
\pgfpathlineto{\pgfqpoint{0.041667in}{0.000000in}}%
\pgfpathmoveto{\pgfqpoint{0.000000in}{-0.041667in}}%
\pgfpathlineto{\pgfqpoint{0.000000in}{0.041667in}}%
\pgfusepath{stroke,fill}%
}%
\begin{pgfscope}%
\pgfsys@transformshift{0.655174in}{1.042857in}%
\pgfsys@useobject{currentmarker}{}%
\end{pgfscope}%
\end{pgfscope}%
\begin{pgfscope}%
\definecolor{textcolor}{rgb}{0.000000,0.000000,0.000000}%
\pgfsetstrokecolor{textcolor}%
\pgfsetfillcolor{textcolor}%
\pgftext[x=0.855174in,y=1.013690in,left,base]{\color{textcolor}\rmfamily\fontsize{8.000000}{9.600000}\selectfont \textsc{Dijkstra}}%
\end{pgfscope}%
\begin{pgfscope}%
\pgfsetbuttcap%
\pgfsetroundjoin%
\definecolor{currentfill}{rgb}{0.121569,0.466667,0.705882}%
\pgfsetfillcolor{currentfill}%
\pgfsetlinewidth{1.505625pt}%
\definecolor{currentstroke}{rgb}{0.121569,0.466667,0.705882}%
\pgfsetstrokecolor{currentstroke}%
\pgfsetdash{}{0pt}%
\pgfsys@defobject{currentmarker}{\pgfqpoint{-0.033333in}{-0.041667in}}{\pgfqpoint{0.033333in}{0.020833in}}{%
\pgfpathmoveto{\pgfqpoint{0.000000in}{0.000000in}}%
\pgfpathlineto{\pgfqpoint{0.000000in}{-0.041667in}}%
\pgfpathmoveto{\pgfqpoint{0.000000in}{0.000000in}}%
\pgfpathlineto{\pgfqpoint{0.033333in}{0.020833in}}%
\pgfpathmoveto{\pgfqpoint{0.000000in}{0.000000in}}%
\pgfpathlineto{\pgfqpoint{-0.033333in}{0.020833in}}%
\pgfusepath{stroke,fill}%
}%
\begin{pgfscope}%
\pgfsys@transformshift{0.655174in}{0.887919in}%
\pgfsys@useobject{currentmarker}{}%
\end{pgfscope}%
\end{pgfscope}%
\begin{pgfscope}%
\definecolor{textcolor}{rgb}{0.000000,0.000000,0.000000}%
\pgfsetstrokecolor{textcolor}%
\pgfsetfillcolor{textcolor}%
\pgftext[x=0.855174in,y=0.858752in,left,base]{\color{textcolor}\rmfamily\fontsize{8.000000}{9.600000}\selectfont \textsc{Stubborn}}%
\end{pgfscope}%
\begin{pgfscope}%
\pgfsetbuttcap%
\pgfsetroundjoin%
\definecolor{currentfill}{rgb}{0.839216,0.152941,0.156863}%
\pgfsetfillcolor{currentfill}%
\pgfsetlinewidth{0.501875pt}%
\definecolor{currentstroke}{rgb}{0.839216,0.152941,0.156863}%
\pgfsetstrokecolor{currentstroke}%
\pgfsetdash{}{0pt}%
\pgfsys@defobject{currentmarker}{\pgfqpoint{-0.039627in}{-0.033709in}}{\pgfqpoint{0.039627in}{0.041667in}}{%
\pgfpathmoveto{\pgfqpoint{0.000000in}{0.041667in}}%
\pgfpathlineto{\pgfqpoint{-0.039627in}{0.012876in}}%
\pgfpathlineto{\pgfqpoint{-0.024491in}{-0.033709in}}%
\pgfpathlineto{\pgfqpoint{0.024491in}{-0.033709in}}%
\pgfpathlineto{\pgfqpoint{0.039627in}{0.012876in}}%
\pgfpathlineto{\pgfqpoint{0.000000in}{0.041667in}}%
\pgfpathclose%
\pgfusepath{stroke,fill}%
}%
\begin{pgfscope}%
\pgfsys@transformshift{0.655174in}{0.732980in}%
\pgfsys@useobject{currentmarker}{}%
\end{pgfscope}%
\end{pgfscope}%
\begin{pgfscope}%
\definecolor{textcolor}{rgb}{0.000000,0.000000,0.000000}%
\pgfsetstrokecolor{textcolor}%
\pgfsetfillcolor{textcolor}%
\pgftext[x=0.855174in,y=0.703814in,left,base]{\color{textcolor}\rmfamily\fontsize{8.000000}{9.600000}\selectfont \textsc{Tenacious}}%
\end{pgfscope}%
\begin{pgfscope}%
\pgfsetbuttcap%
\pgfsetroundjoin%
\definecolor{currentfill}{rgb}{0.172549,0.627451,0.172549}%
\pgfsetfillcolor{currentfill}%
\pgfsetlinewidth{0.501875pt}%
\definecolor{currentstroke}{rgb}{0.172549,0.627451,0.172549}%
\pgfsetstrokecolor{currentstroke}%
\pgfsetdash{}{0pt}%
\pgfsys@defobject{currentmarker}{\pgfqpoint{-0.039627in}{-0.033709in}}{\pgfqpoint{0.039627in}{0.041667in}}{%
\pgfpathmoveto{\pgfqpoint{0.000000in}{0.041667in}}%
\pgfpathlineto{\pgfqpoint{-0.009355in}{0.012876in}}%
\pgfpathlineto{\pgfqpoint{-0.039627in}{0.012876in}}%
\pgfpathlineto{\pgfqpoint{-0.015136in}{-0.004918in}}%
\pgfpathlineto{\pgfqpoint{-0.024491in}{-0.033709in}}%
\pgfpathlineto{\pgfqpoint{-0.000000in}{-0.015915in}}%
\pgfpathlineto{\pgfqpoint{0.024491in}{-0.033709in}}%
\pgfpathlineto{\pgfqpoint{0.015136in}{-0.004918in}}%
\pgfpathlineto{\pgfqpoint{0.039627in}{0.012876in}}%
\pgfpathlineto{\pgfqpoint{0.009355in}{0.012876in}}%
\pgfpathlineto{\pgfqpoint{0.000000in}{0.041667in}}%
\pgfpathclose%
\pgfusepath{stroke,fill}%
}%
\begin{pgfscope}%
\pgfsys@transformshift{0.655174in}{0.578042in}%
\pgfsys@useobject{currentmarker}{}%
\end{pgfscope}%
\end{pgfscope}%
\begin{pgfscope}%
\definecolor{textcolor}{rgb}{0.000000,0.000000,0.000000}%
\pgfsetstrokecolor{textcolor}%
\pgfsetfillcolor{textcolor}%
\pgftext[x=0.855174in,y=0.548875in,left,base]{\color{textcolor}\rmfamily\fontsize{8.000000}{9.600000}\selectfont \textsc{SetCover}}%
\end{pgfscope}%
\end{pgfpicture}%
\makeatother%
\endgroup%

%% file: figures/grid_avg_delay_validity_starlink_delay.pgf
\begingroup%
\makeatletter%
\begin{pgfpicture}%
\pgfpathrectangle{\pgfpointorigin}{\pgfqpoint{2.300000in}{2.300000in}}%
\pgfusepath{use as bounding box, clip}%
\begin{pgfscope}%
\pgfsetbuttcap%
\pgfsetmiterjoin%
\definecolor{currentfill}{rgb}{1.000000,1.000000,1.000000}%
\pgfsetfillcolor{currentfill}%
\pgfsetlinewidth{0.000000pt}%
\definecolor{currentstroke}{rgb}{0.500000,0.500000,0.500000}%
\pgfsetstrokecolor{currentstroke}%
\pgfsetdash{}{0pt}%
\pgfpathmoveto{\pgfqpoint{0.000000in}{0.000000in}}%
\pgfpathlineto{\pgfqpoint{2.300000in}{0.000000in}}%
\pgfpathlineto{\pgfqpoint{2.300000in}{2.300000in}}%
\pgfpathlineto{\pgfqpoint{0.000000in}{2.300000in}}%
\pgfpathlineto{\pgfqpoint{0.000000in}{0.000000in}}%
\pgfpathclose%
\pgfusepath{fill}%
\end{pgfscope}%
\begin{pgfscope}%
\pgfsetbuttcap%
\pgfsetmiterjoin%
\definecolor{currentfill}{rgb}{0.898039,0.898039,0.898039}%
\pgfsetfillcolor{currentfill}%
\pgfsetlinewidth{0.000000pt}%
\definecolor{currentstroke}{rgb}{0.000000,0.000000,0.000000}%
\pgfsetstrokecolor{currentstroke}%
\pgfsetstrokeopacity{0.000000}%
\pgfsetdash{}{0pt}%
\pgfpathmoveto{\pgfqpoint{0.461616in}{0.433022in}}%
\pgfpathlineto{\pgfqpoint{2.255603in}{0.433022in}}%
\pgfpathlineto{\pgfqpoint{2.255603in}{2.227009in}}%
\pgfpathlineto{\pgfqpoint{0.461616in}{2.227009in}}%
\pgfpathlineto{\pgfqpoint{0.461616in}{0.433022in}}%
\pgfpathclose%
\pgfusepath{fill}%
\end{pgfscope}%
\begin{pgfscope}%
\pgfpathrectangle{\pgfqpoint{0.461616in}{0.433022in}}{\pgfqpoint{1.793987in}{1.793987in}}%
\pgfusepath{clip}%
\pgfsetrectcap%
\pgfsetroundjoin%
\pgfsetlinewidth{0.803000pt}%
\definecolor{currentstroke}{rgb}{1.000000,1.000000,1.000000}%
\pgfsetstrokecolor{currentstroke}%
\pgfsetdash{}{0pt}%
\pgfpathmoveto{\pgfqpoint{0.511482in}{0.433022in}}%
\pgfpathlineto{\pgfqpoint{0.511482in}{2.227009in}}%
\pgfusepath{stroke}%
\end{pgfscope}%
\begin{pgfscope}%
\pgfsetbuttcap%
\pgfsetroundjoin%
\definecolor{currentfill}{rgb}{0.333333,0.333333,0.333333}%
\pgfsetfillcolor{currentfill}%
\pgfsetlinewidth{0.803000pt}%
\definecolor{currentstroke}{rgb}{0.333333,0.333333,0.333333}%
\pgfsetstrokecolor{currentstroke}%
\pgfsetdash{}{0pt}%
\pgfsys@defobject{currentmarker}{\pgfqpoint{0.000000in}{-0.048611in}}{\pgfqpoint{0.000000in}{0.000000in}}{%
\pgfpathmoveto{\pgfqpoint{0.000000in}{0.000000in}}%
\pgfpathlineto{\pgfqpoint{0.000000in}{-0.048611in}}%
\pgfusepath{stroke,fill}%
}%
\begin{pgfscope}%
\pgfsys@transformshift{0.511482in}{0.433022in}%
\pgfsys@useobject{currentmarker}{}%
\end{pgfscope}%
\end{pgfscope}%
\begin{pgfscope}%
\definecolor{textcolor}{rgb}{0.333333,0.333333,0.333333}%
\pgfsetstrokecolor{textcolor}%
\pgfsetfillcolor{textcolor}%
\pgftext[x=0.511482in,y=0.335800in,,top]{\color{textcolor}\rmfamily\fontsize{7.000000}{8.400000}\selectfont \(\displaystyle {0}\)}%
\end{pgfscope}%
\begin{pgfscope}%
\pgfpathrectangle{\pgfqpoint{0.461616in}{0.433022in}}{\pgfqpoint{1.793987in}{1.793987in}}%
\pgfusepath{clip}%
\pgfsetrectcap%
\pgfsetroundjoin%
\pgfsetlinewidth{0.803000pt}%
\definecolor{currentstroke}{rgb}{1.000000,1.000000,1.000000}%
\pgfsetstrokecolor{currentstroke}%
\pgfsetdash{}{0pt}%
\pgfpathmoveto{\pgfqpoint{0.844319in}{0.433022in}}%
\pgfpathlineto{\pgfqpoint{0.844319in}{2.227009in}}%
\pgfusepath{stroke}%
\end{pgfscope}%
\begin{pgfscope}%
\pgfsetbuttcap%
\pgfsetroundjoin%
\definecolor{currentfill}{rgb}{0.333333,0.333333,0.333333}%
\pgfsetfillcolor{currentfill}%
\pgfsetlinewidth{0.803000pt}%
\definecolor{currentstroke}{rgb}{0.333333,0.333333,0.333333}%
\pgfsetstrokecolor{currentstroke}%
\pgfsetdash{}{0pt}%
\pgfsys@defobject{currentmarker}{\pgfqpoint{0.000000in}{-0.048611in}}{\pgfqpoint{0.000000in}{0.000000in}}{%
\pgfpathmoveto{\pgfqpoint{0.000000in}{0.000000in}}%
\pgfpathlineto{\pgfqpoint{0.000000in}{-0.048611in}}%
\pgfusepath{stroke,fill}%
}%
\begin{pgfscope}%
\pgfsys@transformshift{0.844319in}{0.433022in}%
\pgfsys@useobject{currentmarker}{}%
\end{pgfscope}%
\end{pgfscope}%
\begin{pgfscope}%
\definecolor{textcolor}{rgb}{0.333333,0.333333,0.333333}%
\pgfsetstrokecolor{textcolor}%
\pgfsetfillcolor{textcolor}%
\pgftext[x=0.844319in,y=0.335800in,,top]{\color{textcolor}\rmfamily\fontsize{7.000000}{8.400000}\selectfont \(\displaystyle {50}\)}%
\end{pgfscope}%
\begin{pgfscope}%
\pgfpathrectangle{\pgfqpoint{0.461616in}{0.433022in}}{\pgfqpoint{1.793987in}{1.793987in}}%
\pgfusepath{clip}%
\pgfsetrectcap%
\pgfsetroundjoin%
\pgfsetlinewidth{0.803000pt}%
\definecolor{currentstroke}{rgb}{1.000000,1.000000,1.000000}%
\pgfsetstrokecolor{currentstroke}%
\pgfsetdash{}{0pt}%
\pgfpathmoveto{\pgfqpoint{1.177155in}{0.433022in}}%
\pgfpathlineto{\pgfqpoint{1.177155in}{2.227009in}}%
\pgfusepath{stroke}%
\end{pgfscope}%
\begin{pgfscope}%
\pgfsetbuttcap%
\pgfsetroundjoin%
\definecolor{currentfill}{rgb}{0.333333,0.333333,0.333333}%
\pgfsetfillcolor{currentfill}%
\pgfsetlinewidth{0.803000pt}%
\definecolor{currentstroke}{rgb}{0.333333,0.333333,0.333333}%
\pgfsetstrokecolor{currentstroke}%
\pgfsetdash{}{0pt}%
\pgfsys@defobject{currentmarker}{\pgfqpoint{0.000000in}{-0.048611in}}{\pgfqpoint{0.000000in}{0.000000in}}{%
\pgfpathmoveto{\pgfqpoint{0.000000in}{0.000000in}}%
\pgfpathlineto{\pgfqpoint{0.000000in}{-0.048611in}}%
\pgfusepath{stroke,fill}%
}%
\begin{pgfscope}%
\pgfsys@transformshift{1.177155in}{0.433022in}%
\pgfsys@useobject{currentmarker}{}%
\end{pgfscope}%
\end{pgfscope}%
\begin{pgfscope}%
\definecolor{textcolor}{rgb}{0.333333,0.333333,0.333333}%
\pgfsetstrokecolor{textcolor}%
\pgfsetfillcolor{textcolor}%
\pgftext[x=1.177155in,y=0.335800in,,top]{\color{textcolor}\rmfamily\fontsize{7.000000}{8.400000}\selectfont \(\displaystyle {100}\)}%
\end{pgfscope}%
\begin{pgfscope}%
\pgfpathrectangle{\pgfqpoint{0.461616in}{0.433022in}}{\pgfqpoint{1.793987in}{1.793987in}}%
\pgfusepath{clip}%
\pgfsetrectcap%
\pgfsetroundjoin%
\pgfsetlinewidth{0.803000pt}%
\definecolor{currentstroke}{rgb}{1.000000,1.000000,1.000000}%
\pgfsetstrokecolor{currentstroke}%
\pgfsetdash{}{0pt}%
\pgfpathmoveto{\pgfqpoint{1.509991in}{0.433022in}}%
\pgfpathlineto{\pgfqpoint{1.509991in}{2.227009in}}%
\pgfusepath{stroke}%
\end{pgfscope}%
\begin{pgfscope}%
\pgfsetbuttcap%
\pgfsetroundjoin%
\definecolor{currentfill}{rgb}{0.333333,0.333333,0.333333}%
\pgfsetfillcolor{currentfill}%
\pgfsetlinewidth{0.803000pt}%
\definecolor{currentstroke}{rgb}{0.333333,0.333333,0.333333}%
\pgfsetstrokecolor{currentstroke}%
\pgfsetdash{}{0pt}%
\pgfsys@defobject{currentmarker}{\pgfqpoint{0.000000in}{-0.048611in}}{\pgfqpoint{0.000000in}{0.000000in}}{%
\pgfpathmoveto{\pgfqpoint{0.000000in}{0.000000in}}%
\pgfpathlineto{\pgfqpoint{0.000000in}{-0.048611in}}%
\pgfusepath{stroke,fill}%
}%
\begin{pgfscope}%
\pgfsys@transformshift{1.509991in}{0.433022in}%
\pgfsys@useobject{currentmarker}{}%
\end{pgfscope}%
\end{pgfscope}%
\begin{pgfscope}%
\definecolor{textcolor}{rgb}{0.333333,0.333333,0.333333}%
\pgfsetstrokecolor{textcolor}%
\pgfsetfillcolor{textcolor}%
\pgftext[x=1.509991in,y=0.335800in,,top]{\color{textcolor}\rmfamily\fontsize{7.000000}{8.400000}\selectfont \(\displaystyle {150}\)}%
\end{pgfscope}%
\begin{pgfscope}%
\pgfpathrectangle{\pgfqpoint{0.461616in}{0.433022in}}{\pgfqpoint{1.793987in}{1.793987in}}%
\pgfusepath{clip}%
\pgfsetrectcap%
\pgfsetroundjoin%
\pgfsetlinewidth{0.803000pt}%
\definecolor{currentstroke}{rgb}{1.000000,1.000000,1.000000}%
\pgfsetstrokecolor{currentstroke}%
\pgfsetdash{}{0pt}%
\pgfpathmoveto{\pgfqpoint{1.842827in}{0.433022in}}%
\pgfpathlineto{\pgfqpoint{1.842827in}{2.227009in}}%
\pgfusepath{stroke}%
\end{pgfscope}%
\begin{pgfscope}%
\pgfsetbuttcap%
\pgfsetroundjoin%
\definecolor{currentfill}{rgb}{0.333333,0.333333,0.333333}%
\pgfsetfillcolor{currentfill}%
\pgfsetlinewidth{0.803000pt}%
\definecolor{currentstroke}{rgb}{0.333333,0.333333,0.333333}%
\pgfsetstrokecolor{currentstroke}%
\pgfsetdash{}{0pt}%
\pgfsys@defobject{currentmarker}{\pgfqpoint{0.000000in}{-0.048611in}}{\pgfqpoint{0.000000in}{0.000000in}}{%
\pgfpathmoveto{\pgfqpoint{0.000000in}{0.000000in}}%
\pgfpathlineto{\pgfqpoint{0.000000in}{-0.048611in}}%
\pgfusepath{stroke,fill}%
}%
\begin{pgfscope}%
\pgfsys@transformshift{1.842827in}{0.433022in}%
\pgfsys@useobject{currentmarker}{}%
\end{pgfscope}%
\end{pgfscope}%
\begin{pgfscope}%
\definecolor{textcolor}{rgb}{0.333333,0.333333,0.333333}%
\pgfsetstrokecolor{textcolor}%
\pgfsetfillcolor{textcolor}%
\pgftext[x=1.842827in,y=0.335800in,,top]{\color{textcolor}\rmfamily\fontsize{7.000000}{8.400000}\selectfont \(\displaystyle {200}\)}%
\end{pgfscope}%
\begin{pgfscope}%
\pgfpathrectangle{\pgfqpoint{0.461616in}{0.433022in}}{\pgfqpoint{1.793987in}{1.793987in}}%
\pgfusepath{clip}%
\pgfsetrectcap%
\pgfsetroundjoin%
\pgfsetlinewidth{0.803000pt}%
\definecolor{currentstroke}{rgb}{1.000000,1.000000,1.000000}%
\pgfsetstrokecolor{currentstroke}%
\pgfsetdash{}{0pt}%
\pgfpathmoveto{\pgfqpoint{2.175663in}{0.433022in}}%
\pgfpathlineto{\pgfqpoint{2.175663in}{2.227009in}}%
\pgfusepath{stroke}%
\end{pgfscope}%
\begin{pgfscope}%
\pgfsetbuttcap%
\pgfsetroundjoin%
\definecolor{currentfill}{rgb}{0.333333,0.333333,0.333333}%
\pgfsetfillcolor{currentfill}%
\pgfsetlinewidth{0.803000pt}%
\definecolor{currentstroke}{rgb}{0.333333,0.333333,0.333333}%
\pgfsetstrokecolor{currentstroke}%
\pgfsetdash{}{0pt}%
\pgfsys@defobject{currentmarker}{\pgfqpoint{0.000000in}{-0.048611in}}{\pgfqpoint{0.000000in}{0.000000in}}{%
\pgfpathmoveto{\pgfqpoint{0.000000in}{0.000000in}}%
\pgfpathlineto{\pgfqpoint{0.000000in}{-0.048611in}}%
\pgfusepath{stroke,fill}%
}%
\begin{pgfscope}%
\pgfsys@transformshift{2.175663in}{0.433022in}%
\pgfsys@useobject{currentmarker}{}%
\end{pgfscope}%
\end{pgfscope}%
\begin{pgfscope}%
\definecolor{textcolor}{rgb}{0.333333,0.333333,0.333333}%
\pgfsetstrokecolor{textcolor}%
\pgfsetfillcolor{textcolor}%
\pgftext[x=2.175663in,y=0.335800in,,top]{\color{textcolor}\rmfamily\fontsize{7.000000}{8.400000}\selectfont \(\displaystyle {250}\)}%
\end{pgfscope}%
\begin{pgfscope}%
\definecolor{textcolor}{rgb}{0.333333,0.333333,0.333333}%
\pgfsetstrokecolor{textcolor}%
\pgfsetfillcolor{textcolor}%
\pgftext[x=1.358610in,y=0.193825in,,top]{\color{textcolor}\rmfamily\fontsize{10.000000}{12.000000}\selectfont Avg. End-to-end Delay [ms]}%
\end{pgfscope}%
\begin{pgfscope}%
\pgfpathrectangle{\pgfqpoint{0.461616in}{0.433022in}}{\pgfqpoint{1.793987in}{1.793987in}}%
\pgfusepath{clip}%
\pgfsetrectcap%
\pgfsetroundjoin%
\pgfsetlinewidth{0.803000pt}%
\definecolor{currentstroke}{rgb}{1.000000,1.000000,1.000000}%
\pgfsetstrokecolor{currentstroke}%
\pgfsetdash{}{0pt}%
\pgfpathmoveto{\pgfqpoint{0.461616in}{0.514567in}}%
\pgfpathlineto{\pgfqpoint{2.255603in}{0.514567in}}%
\pgfusepath{stroke}%
\end{pgfscope}%
\begin{pgfscope}%
\pgfsetbuttcap%
\pgfsetroundjoin%
\definecolor{currentfill}{rgb}{0.333333,0.333333,0.333333}%
\pgfsetfillcolor{currentfill}%
\pgfsetlinewidth{0.803000pt}%
\definecolor{currentstroke}{rgb}{0.333333,0.333333,0.333333}%
\pgfsetstrokecolor{currentstroke}%
\pgfsetdash{}{0pt}%
\pgfsys@defobject{currentmarker}{\pgfqpoint{-0.048611in}{0.000000in}}{\pgfqpoint{-0.000000in}{0.000000in}}{%
\pgfpathmoveto{\pgfqpoint{-0.000000in}{0.000000in}}%
\pgfpathlineto{\pgfqpoint{-0.048611in}{0.000000in}}%
\pgfusepath{stroke,fill}%
}%
\begin{pgfscope}%
\pgfsys@transformshift{0.461616in}{0.514567in}%
\pgfsys@useobject{currentmarker}{}%
\end{pgfscope}%
\end{pgfscope}%
\begin{pgfscope}%
\definecolor{textcolor}{rgb}{0.333333,0.333333,0.333333}%
\pgfsetstrokecolor{textcolor}%
\pgfsetfillcolor{textcolor}%
\pgftext[x=0.220682in, y=0.480809in, left, base]{\color{textcolor}\rmfamily\fontsize{7.000000}{8.400000}\selectfont \(\displaystyle {0.0}\)}%
\end{pgfscope}%
\begin{pgfscope}%
\pgfpathrectangle{\pgfqpoint{0.461616in}{0.433022in}}{\pgfqpoint{1.793987in}{1.793987in}}%
\pgfusepath{clip}%
\pgfsetrectcap%
\pgfsetroundjoin%
\pgfsetlinewidth{0.803000pt}%
\definecolor{currentstroke}{rgb}{1.000000,1.000000,1.000000}%
\pgfsetstrokecolor{currentstroke}%
\pgfsetdash{}{0pt}%
\pgfpathmoveto{\pgfqpoint{0.461616in}{0.840746in}}%
\pgfpathlineto{\pgfqpoint{2.255603in}{0.840746in}}%
\pgfusepath{stroke}%
\end{pgfscope}%
\begin{pgfscope}%
\pgfsetbuttcap%
\pgfsetroundjoin%
\definecolor{currentfill}{rgb}{0.333333,0.333333,0.333333}%
\pgfsetfillcolor{currentfill}%
\pgfsetlinewidth{0.803000pt}%
\definecolor{currentstroke}{rgb}{0.333333,0.333333,0.333333}%
\pgfsetstrokecolor{currentstroke}%
\pgfsetdash{}{0pt}%
\pgfsys@defobject{currentmarker}{\pgfqpoint{-0.048611in}{0.000000in}}{\pgfqpoint{-0.000000in}{0.000000in}}{%
\pgfpathmoveto{\pgfqpoint{-0.000000in}{0.000000in}}%
\pgfpathlineto{\pgfqpoint{-0.048611in}{0.000000in}}%
\pgfusepath{stroke,fill}%
}%
\begin{pgfscope}%
\pgfsys@transformshift{0.461616in}{0.840746in}%
\pgfsys@useobject{currentmarker}{}%
\end{pgfscope}%
\end{pgfscope}%
\begin{pgfscope}%
\definecolor{textcolor}{rgb}{0.333333,0.333333,0.333333}%
\pgfsetstrokecolor{textcolor}%
\pgfsetfillcolor{textcolor}%
\pgftext[x=0.220682in, y=0.806989in, left, base]{\color{textcolor}\rmfamily\fontsize{7.000000}{8.400000}\selectfont \(\displaystyle {0.2}\)}%
\end{pgfscope}%
\begin{pgfscope}%
\pgfpathrectangle{\pgfqpoint{0.461616in}{0.433022in}}{\pgfqpoint{1.793987in}{1.793987in}}%
\pgfusepath{clip}%
\pgfsetrectcap%
\pgfsetroundjoin%
\pgfsetlinewidth{0.803000pt}%
\definecolor{currentstroke}{rgb}{1.000000,1.000000,1.000000}%
\pgfsetstrokecolor{currentstroke}%
\pgfsetdash{}{0pt}%
\pgfpathmoveto{\pgfqpoint{0.461616in}{1.166926in}}%
\pgfpathlineto{\pgfqpoint{2.255603in}{1.166926in}}%
\pgfusepath{stroke}%
\end{pgfscope}%
\begin{pgfscope}%
\pgfsetbuttcap%
\pgfsetroundjoin%
\definecolor{currentfill}{rgb}{0.333333,0.333333,0.333333}%
\pgfsetfillcolor{currentfill}%
\pgfsetlinewidth{0.803000pt}%
\definecolor{currentstroke}{rgb}{0.333333,0.333333,0.333333}%
\pgfsetstrokecolor{currentstroke}%
\pgfsetdash{}{0pt}%
\pgfsys@defobject{currentmarker}{\pgfqpoint{-0.048611in}{0.000000in}}{\pgfqpoint{-0.000000in}{0.000000in}}{%
\pgfpathmoveto{\pgfqpoint{-0.000000in}{0.000000in}}%
\pgfpathlineto{\pgfqpoint{-0.048611in}{0.000000in}}%
\pgfusepath{stroke,fill}%
}%
\begin{pgfscope}%
\pgfsys@transformshift{0.461616in}{1.166926in}%
\pgfsys@useobject{currentmarker}{}%
\end{pgfscope}%
\end{pgfscope}%
\begin{pgfscope}%
\definecolor{textcolor}{rgb}{0.333333,0.333333,0.333333}%
\pgfsetstrokecolor{textcolor}%
\pgfsetfillcolor{textcolor}%
\pgftext[x=0.220682in, y=1.133168in, left, base]{\color{textcolor}\rmfamily\fontsize{7.000000}{8.400000}\selectfont \(\displaystyle {0.4}\)}%
\end{pgfscope}%
\begin{pgfscope}%
\pgfpathrectangle{\pgfqpoint{0.461616in}{0.433022in}}{\pgfqpoint{1.793987in}{1.793987in}}%
\pgfusepath{clip}%
\pgfsetrectcap%
\pgfsetroundjoin%
\pgfsetlinewidth{0.803000pt}%
\definecolor{currentstroke}{rgb}{1.000000,1.000000,1.000000}%
\pgfsetstrokecolor{currentstroke}%
\pgfsetdash{}{0pt}%
\pgfpathmoveto{\pgfqpoint{0.461616in}{1.493105in}}%
\pgfpathlineto{\pgfqpoint{2.255603in}{1.493105in}}%
\pgfusepath{stroke}%
\end{pgfscope}%
\begin{pgfscope}%
\pgfsetbuttcap%
\pgfsetroundjoin%
\definecolor{currentfill}{rgb}{0.333333,0.333333,0.333333}%
\pgfsetfillcolor{currentfill}%
\pgfsetlinewidth{0.803000pt}%
\definecolor{currentstroke}{rgb}{0.333333,0.333333,0.333333}%
\pgfsetstrokecolor{currentstroke}%
\pgfsetdash{}{0pt}%
\pgfsys@defobject{currentmarker}{\pgfqpoint{-0.048611in}{0.000000in}}{\pgfqpoint{-0.000000in}{0.000000in}}{%
\pgfpathmoveto{\pgfqpoint{-0.000000in}{0.000000in}}%
\pgfpathlineto{\pgfqpoint{-0.048611in}{0.000000in}}%
\pgfusepath{stroke,fill}%
}%
\begin{pgfscope}%
\pgfsys@transformshift{0.461616in}{1.493105in}%
\pgfsys@useobject{currentmarker}{}%
\end{pgfscope}%
\end{pgfscope}%
\begin{pgfscope}%
\definecolor{textcolor}{rgb}{0.333333,0.333333,0.333333}%
\pgfsetstrokecolor{textcolor}%
\pgfsetfillcolor{textcolor}%
\pgftext[x=0.220682in, y=1.459348in, left, base]{\color{textcolor}\rmfamily\fontsize{7.000000}{8.400000}\selectfont \(\displaystyle {0.6}\)}%
\end{pgfscope}%
\begin{pgfscope}%
\pgfpathrectangle{\pgfqpoint{0.461616in}{0.433022in}}{\pgfqpoint{1.793987in}{1.793987in}}%
\pgfusepath{clip}%
\pgfsetrectcap%
\pgfsetroundjoin%
\pgfsetlinewidth{0.803000pt}%
\definecolor{currentstroke}{rgb}{1.000000,1.000000,1.000000}%
\pgfsetstrokecolor{currentstroke}%
\pgfsetdash{}{0pt}%
\pgfpathmoveto{\pgfqpoint{0.461616in}{1.819285in}}%
\pgfpathlineto{\pgfqpoint{2.255603in}{1.819285in}}%
\pgfusepath{stroke}%
\end{pgfscope}%
\begin{pgfscope}%
\pgfsetbuttcap%
\pgfsetroundjoin%
\definecolor{currentfill}{rgb}{0.333333,0.333333,0.333333}%
\pgfsetfillcolor{currentfill}%
\pgfsetlinewidth{0.803000pt}%
\definecolor{currentstroke}{rgb}{0.333333,0.333333,0.333333}%
\pgfsetstrokecolor{currentstroke}%
\pgfsetdash{}{0pt}%
\pgfsys@defobject{currentmarker}{\pgfqpoint{-0.048611in}{0.000000in}}{\pgfqpoint{-0.000000in}{0.000000in}}{%
\pgfpathmoveto{\pgfqpoint{-0.000000in}{0.000000in}}%
\pgfpathlineto{\pgfqpoint{-0.048611in}{0.000000in}}%
\pgfusepath{stroke,fill}%
}%
\begin{pgfscope}%
\pgfsys@transformshift{0.461616in}{1.819285in}%
\pgfsys@useobject{currentmarker}{}%
\end{pgfscope}%
\end{pgfscope}%
\begin{pgfscope}%
\definecolor{textcolor}{rgb}{0.333333,0.333333,0.333333}%
\pgfsetstrokecolor{textcolor}%
\pgfsetfillcolor{textcolor}%
\pgftext[x=0.220682in, y=1.785527in, left, base]{\color{textcolor}\rmfamily\fontsize{7.000000}{8.400000}\selectfont \(\displaystyle {0.8}\)}%
\end{pgfscope}%
\begin{pgfscope}%
\pgfpathrectangle{\pgfqpoint{0.461616in}{0.433022in}}{\pgfqpoint{1.793987in}{1.793987in}}%
\pgfusepath{clip}%
\pgfsetrectcap%
\pgfsetroundjoin%
\pgfsetlinewidth{0.803000pt}%
\definecolor{currentstroke}{rgb}{1.000000,1.000000,1.000000}%
\pgfsetstrokecolor{currentstroke}%
\pgfsetdash{}{0pt}%
\pgfpathmoveto{\pgfqpoint{0.461616in}{2.145464in}}%
\pgfpathlineto{\pgfqpoint{2.255603in}{2.145464in}}%
\pgfusepath{stroke}%
\end{pgfscope}%
\begin{pgfscope}%
\pgfsetbuttcap%
\pgfsetroundjoin%
\definecolor{currentfill}{rgb}{0.333333,0.333333,0.333333}%
\pgfsetfillcolor{currentfill}%
\pgfsetlinewidth{0.803000pt}%
\definecolor{currentstroke}{rgb}{0.333333,0.333333,0.333333}%
\pgfsetstrokecolor{currentstroke}%
\pgfsetdash{}{0pt}%
\pgfsys@defobject{currentmarker}{\pgfqpoint{-0.048611in}{0.000000in}}{\pgfqpoint{-0.000000in}{0.000000in}}{%
\pgfpathmoveto{\pgfqpoint{-0.000000in}{0.000000in}}%
\pgfpathlineto{\pgfqpoint{-0.048611in}{0.000000in}}%
\pgfusepath{stroke,fill}%
}%
\begin{pgfscope}%
\pgfsys@transformshift{0.461616in}{2.145464in}%
\pgfsys@useobject{currentmarker}{}%
\end{pgfscope}%
\end{pgfscope}%
\begin{pgfscope}%
\definecolor{textcolor}{rgb}{0.333333,0.333333,0.333333}%
\pgfsetstrokecolor{textcolor}%
\pgfsetfillcolor{textcolor}%
\pgftext[x=0.220682in, y=2.111706in, left, base]{\color{textcolor}\rmfamily\fontsize{7.000000}{8.400000}\selectfont \(\displaystyle {1.0}\)}%
\end{pgfscope}%
\begin{pgfscope}%
\definecolor{textcolor}{rgb}{0.333333,0.333333,0.333333}%
\pgfsetstrokecolor{textcolor}%
\pgfsetfillcolor{textcolor}%
\pgftext[x=0.165127in,y=1.330015in,,bottom,rotate=90.000000]{\color{textcolor}\rmfamily\fontsize{10.000000}{12.000000}\selectfont Cumulative Probability}%
\end{pgfscope}%
\begin{pgfscope}%
\pgfpathrectangle{\pgfqpoint{0.461616in}{0.433022in}}{\pgfqpoint{1.793987in}{1.793987in}}%
\pgfusepath{clip}%
\pgfsetrectcap%
\pgfsetroundjoin%
\pgfsetlinewidth{1.003750pt}%
\definecolor{currentstroke}{rgb}{1.000000,0.498039,0.054902}%
\pgfsetstrokecolor{currentstroke}%
\pgfsetdash{}{0pt}%
\pgfpathmoveto{\pgfqpoint{0.543161in}{0.514567in}}%
\pgfpathlineto{\pgfqpoint{0.549818in}{0.514567in}}%
\pgfpathlineto{\pgfqpoint{0.556475in}{0.519910in}}%
\pgfpathlineto{\pgfqpoint{0.563131in}{0.519910in}}%
\pgfpathlineto{\pgfqpoint{0.569788in}{0.521245in}}%
\pgfpathlineto{\pgfqpoint{0.589758in}{0.521245in}}%
\pgfpathlineto{\pgfqpoint{0.596415in}{0.526588in}}%
\pgfpathlineto{\pgfqpoint{0.616385in}{0.534602in}}%
\pgfpathlineto{\pgfqpoint{0.623042in}{0.535938in}}%
\pgfpathlineto{\pgfqpoint{0.636355in}{0.554638in}}%
\pgfpathlineto{\pgfqpoint{0.643012in}{0.561317in}}%
\pgfpathlineto{\pgfqpoint{0.649669in}{0.563988in}}%
\pgfpathlineto{\pgfqpoint{0.656326in}{0.572670in}}%
\pgfpathlineto{\pgfqpoint{0.662982in}{0.574006in}}%
\pgfpathlineto{\pgfqpoint{0.676296in}{0.581352in}}%
\pgfpathlineto{\pgfqpoint{0.682952in}{0.588699in}}%
\pgfpathlineto{\pgfqpoint{0.696266in}{0.594041in}}%
\pgfpathlineto{\pgfqpoint{0.702923in}{0.600052in}}%
\pgfpathlineto{\pgfqpoint{0.709579in}{0.622091in}}%
\pgfpathlineto{\pgfqpoint{0.722893in}{0.628102in}}%
\pgfpathlineto{\pgfqpoint{0.749520in}{0.654148in}}%
\pgfpathlineto{\pgfqpoint{0.756176in}{0.655484in}}%
\pgfpathlineto{\pgfqpoint{0.762833in}{0.664166in}}%
\pgfpathlineto{\pgfqpoint{0.769490in}{0.668841in}}%
\pgfpathlineto{\pgfqpoint{0.776147in}{0.676855in}}%
\pgfpathlineto{\pgfqpoint{0.782803in}{0.690880in}}%
\pgfpathlineto{\pgfqpoint{0.796117in}{0.710248in}}%
\pgfpathlineto{\pgfqpoint{0.802773in}{0.716258in}}%
\pgfpathlineto{\pgfqpoint{0.809430in}{0.726276in}}%
\pgfpathlineto{\pgfqpoint{0.816087in}{0.738965in}}%
\pgfpathlineto{\pgfqpoint{0.822744in}{0.746980in}}%
\pgfpathlineto{\pgfqpoint{0.836057in}{0.775697in}}%
\pgfpathlineto{\pgfqpoint{0.842714in}{0.789722in}}%
\pgfpathlineto{\pgfqpoint{0.856027in}{0.826454in}}%
\pgfpathlineto{\pgfqpoint{0.862684in}{0.837140in}}%
\pgfpathlineto{\pgfqpoint{0.869341in}{0.854504in}}%
\pgfpathlineto{\pgfqpoint{0.875997in}{0.865190in}}%
\pgfpathlineto{\pgfqpoint{0.882654in}{0.882554in}}%
\pgfpathlineto{\pgfqpoint{0.889311in}{0.892572in}}%
\pgfpathlineto{\pgfqpoint{0.895968in}{0.913275in}}%
\pgfpathlineto{\pgfqpoint{0.902624in}{0.919954in}}%
\pgfpathlineto{\pgfqpoint{0.909281in}{0.943329in}}%
\pgfpathlineto{\pgfqpoint{0.915938in}{0.960025in}}%
\pgfpathlineto{\pgfqpoint{0.929251in}{1.015457in}}%
\pgfpathlineto{\pgfqpoint{0.935908in}{1.029482in}}%
\pgfpathlineto{\pgfqpoint{0.942565in}{1.051521in}}%
\pgfpathlineto{\pgfqpoint{0.949221in}{1.064210in}}%
\pgfpathlineto{\pgfqpoint{0.969192in}{1.129660in}}%
\pgfpathlineto{\pgfqpoint{0.989162in}{1.217816in}}%
\pgfpathlineto{\pgfqpoint{0.995818in}{1.250541in}}%
\pgfpathlineto{\pgfqpoint{1.002475in}{1.276587in}}%
\pgfpathlineto{\pgfqpoint{1.015789in}{1.341369in}}%
\pgfpathlineto{\pgfqpoint{1.022445in}{1.366080in}}%
\pgfpathlineto{\pgfqpoint{1.029102in}{1.395465in}}%
\pgfpathlineto{\pgfqpoint{1.042416in}{1.434868in}}%
\pgfpathlineto{\pgfqpoint{1.049072in}{1.465590in}}%
\pgfpathlineto{\pgfqpoint{1.055729in}{1.469597in}}%
\pgfpathlineto{\pgfqpoint{1.062386in}{1.493639in}}%
\pgfpathlineto{\pgfqpoint{1.075699in}{1.520354in}}%
\pgfpathlineto{\pgfqpoint{1.082356in}{1.549071in}}%
\pgfpathlineto{\pgfqpoint{1.089013in}{1.560425in}}%
\pgfpathlineto{\pgfqpoint{1.095669in}{1.583800in}}%
\pgfpathlineto{\pgfqpoint{1.108983in}{1.619196in}}%
\pgfpathlineto{\pgfqpoint{1.122296in}{1.641235in}}%
\pgfpathlineto{\pgfqpoint{1.128953in}{1.651921in}}%
\pgfpathlineto{\pgfqpoint{1.135610in}{1.669953in}}%
\pgfpathlineto{\pgfqpoint{1.142266in}{1.684645in}}%
\pgfpathlineto{\pgfqpoint{1.148923in}{1.693328in}}%
\pgfpathlineto{\pgfqpoint{1.155580in}{1.710024in}}%
\pgfpathlineto{\pgfqpoint{1.162237in}{1.715367in}}%
\pgfpathlineto{\pgfqpoint{1.168893in}{1.730059in}}%
\pgfpathlineto{\pgfqpoint{1.175550in}{1.738742in}}%
\pgfpathlineto{\pgfqpoint{1.195520in}{1.772802in}}%
\pgfpathlineto{\pgfqpoint{1.202177in}{1.777477in}}%
\pgfpathlineto{\pgfqpoint{1.208834in}{1.787495in}}%
\pgfpathlineto{\pgfqpoint{1.222147in}{1.816880in}}%
\pgfpathlineto{\pgfqpoint{1.228804in}{1.820888in}}%
\pgfpathlineto{\pgfqpoint{1.235461in}{1.832909in}}%
\pgfpathlineto{\pgfqpoint{1.242117in}{1.852277in}}%
\pgfpathlineto{\pgfqpoint{1.248774in}{1.862962in}}%
\pgfpathlineto{\pgfqpoint{1.255431in}{1.868973in}}%
\pgfpathlineto{\pgfqpoint{1.262087in}{1.885001in}}%
\pgfpathlineto{\pgfqpoint{1.268744in}{1.889009in}}%
\pgfpathlineto{\pgfqpoint{1.275401in}{1.897691in}}%
\pgfpathlineto{\pgfqpoint{1.288714in}{1.918394in}}%
\pgfpathlineto{\pgfqpoint{1.295371in}{1.923737in}}%
\pgfpathlineto{\pgfqpoint{1.308684in}{1.944440in}}%
\pgfpathlineto{\pgfqpoint{1.315341in}{1.953122in}}%
\pgfpathlineto{\pgfqpoint{1.321998in}{1.967815in}}%
\pgfpathlineto{\pgfqpoint{1.328655in}{1.975162in}}%
\pgfpathlineto{\pgfqpoint{1.335311in}{1.985179in}}%
\pgfpathlineto{\pgfqpoint{1.348625in}{1.990522in}}%
\pgfpathlineto{\pgfqpoint{1.355282in}{2.005215in}}%
\pgfpathlineto{\pgfqpoint{1.361938in}{2.007886in}}%
\pgfpathlineto{\pgfqpoint{1.368595in}{2.021911in}}%
\pgfpathlineto{\pgfqpoint{1.375252in}{2.027922in}}%
\pgfpathlineto{\pgfqpoint{1.381908in}{2.038608in}}%
\pgfpathlineto{\pgfqpoint{1.388565in}{2.041279in}}%
\pgfpathlineto{\pgfqpoint{1.395222in}{2.042615in}}%
\pgfpathlineto{\pgfqpoint{1.401879in}{2.052633in}}%
\pgfpathlineto{\pgfqpoint{1.415192in}{2.063986in}}%
\pgfpathlineto{\pgfqpoint{1.421849in}{2.067993in}}%
\pgfpathlineto{\pgfqpoint{1.428505in}{2.074672in}}%
\pgfpathlineto{\pgfqpoint{1.435162in}{2.076675in}}%
\pgfpathlineto{\pgfqpoint{1.455132in}{2.086693in}}%
\pgfpathlineto{\pgfqpoint{1.468446in}{2.102722in}}%
\pgfpathlineto{\pgfqpoint{1.475103in}{2.107396in}}%
\pgfpathlineto{\pgfqpoint{1.481759in}{2.107396in}}%
\pgfpathlineto{\pgfqpoint{1.488416in}{2.108732in}}%
\pgfpathlineto{\pgfqpoint{1.495073in}{2.120086in}}%
\pgfpathlineto{\pgfqpoint{1.501729in}{2.120086in}}%
\pgfpathlineto{\pgfqpoint{1.508386in}{2.122089in}}%
\pgfpathlineto{\pgfqpoint{1.515043in}{2.126096in}}%
\pgfpathlineto{\pgfqpoint{1.521700in}{2.126096in}}%
\pgfpathlineto{\pgfqpoint{1.528356in}{2.128100in}}%
\pgfpathlineto{\pgfqpoint{1.535013in}{2.132107in}}%
\pgfpathlineto{\pgfqpoint{1.548327in}{2.132107in}}%
\pgfpathlineto{\pgfqpoint{1.554983in}{2.133443in}}%
\pgfpathlineto{\pgfqpoint{1.561640in}{2.133443in}}%
\pgfpathlineto{\pgfqpoint{1.568297in}{2.134778in}}%
\pgfpathlineto{\pgfqpoint{1.594924in}{2.134778in}}%
\pgfpathlineto{\pgfqpoint{1.601580in}{2.137450in}}%
\pgfpathlineto{\pgfqpoint{1.648177in}{2.137450in}}%
\pgfpathlineto{\pgfqpoint{1.661491in}{2.140121in}}%
\pgfpathlineto{\pgfqpoint{1.694774in}{2.140121in}}%
\pgfpathlineto{\pgfqpoint{1.701431in}{2.141457in}}%
\pgfpathlineto{\pgfqpoint{1.827909in}{2.141457in}}%
\pgfpathlineto{\pgfqpoint{1.834566in}{2.142793in}}%
\pgfpathlineto{\pgfqpoint{1.947730in}{2.142793in}}%
\pgfpathlineto{\pgfqpoint{1.954387in}{2.144128in}}%
\pgfpathlineto{\pgfqpoint{2.000984in}{2.144128in}}%
\pgfpathlineto{\pgfqpoint{2.007640in}{2.145464in}}%
\pgfpathlineto{\pgfqpoint{2.174059in}{2.145464in}}%
\pgfpathlineto{\pgfqpoint{2.174059in}{2.145464in}}%
\pgfusepath{stroke}%
\end{pgfscope}%
\begin{pgfscope}%
\pgfpathrectangle{\pgfqpoint{0.461616in}{0.433022in}}{\pgfqpoint{1.793987in}{1.793987in}}%
\pgfusepath{clip}%
\pgfsetbuttcap%
\pgfsetroundjoin%
\pgfsetlinewidth{1.003750pt}%
\definecolor{currentstroke}{rgb}{0.121569,0.466667,0.705882}%
\pgfsetstrokecolor{currentstroke}%
\pgfsetdash{{3.700000pt}{1.600000pt}}{0.000000pt}%
\pgfpathmoveto{\pgfqpoint{0.543161in}{0.514567in}}%
\pgfpathlineto{\pgfqpoint{0.549818in}{0.514567in}}%
\pgfpathlineto{\pgfqpoint{0.556475in}{0.519910in}}%
\pgfpathlineto{\pgfqpoint{0.589758in}{0.519910in}}%
\pgfpathlineto{\pgfqpoint{0.616385in}{0.530595in}}%
\pgfpathlineto{\pgfqpoint{0.623042in}{0.530595in}}%
\pgfpathlineto{\pgfqpoint{0.629699in}{0.533267in}}%
\pgfpathlineto{\pgfqpoint{0.636355in}{0.538610in}}%
\pgfpathlineto{\pgfqpoint{0.643012in}{0.539277in}}%
\pgfpathlineto{\pgfqpoint{0.649669in}{0.544620in}}%
\pgfpathlineto{\pgfqpoint{0.656326in}{0.551967in}}%
\pgfpathlineto{\pgfqpoint{0.689609in}{0.574006in}}%
\pgfpathlineto{\pgfqpoint{0.696266in}{0.584691in}}%
\pgfpathlineto{\pgfqpoint{0.709579in}{0.587363in}}%
\pgfpathlineto{\pgfqpoint{0.716236in}{0.592706in}}%
\pgfpathlineto{\pgfqpoint{0.736206in}{0.600720in}}%
\pgfpathlineto{\pgfqpoint{0.742863in}{0.606731in}}%
\pgfpathlineto{\pgfqpoint{0.749520in}{0.609402in}}%
\pgfpathlineto{\pgfqpoint{0.762833in}{0.619420in}}%
\pgfpathlineto{\pgfqpoint{0.769490in}{0.622759in}}%
\pgfpathlineto{\pgfqpoint{0.776147in}{0.633445in}}%
\pgfpathlineto{\pgfqpoint{0.782803in}{0.638788in}}%
\pgfpathlineto{\pgfqpoint{0.789460in}{0.650141in}}%
\pgfpathlineto{\pgfqpoint{0.796117in}{0.652145in}}%
\pgfpathlineto{\pgfqpoint{0.802773in}{0.660827in}}%
\pgfpathlineto{\pgfqpoint{0.809430in}{0.664834in}}%
\pgfpathlineto{\pgfqpoint{0.822744in}{0.670177in}}%
\pgfpathlineto{\pgfqpoint{0.829400in}{0.679527in}}%
\pgfpathlineto{\pgfqpoint{0.836057in}{0.686873in}}%
\pgfpathlineto{\pgfqpoint{0.842714in}{0.689544in}}%
\pgfpathlineto{\pgfqpoint{0.849371in}{0.697559in}}%
\pgfpathlineto{\pgfqpoint{0.856027in}{0.701566in}}%
\pgfpathlineto{\pgfqpoint{0.869341in}{0.711584in}}%
\pgfpathlineto{\pgfqpoint{0.875997in}{0.724941in}}%
\pgfpathlineto{\pgfqpoint{0.882654in}{0.735626in}}%
\pgfpathlineto{\pgfqpoint{0.889311in}{0.741637in}}%
\pgfpathlineto{\pgfqpoint{0.902624in}{0.763676in}}%
\pgfpathlineto{\pgfqpoint{0.909281in}{0.766347in}}%
\pgfpathlineto{\pgfqpoint{0.922594in}{0.778369in}}%
\pgfpathlineto{\pgfqpoint{0.935908in}{0.801744in}}%
\pgfpathlineto{\pgfqpoint{0.949221in}{0.836472in}}%
\pgfpathlineto{\pgfqpoint{0.975848in}{0.901922in}}%
\pgfpathlineto{\pgfqpoint{0.989162in}{0.943996in}}%
\pgfpathlineto{\pgfqpoint{0.995818in}{0.969375in}}%
\pgfpathlineto{\pgfqpoint{1.009132in}{1.032821in}}%
\pgfpathlineto{\pgfqpoint{1.015789in}{1.070221in}}%
\pgfpathlineto{\pgfqpoint{1.035759in}{1.159713in}}%
\pgfpathlineto{\pgfqpoint{1.042416in}{1.201788in}}%
\pgfpathlineto{\pgfqpoint{1.049072in}{1.231841in}}%
\pgfpathlineto{\pgfqpoint{1.055729in}{1.271912in}}%
\pgfpathlineto{\pgfqpoint{1.062386in}{1.293284in}}%
\pgfpathlineto{\pgfqpoint{1.069042in}{1.330683in}}%
\pgfpathlineto{\pgfqpoint{1.075699in}{1.356730in}}%
\pgfpathlineto{\pgfqpoint{1.082356in}{1.372090in}}%
\pgfpathlineto{\pgfqpoint{1.089013in}{1.406151in}}%
\pgfpathlineto{\pgfqpoint{1.095669in}{1.420176in}}%
\pgfpathlineto{\pgfqpoint{1.102326in}{1.431529in}}%
\pgfpathlineto{\pgfqpoint{1.108983in}{1.450229in}}%
\pgfpathlineto{\pgfqpoint{1.115639in}{1.472936in}}%
\pgfpathlineto{\pgfqpoint{1.122296in}{1.490968in}}%
\pgfpathlineto{\pgfqpoint{1.128953in}{1.515011in}}%
\pgfpathlineto{\pgfqpoint{1.135610in}{1.531707in}}%
\pgfpathlineto{\pgfqpoint{1.142266in}{1.539721in}}%
\pgfpathlineto{\pgfqpoint{1.148923in}{1.552411in}}%
\pgfpathlineto{\pgfqpoint{1.155580in}{1.579793in}}%
\pgfpathlineto{\pgfqpoint{1.162237in}{1.588475in}}%
\pgfpathlineto{\pgfqpoint{1.168893in}{1.602500in}}%
\pgfpathlineto{\pgfqpoint{1.182207in}{1.634557in}}%
\pgfpathlineto{\pgfqpoint{1.202177in}{1.665278in}}%
\pgfpathlineto{\pgfqpoint{1.208834in}{1.681306in}}%
\pgfpathlineto{\pgfqpoint{1.215490in}{1.693995in}}%
\pgfpathlineto{\pgfqpoint{1.222147in}{1.701342in}}%
\pgfpathlineto{\pgfqpoint{1.228804in}{1.717370in}}%
\pgfpathlineto{\pgfqpoint{1.235461in}{1.724717in}}%
\pgfpathlineto{\pgfqpoint{1.262087in}{1.780816in}}%
\pgfpathlineto{\pgfqpoint{1.268744in}{1.790834in}}%
\pgfpathlineto{\pgfqpoint{1.275401in}{1.803523in}}%
\pgfpathlineto{\pgfqpoint{1.302028in}{1.834245in}}%
\pgfpathlineto{\pgfqpoint{1.308684in}{1.845598in}}%
\pgfpathlineto{\pgfqpoint{1.315341in}{1.866302in}}%
\pgfpathlineto{\pgfqpoint{1.341968in}{1.919062in}}%
\pgfpathlineto{\pgfqpoint{1.355282in}{1.931751in}}%
\pgfpathlineto{\pgfqpoint{1.368595in}{1.941769in}}%
\pgfpathlineto{\pgfqpoint{1.375252in}{1.957797in}}%
\pgfpathlineto{\pgfqpoint{1.381908in}{1.969151in}}%
\pgfpathlineto{\pgfqpoint{1.388565in}{1.984512in}}%
\pgfpathlineto{\pgfqpoint{1.395222in}{1.990522in}}%
\pgfpathlineto{\pgfqpoint{1.401879in}{1.998536in}}%
\pgfpathlineto{\pgfqpoint{1.408535in}{2.010558in}}%
\pgfpathlineto{\pgfqpoint{1.428505in}{2.033933in}}%
\pgfpathlineto{\pgfqpoint{1.435162in}{2.045286in}}%
\pgfpathlineto{\pgfqpoint{1.448476in}{2.059311in}}%
\pgfpathlineto{\pgfqpoint{1.455132in}{2.062650in}}%
\pgfpathlineto{\pgfqpoint{1.461789in}{2.072668in}}%
\pgfpathlineto{\pgfqpoint{1.468446in}{2.076675in}}%
\pgfpathlineto{\pgfqpoint{1.475103in}{2.082018in}}%
\pgfpathlineto{\pgfqpoint{1.481759in}{2.091368in}}%
\pgfpathlineto{\pgfqpoint{1.488416in}{2.093372in}}%
\pgfpathlineto{\pgfqpoint{1.495073in}{2.104057in}}%
\pgfpathlineto{\pgfqpoint{1.501729in}{2.111404in}}%
\pgfpathlineto{\pgfqpoint{1.508386in}{2.114075in}}%
\pgfpathlineto{\pgfqpoint{1.515043in}{2.118082in}}%
\pgfpathlineto{\pgfqpoint{1.521700in}{2.120754in}}%
\pgfpathlineto{\pgfqpoint{1.528356in}{2.122089in}}%
\pgfpathlineto{\pgfqpoint{1.535013in}{2.124761in}}%
\pgfpathlineto{\pgfqpoint{1.541670in}{2.125429in}}%
\pgfpathlineto{\pgfqpoint{1.561640in}{2.130771in}}%
\pgfpathlineto{\pgfqpoint{1.574953in}{2.130771in}}%
\pgfpathlineto{\pgfqpoint{1.581610in}{2.132107in}}%
\pgfpathlineto{\pgfqpoint{1.594924in}{2.132107in}}%
\pgfpathlineto{\pgfqpoint{1.601580in}{2.133443in}}%
\pgfpathlineto{\pgfqpoint{1.614894in}{2.133443in}}%
\pgfpathlineto{\pgfqpoint{1.621550in}{2.134778in}}%
\pgfpathlineto{\pgfqpoint{1.648177in}{2.134778in}}%
\pgfpathlineto{\pgfqpoint{1.654834in}{2.136114in}}%
\pgfpathlineto{\pgfqpoint{1.661491in}{2.136114in}}%
\pgfpathlineto{\pgfqpoint{1.668148in}{2.137450in}}%
\pgfpathlineto{\pgfqpoint{1.681461in}{2.137450in}}%
\pgfpathlineto{\pgfqpoint{1.688118in}{2.140121in}}%
\pgfpathlineto{\pgfqpoint{1.741371in}{2.140121in}}%
\pgfpathlineto{\pgfqpoint{1.748028in}{2.141457in}}%
\pgfpathlineto{\pgfqpoint{1.861193in}{2.141457in}}%
\pgfpathlineto{\pgfqpoint{1.867849in}{2.142793in}}%
\pgfpathlineto{\pgfqpoint{1.994327in}{2.143461in}}%
\pgfpathlineto{\pgfqpoint{2.007640in}{2.144128in}}%
\pgfpathlineto{\pgfqpoint{2.027611in}{2.144128in}}%
\pgfpathlineto{\pgfqpoint{2.034267in}{2.145464in}}%
\pgfpathlineto{\pgfqpoint{2.174059in}{2.145464in}}%
\pgfpathlineto{\pgfqpoint{2.174059in}{2.145464in}}%
\pgfusepath{stroke}%
\end{pgfscope}%
\begin{pgfscope}%
\pgfpathrectangle{\pgfqpoint{0.461616in}{0.433022in}}{\pgfqpoint{1.793987in}{1.793987in}}%
\pgfusepath{clip}%
\pgfsetbuttcap%
\pgfsetroundjoin%
\pgfsetlinewidth{1.003750pt}%
\definecolor{currentstroke}{rgb}{0.839216,0.152941,0.156863}%
\pgfsetstrokecolor{currentstroke}%
\pgfsetdash{{6.400000pt}{1.600000pt}{1.000000pt}{1.600000pt}}{0.000000pt}%
\pgfpathmoveto{\pgfqpoint{0.543161in}{0.514567in}}%
\pgfpathlineto{\pgfqpoint{0.549818in}{0.514567in}}%
\pgfpathlineto{\pgfqpoint{0.563131in}{0.518574in}}%
\pgfpathlineto{\pgfqpoint{0.616385in}{0.518574in}}%
\pgfpathlineto{\pgfqpoint{0.623042in}{0.519910in}}%
\pgfpathlineto{\pgfqpoint{0.643012in}{0.519910in}}%
\pgfpathlineto{\pgfqpoint{0.649669in}{0.521913in}}%
\pgfpathlineto{\pgfqpoint{0.656326in}{0.521913in}}%
\pgfpathlineto{\pgfqpoint{0.669639in}{0.529927in}}%
\pgfpathlineto{\pgfqpoint{0.676296in}{0.531931in}}%
\pgfpathlineto{\pgfqpoint{0.682952in}{0.532599in}}%
\pgfpathlineto{\pgfqpoint{0.689609in}{0.539945in}}%
\pgfpathlineto{\pgfqpoint{0.702923in}{0.547960in}}%
\pgfpathlineto{\pgfqpoint{0.709579in}{0.551299in}}%
\pgfpathlineto{\pgfqpoint{0.722893in}{0.561984in}}%
\pgfpathlineto{\pgfqpoint{0.736206in}{0.572670in}}%
\pgfpathlineto{\pgfqpoint{0.742863in}{0.574006in}}%
\pgfpathlineto{\pgfqpoint{0.749520in}{0.574006in}}%
\pgfpathlineto{\pgfqpoint{0.756176in}{0.578013in}}%
\pgfpathlineto{\pgfqpoint{0.762833in}{0.580016in}}%
\pgfpathlineto{\pgfqpoint{0.776147in}{0.590702in}}%
\pgfpathlineto{\pgfqpoint{0.802773in}{0.596045in}}%
\pgfpathlineto{\pgfqpoint{0.816087in}{0.610070in}}%
\pgfpathlineto{\pgfqpoint{0.822744in}{0.613409in}}%
\pgfpathlineto{\pgfqpoint{0.829400in}{0.618752in}}%
\pgfpathlineto{\pgfqpoint{0.842714in}{0.636784in}}%
\pgfpathlineto{\pgfqpoint{0.849371in}{0.642795in}}%
\pgfpathlineto{\pgfqpoint{0.856027in}{0.651477in}}%
\pgfpathlineto{\pgfqpoint{0.875997in}{0.668841in}}%
\pgfpathlineto{\pgfqpoint{0.882654in}{0.670177in}}%
\pgfpathlineto{\pgfqpoint{0.895968in}{0.680194in}}%
\pgfpathlineto{\pgfqpoint{0.935908in}{0.719598in}}%
\pgfpathlineto{\pgfqpoint{0.962535in}{0.740969in}}%
\pgfpathlineto{\pgfqpoint{0.969192in}{0.750319in}}%
\pgfpathlineto{\pgfqpoint{0.975848in}{0.761672in}}%
\pgfpathlineto{\pgfqpoint{0.982505in}{0.770355in}}%
\pgfpathlineto{\pgfqpoint{0.995818in}{0.795733in}}%
\pgfpathlineto{\pgfqpoint{1.015789in}{0.847158in}}%
\pgfpathlineto{\pgfqpoint{1.022445in}{0.870533in}}%
\pgfpathlineto{\pgfqpoint{1.035759in}{0.909268in}}%
\pgfpathlineto{\pgfqpoint{1.049072in}{0.961361in}}%
\pgfpathlineto{\pgfqpoint{1.055729in}{0.981396in}}%
\pgfpathlineto{\pgfqpoint{1.062386in}{1.006107in}}%
\pgfpathlineto{\pgfqpoint{1.069042in}{1.041503in}}%
\pgfpathlineto{\pgfqpoint{1.075699in}{1.066214in}}%
\pgfpathlineto{\pgfqpoint{1.082356in}{1.084913in}}%
\pgfpathlineto{\pgfqpoint{1.089013in}{1.112295in}}%
\pgfpathlineto{\pgfqpoint{1.095669in}{1.150363in}}%
\pgfpathlineto{\pgfqpoint{1.102326in}{1.163052in}}%
\pgfpathlineto{\pgfqpoint{1.122296in}{1.215813in}}%
\pgfpathlineto{\pgfqpoint{1.142266in}{1.261227in}}%
\pgfpathlineto{\pgfqpoint{1.155580in}{1.297291in}}%
\pgfpathlineto{\pgfqpoint{1.168893in}{1.342037in}}%
\pgfpathlineto{\pgfqpoint{1.175550in}{1.358733in}}%
\pgfpathlineto{\pgfqpoint{1.182207in}{1.371422in}}%
\pgfpathlineto{\pgfqpoint{1.222147in}{1.469597in}}%
\pgfpathlineto{\pgfqpoint{1.228804in}{1.480282in}}%
\pgfpathlineto{\pgfqpoint{1.242117in}{1.507664in}}%
\pgfpathlineto{\pgfqpoint{1.255431in}{1.547068in}}%
\pgfpathlineto{\pgfqpoint{1.262087in}{1.561761in}}%
\pgfpathlineto{\pgfqpoint{1.268744in}{1.583800in}}%
\pgfpathlineto{\pgfqpoint{1.275401in}{1.589810in}}%
\pgfpathlineto{\pgfqpoint{1.282058in}{1.601164in}}%
\pgfpathlineto{\pgfqpoint{1.288714in}{1.615857in}}%
\pgfpathlineto{\pgfqpoint{1.295371in}{1.635892in}}%
\pgfpathlineto{\pgfqpoint{1.308684in}{1.667281in}}%
\pgfpathlineto{\pgfqpoint{1.321998in}{1.702010in}}%
\pgfpathlineto{\pgfqpoint{1.328655in}{1.713363in}}%
\pgfpathlineto{\pgfqpoint{1.341968in}{1.744752in}}%
\pgfpathlineto{\pgfqpoint{1.348625in}{1.763452in}}%
\pgfpathlineto{\pgfqpoint{1.388565in}{1.838920in}}%
\pgfpathlineto{\pgfqpoint{1.428505in}{1.935090in}}%
\pgfpathlineto{\pgfqpoint{1.448476in}{1.965812in}}%
\pgfpathlineto{\pgfqpoint{1.461789in}{1.977165in}}%
\pgfpathlineto{\pgfqpoint{1.475103in}{1.997869in}}%
\pgfpathlineto{\pgfqpoint{1.481759in}{2.002544in}}%
\pgfpathlineto{\pgfqpoint{1.501729in}{2.027922in}}%
\pgfpathlineto{\pgfqpoint{1.508386in}{2.034601in}}%
\pgfpathlineto{\pgfqpoint{1.528356in}{2.061982in}}%
\pgfpathlineto{\pgfqpoint{1.535013in}{2.065322in}}%
\pgfpathlineto{\pgfqpoint{1.541670in}{2.069997in}}%
\pgfpathlineto{\pgfqpoint{1.548327in}{2.080015in}}%
\pgfpathlineto{\pgfqpoint{1.568297in}{2.098047in}}%
\pgfpathlineto{\pgfqpoint{1.574953in}{2.099382in}}%
\pgfpathlineto{\pgfqpoint{1.594924in}{2.112739in}}%
\pgfpathlineto{\pgfqpoint{1.634864in}{2.120086in}}%
\pgfpathlineto{\pgfqpoint{1.641521in}{2.120754in}}%
\pgfpathlineto{\pgfqpoint{1.654834in}{2.124761in}}%
\pgfpathlineto{\pgfqpoint{1.734715in}{2.125429in}}%
\pgfpathlineto{\pgfqpoint{1.767998in}{2.128100in}}%
\pgfpathlineto{\pgfqpoint{1.787969in}{2.128768in}}%
\pgfpathlineto{\pgfqpoint{1.801282in}{2.129436in}}%
\pgfpathlineto{\pgfqpoint{1.827909in}{2.129436in}}%
\pgfpathlineto{\pgfqpoint{1.834566in}{2.130771in}}%
\pgfpathlineto{\pgfqpoint{1.854536in}{2.131439in}}%
\pgfpathlineto{\pgfqpoint{1.867849in}{2.132775in}}%
\pgfpathlineto{\pgfqpoint{1.874506in}{2.134778in}}%
\pgfpathlineto{\pgfqpoint{1.887819in}{2.134778in}}%
\pgfpathlineto{\pgfqpoint{1.894476in}{2.136114in}}%
\pgfpathlineto{\pgfqpoint{1.921103in}{2.136782in}}%
\pgfpathlineto{\pgfqpoint{1.934416in}{2.137450in}}%
\pgfpathlineto{\pgfqpoint{1.941073in}{2.140121in}}%
\pgfpathlineto{\pgfqpoint{1.947730in}{2.141457in}}%
\pgfpathlineto{\pgfqpoint{1.994327in}{2.141457in}}%
\pgfpathlineto{\pgfqpoint{2.000984in}{2.142793in}}%
\pgfpathlineto{\pgfqpoint{2.054238in}{2.142793in}}%
\pgfpathlineto{\pgfqpoint{2.067551in}{2.145464in}}%
\pgfpathlineto{\pgfqpoint{2.174059in}{2.145464in}}%
\pgfpathlineto{\pgfqpoint{2.174059in}{2.145464in}}%
\pgfusepath{stroke}%
\end{pgfscope}%
\begin{pgfscope}%
\pgfpathrectangle{\pgfqpoint{0.461616in}{0.433022in}}{\pgfqpoint{1.793987in}{1.793987in}}%
\pgfusepath{clip}%
\pgfsetbuttcap%
\pgfsetroundjoin%
\pgfsetlinewidth{1.003750pt}%
\definecolor{currentstroke}{rgb}{0.172549,0.627451,0.172549}%
\pgfsetstrokecolor{currentstroke}%
\pgfsetdash{{1.000000pt}{1.650000pt}}{0.000000pt}%
\pgfpathmoveto{\pgfqpoint{0.543161in}{0.514567in}}%
\pgfpathlineto{\pgfqpoint{0.596415in}{0.515235in}}%
\pgfpathlineto{\pgfqpoint{0.609728in}{0.515903in}}%
\pgfpathlineto{\pgfqpoint{0.662982in}{0.516571in}}%
\pgfpathlineto{\pgfqpoint{0.676296in}{0.517239in}}%
\pgfpathlineto{\pgfqpoint{0.749520in}{0.517907in}}%
\pgfpathlineto{\pgfqpoint{0.782803in}{0.520580in}}%
\pgfpathlineto{\pgfqpoint{0.789460in}{0.520580in}}%
\pgfpathlineto{\pgfqpoint{0.802773in}{0.522584in}}%
\pgfpathlineto{\pgfqpoint{0.816087in}{0.523252in}}%
\pgfpathlineto{\pgfqpoint{0.822744in}{0.524589in}}%
\pgfpathlineto{\pgfqpoint{0.875997in}{0.525257in}}%
\pgfpathlineto{\pgfqpoint{0.889311in}{0.525925in}}%
\pgfpathlineto{\pgfqpoint{0.922594in}{0.526593in}}%
\pgfpathlineto{\pgfqpoint{1.035759in}{0.537951in}}%
\pgfpathlineto{\pgfqpoint{1.042416in}{0.540624in}}%
\pgfpathlineto{\pgfqpoint{1.055729in}{0.542628in}}%
\pgfpathlineto{\pgfqpoint{1.075699in}{0.544633in}}%
\pgfpathlineto{\pgfqpoint{1.089013in}{0.548641in}}%
\pgfpathlineto{\pgfqpoint{1.095669in}{0.549309in}}%
\pgfpathlineto{\pgfqpoint{1.108983in}{0.559331in}}%
\pgfpathlineto{\pgfqpoint{1.115639in}{0.567349in}}%
\pgfpathlineto{\pgfqpoint{1.128953in}{0.586725in}}%
\pgfpathlineto{\pgfqpoint{1.135610in}{0.595410in}}%
\pgfpathlineto{\pgfqpoint{1.142266in}{0.598751in}}%
\pgfpathlineto{\pgfqpoint{1.155580in}{0.608773in}}%
\pgfpathlineto{\pgfqpoint{1.162237in}{0.616122in}}%
\pgfpathlineto{\pgfqpoint{1.182207in}{0.630153in}}%
\pgfpathlineto{\pgfqpoint{1.188863in}{0.632825in}}%
\pgfpathlineto{\pgfqpoint{1.195520in}{0.637502in}}%
\pgfpathlineto{\pgfqpoint{1.222147in}{0.663559in}}%
\pgfpathlineto{\pgfqpoint{1.235461in}{0.674249in}}%
\pgfpathlineto{\pgfqpoint{1.262087in}{0.703647in}}%
\pgfpathlineto{\pgfqpoint{1.268744in}{0.706987in}}%
\pgfpathlineto{\pgfqpoint{1.282058in}{0.723691in}}%
\pgfpathlineto{\pgfqpoint{1.288714in}{0.728367in}}%
\pgfpathlineto{\pgfqpoint{1.295371in}{0.735049in}}%
\pgfpathlineto{\pgfqpoint{1.321998in}{0.771128in}}%
\pgfpathlineto{\pgfqpoint{1.328655in}{0.779145in}}%
\pgfpathlineto{\pgfqpoint{1.335311in}{0.789167in}}%
\pgfpathlineto{\pgfqpoint{1.355282in}{0.831259in}}%
\pgfpathlineto{\pgfqpoint{1.361938in}{0.841281in}}%
\pgfpathlineto{\pgfqpoint{1.375252in}{0.870010in}}%
\pgfpathlineto{\pgfqpoint{1.388565in}{0.896067in}}%
\pgfpathlineto{\pgfqpoint{1.395222in}{0.915443in}}%
\pgfpathlineto{\pgfqpoint{1.401879in}{0.930142in}}%
\pgfpathlineto{\pgfqpoint{1.408535in}{0.949517in}}%
\pgfpathlineto{\pgfqpoint{1.415192in}{0.973570in}}%
\pgfpathlineto{\pgfqpoint{1.441819in}{1.037710in}}%
\pgfpathlineto{\pgfqpoint{1.448476in}{1.048400in}}%
\pgfpathlineto{\pgfqpoint{1.461789in}{1.079134in}}%
\pgfpathlineto{\pgfqpoint{1.468446in}{1.100514in}}%
\pgfpathlineto{\pgfqpoint{1.481759in}{1.131916in}}%
\pgfpathlineto{\pgfqpoint{1.495073in}{1.161982in}}%
\pgfpathlineto{\pgfqpoint{1.508386in}{1.198729in}}%
\pgfpathlineto{\pgfqpoint{1.515043in}{1.212759in}}%
\pgfpathlineto{\pgfqpoint{1.548327in}{1.336363in}}%
\pgfpathlineto{\pgfqpoint{1.554983in}{1.352398in}}%
\pgfpathlineto{\pgfqpoint{1.561640in}{1.383800in}}%
\pgfpathlineto{\pgfqpoint{1.574953in}{1.429900in}}%
\pgfpathlineto{\pgfqpoint{1.588267in}{1.494709in}}%
\pgfpathlineto{\pgfqpoint{1.594924in}{1.520098in}}%
\pgfpathlineto{\pgfqpoint{1.601580in}{1.556845in}}%
\pgfpathlineto{\pgfqpoint{1.621550in}{1.618980in}}%
\pgfpathlineto{\pgfqpoint{1.641521in}{1.665081in}}%
\pgfpathlineto{\pgfqpoint{1.661491in}{1.727885in}}%
\pgfpathlineto{\pgfqpoint{1.674804in}{1.763296in}}%
\pgfpathlineto{\pgfqpoint{1.681461in}{1.777326in}}%
\pgfpathlineto{\pgfqpoint{1.701431in}{1.828772in}}%
\pgfpathlineto{\pgfqpoint{1.721401in}{1.870864in}}%
\pgfpathlineto{\pgfqpoint{1.741371in}{1.902934in}}%
\pgfpathlineto{\pgfqpoint{1.748028in}{1.910952in}}%
\pgfpathlineto{\pgfqpoint{1.761342in}{1.933000in}}%
\pgfpathlineto{\pgfqpoint{1.767998in}{1.939681in}}%
\pgfpathlineto{\pgfqpoint{1.774655in}{1.948367in}}%
\pgfpathlineto{\pgfqpoint{1.794625in}{1.983109in}}%
\pgfpathlineto{\pgfqpoint{1.801282in}{1.990459in}}%
\pgfpathlineto{\pgfqpoint{1.807939in}{1.995804in}}%
\pgfpathlineto{\pgfqpoint{1.821252in}{2.018520in}}%
\pgfpathlineto{\pgfqpoint{1.841222in}{2.035891in}}%
\pgfpathlineto{\pgfqpoint{1.847879in}{2.043909in}}%
\pgfpathlineto{\pgfqpoint{1.867849in}{2.057940in}}%
\pgfpathlineto{\pgfqpoint{1.874506in}{2.064621in}}%
\pgfpathlineto{\pgfqpoint{1.887819in}{2.073306in}}%
\pgfpathlineto{\pgfqpoint{1.894476in}{2.075311in}}%
\pgfpathlineto{\pgfqpoint{1.914446in}{2.096691in}}%
\pgfpathlineto{\pgfqpoint{1.947730in}{2.114062in}}%
\pgfpathlineto{\pgfqpoint{1.961043in}{2.121412in}}%
\pgfpathlineto{\pgfqpoint{1.967700in}{2.122080in}}%
\pgfpathlineto{\pgfqpoint{1.974357in}{2.124752in}}%
\pgfpathlineto{\pgfqpoint{1.994327in}{2.125420in}}%
\pgfpathlineto{\pgfqpoint{2.000984in}{2.128093in}}%
\pgfpathlineto{\pgfqpoint{2.047581in}{2.132770in}}%
\pgfpathlineto{\pgfqpoint{2.054238in}{2.134774in}}%
\pgfpathlineto{\pgfqpoint{2.060894in}{2.135442in}}%
\pgfpathlineto{\pgfqpoint{2.087521in}{2.142792in}}%
\pgfpathlineto{\pgfqpoint{2.114148in}{2.143460in}}%
\pgfpathlineto{\pgfqpoint{2.127461in}{2.145464in}}%
\pgfpathlineto{\pgfqpoint{2.174059in}{2.145464in}}%
\pgfpathlineto{\pgfqpoint{2.174059in}{2.145464in}}%
\pgfusepath{stroke}%
\end{pgfscope}%
\begin{pgfscope}%
\pgfsetrectcap%
\pgfsetmiterjoin%
\pgfsetlinewidth{1.003750pt}%
\definecolor{currentstroke}{rgb}{1.000000,1.000000,1.000000}%
\pgfsetstrokecolor{currentstroke}%
\pgfsetdash{}{0pt}%
\pgfpathmoveto{\pgfqpoint{0.461616in}{0.433022in}}%
\pgfpathlineto{\pgfqpoint{0.461616in}{2.227009in}}%
\pgfusepath{stroke}%
\end{pgfscope}%
\begin{pgfscope}%
\pgfsetrectcap%
\pgfsetmiterjoin%
\pgfsetlinewidth{1.003750pt}%
\definecolor{currentstroke}{rgb}{1.000000,1.000000,1.000000}%
\pgfsetstrokecolor{currentstroke}%
\pgfsetdash{}{0pt}%
\pgfpathmoveto{\pgfqpoint{2.255603in}{0.433022in}}%
\pgfpathlineto{\pgfqpoint{2.255603in}{2.227009in}}%
\pgfusepath{stroke}%
\end{pgfscope}%
\begin{pgfscope}%
\pgfsetrectcap%
\pgfsetmiterjoin%
\pgfsetlinewidth{1.003750pt}%
\definecolor{currentstroke}{rgb}{1.000000,1.000000,1.000000}%
\pgfsetstrokecolor{currentstroke}%
\pgfsetdash{}{0pt}%
\pgfpathmoveto{\pgfqpoint{0.461616in}{0.433022in}}%
\pgfpathlineto{\pgfqpoint{2.255603in}{0.433022in}}%
\pgfusepath{stroke}%
\end{pgfscope}%
\begin{pgfscope}%
\pgfsetrectcap%
\pgfsetmiterjoin%
\pgfsetlinewidth{1.003750pt}%
\definecolor{currentstroke}{rgb}{1.000000,1.000000,1.000000}%
\pgfsetstrokecolor{currentstroke}%
\pgfsetdash{}{0pt}%
\pgfpathmoveto{\pgfqpoint{0.461616in}{2.227009in}}%
\pgfpathlineto{\pgfqpoint{2.255603in}{2.227009in}}%
\pgfusepath{stroke}%
\end{pgfscope}%
\end{pgfpicture}%
\makeatother%
\endgroup%

%% file: figures/grid_avg_delay_validity_starlink_valid.pgf
\begingroup%
\makeatletter%
\begin{pgfpicture}%
\pgfpathrectangle{\pgfpointorigin}{\pgfqpoint{2.300000in}{2.300000in}}%
\pgfusepath{use as bounding box, clip}%
\begin{pgfscope}%
\pgfsetbuttcap%
\pgfsetmiterjoin%
\definecolor{currentfill}{rgb}{1.000000,1.000000,1.000000}%
\pgfsetfillcolor{currentfill}%
\pgfsetlinewidth{0.000000pt}%
\definecolor{currentstroke}{rgb}{0.500000,0.500000,0.500000}%
\pgfsetstrokecolor{currentstroke}%
\pgfsetdash{}{0pt}%
\pgfpathmoveto{\pgfqpoint{0.000000in}{0.000000in}}%
\pgfpathlineto{\pgfqpoint{2.300000in}{0.000000in}}%
\pgfpathlineto{\pgfqpoint{2.300000in}{2.300000in}}%
\pgfpathlineto{\pgfqpoint{0.000000in}{2.300000in}}%
\pgfpathlineto{\pgfqpoint{0.000000in}{0.000000in}}%
\pgfpathclose%
\pgfusepath{fill}%
\end{pgfscope}%
\begin{pgfscope}%
\pgfsetbuttcap%
\pgfsetmiterjoin%
\definecolor{currentfill}{rgb}{0.898039,0.898039,0.898039}%
\pgfsetfillcolor{currentfill}%
\pgfsetlinewidth{0.000000pt}%
\definecolor{currentstroke}{rgb}{0.000000,0.000000,0.000000}%
\pgfsetstrokecolor{currentstroke}%
\pgfsetstrokeopacity{0.000000}%
\pgfsetdash{}{0pt}%
\pgfpathmoveto{\pgfqpoint{0.461616in}{0.431659in}}%
\pgfpathlineto{\pgfqpoint{2.258330in}{0.431659in}}%
\pgfpathlineto{\pgfqpoint{2.258330in}{2.228372in}}%
\pgfpathlineto{\pgfqpoint{0.461616in}{2.228372in}}%
\pgfpathlineto{\pgfqpoint{0.461616in}{0.431659in}}%
\pgfpathclose%
\pgfusepath{fill}%
\end{pgfscope}%
\begin{pgfscope}%
\pgfpathrectangle{\pgfqpoint{0.461616in}{0.431659in}}{\pgfqpoint{1.796714in}{1.796714in}}%
\pgfusepath{clip}%
\pgfsetrectcap%
\pgfsetroundjoin%
\pgfsetlinewidth{0.803000pt}%
\definecolor{currentstroke}{rgb}{1.000000,1.000000,1.000000}%
\pgfsetstrokecolor{currentstroke}%
\pgfsetdash{}{0pt}%
\pgfpathmoveto{\pgfqpoint{0.530241in}{0.431659in}}%
\pgfpathlineto{\pgfqpoint{0.530241in}{2.228372in}}%
\pgfusepath{stroke}%
\end{pgfscope}%
\begin{pgfscope}%
\pgfsetbuttcap%
\pgfsetroundjoin%
\definecolor{currentfill}{rgb}{0.333333,0.333333,0.333333}%
\pgfsetfillcolor{currentfill}%
\pgfsetlinewidth{0.803000pt}%
\definecolor{currentstroke}{rgb}{0.333333,0.333333,0.333333}%
\pgfsetstrokecolor{currentstroke}%
\pgfsetdash{}{0pt}%
\pgfsys@defobject{currentmarker}{\pgfqpoint{0.000000in}{-0.048611in}}{\pgfqpoint{0.000000in}{0.000000in}}{%
\pgfpathmoveto{\pgfqpoint{0.000000in}{0.000000in}}%
\pgfpathlineto{\pgfqpoint{0.000000in}{-0.048611in}}%
\pgfusepath{stroke,fill}%
}%
\begin{pgfscope}%
\pgfsys@transformshift{0.530241in}{0.431659in}%
\pgfsys@useobject{currentmarker}{}%
\end{pgfscope}%
\end{pgfscope}%
\begin{pgfscope}%
\definecolor{textcolor}{rgb}{0.333333,0.333333,0.333333}%
\pgfsetstrokecolor{textcolor}%
\pgfsetfillcolor{textcolor}%
\pgftext[x=0.530241in,y=0.334436in,,top]{\color{textcolor}\rmfamily\fontsize{7.000000}{8.400000}\selectfont \(\displaystyle {0}\)}%
\end{pgfscope}%
\begin{pgfscope}%
\pgfpathrectangle{\pgfqpoint{0.461616in}{0.431659in}}{\pgfqpoint{1.796714in}{1.796714in}}%
\pgfusepath{clip}%
\pgfsetrectcap%
\pgfsetroundjoin%
\pgfsetlinewidth{0.803000pt}%
\definecolor{currentstroke}{rgb}{1.000000,1.000000,1.000000}%
\pgfsetstrokecolor{currentstroke}%
\pgfsetdash{}{0pt}%
\pgfpathmoveto{\pgfqpoint{1.057136in}{0.431659in}}%
\pgfpathlineto{\pgfqpoint{1.057136in}{2.228372in}}%
\pgfusepath{stroke}%
\end{pgfscope}%
\begin{pgfscope}%
\pgfsetbuttcap%
\pgfsetroundjoin%
\definecolor{currentfill}{rgb}{0.333333,0.333333,0.333333}%
\pgfsetfillcolor{currentfill}%
\pgfsetlinewidth{0.803000pt}%
\definecolor{currentstroke}{rgb}{0.333333,0.333333,0.333333}%
\pgfsetstrokecolor{currentstroke}%
\pgfsetdash{}{0pt}%
\pgfsys@defobject{currentmarker}{\pgfqpoint{0.000000in}{-0.048611in}}{\pgfqpoint{0.000000in}{0.000000in}}{%
\pgfpathmoveto{\pgfqpoint{0.000000in}{0.000000in}}%
\pgfpathlineto{\pgfqpoint{0.000000in}{-0.048611in}}%
\pgfusepath{stroke,fill}%
}%
\begin{pgfscope}%
\pgfsys@transformshift{1.057136in}{0.431659in}%
\pgfsys@useobject{currentmarker}{}%
\end{pgfscope}%
\end{pgfscope}%
\begin{pgfscope}%
\definecolor{textcolor}{rgb}{0.333333,0.333333,0.333333}%
\pgfsetstrokecolor{textcolor}%
\pgfsetfillcolor{textcolor}%
\pgftext[x=1.057136in,y=0.334436in,,top]{\color{textcolor}\rmfamily\fontsize{7.000000}{8.400000}\selectfont \(\displaystyle {50}\)}%
\end{pgfscope}%
\begin{pgfscope}%
\pgfpathrectangle{\pgfqpoint{0.461616in}{0.431659in}}{\pgfqpoint{1.796714in}{1.796714in}}%
\pgfusepath{clip}%
\pgfsetrectcap%
\pgfsetroundjoin%
\pgfsetlinewidth{0.803000pt}%
\definecolor{currentstroke}{rgb}{1.000000,1.000000,1.000000}%
\pgfsetstrokecolor{currentstroke}%
\pgfsetdash{}{0pt}%
\pgfpathmoveto{\pgfqpoint{1.584032in}{0.431659in}}%
\pgfpathlineto{\pgfqpoint{1.584032in}{2.228372in}}%
\pgfusepath{stroke}%
\end{pgfscope}%
\begin{pgfscope}%
\pgfsetbuttcap%
\pgfsetroundjoin%
\definecolor{currentfill}{rgb}{0.333333,0.333333,0.333333}%
\pgfsetfillcolor{currentfill}%
\pgfsetlinewidth{0.803000pt}%
\definecolor{currentstroke}{rgb}{0.333333,0.333333,0.333333}%
\pgfsetstrokecolor{currentstroke}%
\pgfsetdash{}{0pt}%
\pgfsys@defobject{currentmarker}{\pgfqpoint{0.000000in}{-0.048611in}}{\pgfqpoint{0.000000in}{0.000000in}}{%
\pgfpathmoveto{\pgfqpoint{0.000000in}{0.000000in}}%
\pgfpathlineto{\pgfqpoint{0.000000in}{-0.048611in}}%
\pgfusepath{stroke,fill}%
}%
\begin{pgfscope}%
\pgfsys@transformshift{1.584032in}{0.431659in}%
\pgfsys@useobject{currentmarker}{}%
\end{pgfscope}%
\end{pgfscope}%
\begin{pgfscope}%
\definecolor{textcolor}{rgb}{0.333333,0.333333,0.333333}%
\pgfsetstrokecolor{textcolor}%
\pgfsetfillcolor{textcolor}%
\pgftext[x=1.584032in,y=0.334436in,,top]{\color{textcolor}\rmfamily\fontsize{7.000000}{8.400000}\selectfont \(\displaystyle {100}\)}%
\end{pgfscope}%
\begin{pgfscope}%
\pgfpathrectangle{\pgfqpoint{0.461616in}{0.431659in}}{\pgfqpoint{1.796714in}{1.796714in}}%
\pgfusepath{clip}%
\pgfsetrectcap%
\pgfsetroundjoin%
\pgfsetlinewidth{0.803000pt}%
\definecolor{currentstroke}{rgb}{1.000000,1.000000,1.000000}%
\pgfsetstrokecolor{currentstroke}%
\pgfsetdash{}{0pt}%
\pgfpathmoveto{\pgfqpoint{2.110927in}{0.431659in}}%
\pgfpathlineto{\pgfqpoint{2.110927in}{2.228372in}}%
\pgfusepath{stroke}%
\end{pgfscope}%
\begin{pgfscope}%
\pgfsetbuttcap%
\pgfsetroundjoin%
\definecolor{currentfill}{rgb}{0.333333,0.333333,0.333333}%
\pgfsetfillcolor{currentfill}%
\pgfsetlinewidth{0.803000pt}%
\definecolor{currentstroke}{rgb}{0.333333,0.333333,0.333333}%
\pgfsetstrokecolor{currentstroke}%
\pgfsetdash{}{0pt}%
\pgfsys@defobject{currentmarker}{\pgfqpoint{0.000000in}{-0.048611in}}{\pgfqpoint{0.000000in}{0.000000in}}{%
\pgfpathmoveto{\pgfqpoint{0.000000in}{0.000000in}}%
\pgfpathlineto{\pgfqpoint{0.000000in}{-0.048611in}}%
\pgfusepath{stroke,fill}%
}%
\begin{pgfscope}%
\pgfsys@transformshift{2.110927in}{0.431659in}%
\pgfsys@useobject{currentmarker}{}%
\end{pgfscope}%
\end{pgfscope}%
\begin{pgfscope}%
\definecolor{textcolor}{rgb}{0.333333,0.333333,0.333333}%
\pgfsetstrokecolor{textcolor}%
\pgfsetfillcolor{textcolor}%
\pgftext[x=2.110927in,y=0.334436in,,top]{\color{textcolor}\rmfamily\fontsize{7.000000}{8.400000}\selectfont \(\displaystyle {150}\)}%
\end{pgfscope}%
\begin{pgfscope}%
\definecolor{textcolor}{rgb}{0.333333,0.333333,0.333333}%
\pgfsetstrokecolor{textcolor}%
\pgfsetfillcolor{textcolor}%
\pgftext[x=1.359973in,y=0.192461in,,top]{\color{textcolor}\rmfamily\fontsize{10.000000}{12.000000}\selectfont Avg. Route Validity [s]}%
\end{pgfscope}%
\begin{pgfscope}%
\pgfpathrectangle{\pgfqpoint{0.461616in}{0.431659in}}{\pgfqpoint{1.796714in}{1.796714in}}%
\pgfusepath{clip}%
\pgfsetrectcap%
\pgfsetroundjoin%
\pgfsetlinewidth{0.803000pt}%
\definecolor{currentstroke}{rgb}{1.000000,1.000000,1.000000}%
\pgfsetstrokecolor{currentstroke}%
\pgfsetdash{}{0pt}%
\pgfpathmoveto{\pgfqpoint{0.461616in}{0.513327in}}%
\pgfpathlineto{\pgfqpoint{2.258330in}{0.513327in}}%
\pgfusepath{stroke}%
\end{pgfscope}%
\begin{pgfscope}%
\pgfsetbuttcap%
\pgfsetroundjoin%
\definecolor{currentfill}{rgb}{0.333333,0.333333,0.333333}%
\pgfsetfillcolor{currentfill}%
\pgfsetlinewidth{0.803000pt}%
\definecolor{currentstroke}{rgb}{0.333333,0.333333,0.333333}%
\pgfsetstrokecolor{currentstroke}%
\pgfsetdash{}{0pt}%
\pgfsys@defobject{currentmarker}{\pgfqpoint{-0.048611in}{0.000000in}}{\pgfqpoint{-0.000000in}{0.000000in}}{%
\pgfpathmoveto{\pgfqpoint{-0.000000in}{0.000000in}}%
\pgfpathlineto{\pgfqpoint{-0.048611in}{0.000000in}}%
\pgfusepath{stroke,fill}%
}%
\begin{pgfscope}%
\pgfsys@transformshift{0.461616in}{0.513327in}%
\pgfsys@useobject{currentmarker}{}%
\end{pgfscope}%
\end{pgfscope}%
\begin{pgfscope}%
\definecolor{textcolor}{rgb}{0.333333,0.333333,0.333333}%
\pgfsetstrokecolor{textcolor}%
\pgfsetfillcolor{textcolor}%
\pgftext[x=0.220682in, y=0.479570in, left, base]{\color{textcolor}\rmfamily\fontsize{7.000000}{8.400000}\selectfont \(\displaystyle {0.0}\)}%
\end{pgfscope}%
\begin{pgfscope}%
\pgfpathrectangle{\pgfqpoint{0.461616in}{0.431659in}}{\pgfqpoint{1.796714in}{1.796714in}}%
\pgfusepath{clip}%
\pgfsetrectcap%
\pgfsetroundjoin%
\pgfsetlinewidth{0.803000pt}%
\definecolor{currentstroke}{rgb}{1.000000,1.000000,1.000000}%
\pgfsetstrokecolor{currentstroke}%
\pgfsetdash{}{0pt}%
\pgfpathmoveto{\pgfqpoint{0.461616in}{0.840003in}}%
\pgfpathlineto{\pgfqpoint{2.258330in}{0.840003in}}%
\pgfusepath{stroke}%
\end{pgfscope}%
\begin{pgfscope}%
\pgfsetbuttcap%
\pgfsetroundjoin%
\definecolor{currentfill}{rgb}{0.333333,0.333333,0.333333}%
\pgfsetfillcolor{currentfill}%
\pgfsetlinewidth{0.803000pt}%
\definecolor{currentstroke}{rgb}{0.333333,0.333333,0.333333}%
\pgfsetstrokecolor{currentstroke}%
\pgfsetdash{}{0pt}%
\pgfsys@defobject{currentmarker}{\pgfqpoint{-0.048611in}{0.000000in}}{\pgfqpoint{-0.000000in}{0.000000in}}{%
\pgfpathmoveto{\pgfqpoint{-0.000000in}{0.000000in}}%
\pgfpathlineto{\pgfqpoint{-0.048611in}{0.000000in}}%
\pgfusepath{stroke,fill}%
}%
\begin{pgfscope}%
\pgfsys@transformshift{0.461616in}{0.840003in}%
\pgfsys@useobject{currentmarker}{}%
\end{pgfscope}%
\end{pgfscope}%
\begin{pgfscope}%
\definecolor{textcolor}{rgb}{0.333333,0.333333,0.333333}%
\pgfsetstrokecolor{textcolor}%
\pgfsetfillcolor{textcolor}%
\pgftext[x=0.220682in, y=0.806245in, left, base]{\color{textcolor}\rmfamily\fontsize{7.000000}{8.400000}\selectfont \(\displaystyle {0.2}\)}%
\end{pgfscope}%
\begin{pgfscope}%
\pgfpathrectangle{\pgfqpoint{0.461616in}{0.431659in}}{\pgfqpoint{1.796714in}{1.796714in}}%
\pgfusepath{clip}%
\pgfsetrectcap%
\pgfsetroundjoin%
\pgfsetlinewidth{0.803000pt}%
\definecolor{currentstroke}{rgb}{1.000000,1.000000,1.000000}%
\pgfsetstrokecolor{currentstroke}%
\pgfsetdash{}{0pt}%
\pgfpathmoveto{\pgfqpoint{0.461616in}{1.166678in}}%
\pgfpathlineto{\pgfqpoint{2.258330in}{1.166678in}}%
\pgfusepath{stroke}%
\end{pgfscope}%
\begin{pgfscope}%
\pgfsetbuttcap%
\pgfsetroundjoin%
\definecolor{currentfill}{rgb}{0.333333,0.333333,0.333333}%
\pgfsetfillcolor{currentfill}%
\pgfsetlinewidth{0.803000pt}%
\definecolor{currentstroke}{rgb}{0.333333,0.333333,0.333333}%
\pgfsetstrokecolor{currentstroke}%
\pgfsetdash{}{0pt}%
\pgfsys@defobject{currentmarker}{\pgfqpoint{-0.048611in}{0.000000in}}{\pgfqpoint{-0.000000in}{0.000000in}}{%
\pgfpathmoveto{\pgfqpoint{-0.000000in}{0.000000in}}%
\pgfpathlineto{\pgfqpoint{-0.048611in}{0.000000in}}%
\pgfusepath{stroke,fill}%
}%
\begin{pgfscope}%
\pgfsys@transformshift{0.461616in}{1.166678in}%
\pgfsys@useobject{currentmarker}{}%
\end{pgfscope}%
\end{pgfscope}%
\begin{pgfscope}%
\definecolor{textcolor}{rgb}{0.333333,0.333333,0.333333}%
\pgfsetstrokecolor{textcolor}%
\pgfsetfillcolor{textcolor}%
\pgftext[x=0.220682in, y=1.132920in, left, base]{\color{textcolor}\rmfamily\fontsize{7.000000}{8.400000}\selectfont \(\displaystyle {0.4}\)}%
\end{pgfscope}%
\begin{pgfscope}%
\pgfpathrectangle{\pgfqpoint{0.461616in}{0.431659in}}{\pgfqpoint{1.796714in}{1.796714in}}%
\pgfusepath{clip}%
\pgfsetrectcap%
\pgfsetroundjoin%
\pgfsetlinewidth{0.803000pt}%
\definecolor{currentstroke}{rgb}{1.000000,1.000000,1.000000}%
\pgfsetstrokecolor{currentstroke}%
\pgfsetdash{}{0pt}%
\pgfpathmoveto{\pgfqpoint{0.461616in}{1.493353in}}%
\pgfpathlineto{\pgfqpoint{2.258330in}{1.493353in}}%
\pgfusepath{stroke}%
\end{pgfscope}%
\begin{pgfscope}%
\pgfsetbuttcap%
\pgfsetroundjoin%
\definecolor{currentfill}{rgb}{0.333333,0.333333,0.333333}%
\pgfsetfillcolor{currentfill}%
\pgfsetlinewidth{0.803000pt}%
\definecolor{currentstroke}{rgb}{0.333333,0.333333,0.333333}%
\pgfsetstrokecolor{currentstroke}%
\pgfsetdash{}{0pt}%
\pgfsys@defobject{currentmarker}{\pgfqpoint{-0.048611in}{0.000000in}}{\pgfqpoint{-0.000000in}{0.000000in}}{%
\pgfpathmoveto{\pgfqpoint{-0.000000in}{0.000000in}}%
\pgfpathlineto{\pgfqpoint{-0.048611in}{0.000000in}}%
\pgfusepath{stroke,fill}%
}%
\begin{pgfscope}%
\pgfsys@transformshift{0.461616in}{1.493353in}%
\pgfsys@useobject{currentmarker}{}%
\end{pgfscope}%
\end{pgfscope}%
\begin{pgfscope}%
\definecolor{textcolor}{rgb}{0.333333,0.333333,0.333333}%
\pgfsetstrokecolor{textcolor}%
\pgfsetfillcolor{textcolor}%
\pgftext[x=0.220682in, y=1.459595in, left, base]{\color{textcolor}\rmfamily\fontsize{7.000000}{8.400000}\selectfont \(\displaystyle {0.6}\)}%
\end{pgfscope}%
\begin{pgfscope}%
\pgfpathrectangle{\pgfqpoint{0.461616in}{0.431659in}}{\pgfqpoint{1.796714in}{1.796714in}}%
\pgfusepath{clip}%
\pgfsetrectcap%
\pgfsetroundjoin%
\pgfsetlinewidth{0.803000pt}%
\definecolor{currentstroke}{rgb}{1.000000,1.000000,1.000000}%
\pgfsetstrokecolor{currentstroke}%
\pgfsetdash{}{0pt}%
\pgfpathmoveto{\pgfqpoint{0.461616in}{1.820028in}}%
\pgfpathlineto{\pgfqpoint{2.258330in}{1.820028in}}%
\pgfusepath{stroke}%
\end{pgfscope}%
\begin{pgfscope}%
\pgfsetbuttcap%
\pgfsetroundjoin%
\definecolor{currentfill}{rgb}{0.333333,0.333333,0.333333}%
\pgfsetfillcolor{currentfill}%
\pgfsetlinewidth{0.803000pt}%
\definecolor{currentstroke}{rgb}{0.333333,0.333333,0.333333}%
\pgfsetstrokecolor{currentstroke}%
\pgfsetdash{}{0pt}%
\pgfsys@defobject{currentmarker}{\pgfqpoint{-0.048611in}{0.000000in}}{\pgfqpoint{-0.000000in}{0.000000in}}{%
\pgfpathmoveto{\pgfqpoint{-0.000000in}{0.000000in}}%
\pgfpathlineto{\pgfqpoint{-0.048611in}{0.000000in}}%
\pgfusepath{stroke,fill}%
}%
\begin{pgfscope}%
\pgfsys@transformshift{0.461616in}{1.820028in}%
\pgfsys@useobject{currentmarker}{}%
\end{pgfscope}%
\end{pgfscope}%
\begin{pgfscope}%
\definecolor{textcolor}{rgb}{0.333333,0.333333,0.333333}%
\pgfsetstrokecolor{textcolor}%
\pgfsetfillcolor{textcolor}%
\pgftext[x=0.220682in, y=1.786271in, left, base]{\color{textcolor}\rmfamily\fontsize{7.000000}{8.400000}\selectfont \(\displaystyle {0.8}\)}%
\end{pgfscope}%
\begin{pgfscope}%
\pgfpathrectangle{\pgfqpoint{0.461616in}{0.431659in}}{\pgfqpoint{1.796714in}{1.796714in}}%
\pgfusepath{clip}%
\pgfsetrectcap%
\pgfsetroundjoin%
\pgfsetlinewidth{0.803000pt}%
\definecolor{currentstroke}{rgb}{1.000000,1.000000,1.000000}%
\pgfsetstrokecolor{currentstroke}%
\pgfsetdash{}{0pt}%
\pgfpathmoveto{\pgfqpoint{0.461616in}{2.146703in}}%
\pgfpathlineto{\pgfqpoint{2.258330in}{2.146703in}}%
\pgfusepath{stroke}%
\end{pgfscope}%
\begin{pgfscope}%
\pgfsetbuttcap%
\pgfsetroundjoin%
\definecolor{currentfill}{rgb}{0.333333,0.333333,0.333333}%
\pgfsetfillcolor{currentfill}%
\pgfsetlinewidth{0.803000pt}%
\definecolor{currentstroke}{rgb}{0.333333,0.333333,0.333333}%
\pgfsetstrokecolor{currentstroke}%
\pgfsetdash{}{0pt}%
\pgfsys@defobject{currentmarker}{\pgfqpoint{-0.048611in}{0.000000in}}{\pgfqpoint{-0.000000in}{0.000000in}}{%
\pgfpathmoveto{\pgfqpoint{-0.000000in}{0.000000in}}%
\pgfpathlineto{\pgfqpoint{-0.048611in}{0.000000in}}%
\pgfusepath{stroke,fill}%
}%
\begin{pgfscope}%
\pgfsys@transformshift{0.461616in}{2.146703in}%
\pgfsys@useobject{currentmarker}{}%
\end{pgfscope}%
\end{pgfscope}%
\begin{pgfscope}%
\definecolor{textcolor}{rgb}{0.333333,0.333333,0.333333}%
\pgfsetstrokecolor{textcolor}%
\pgfsetfillcolor{textcolor}%
\pgftext[x=0.220682in, y=2.112946in, left, base]{\color{textcolor}\rmfamily\fontsize{7.000000}{8.400000}\selectfont \(\displaystyle {1.0}\)}%
\end{pgfscope}%
\begin{pgfscope}%
\definecolor{textcolor}{rgb}{0.333333,0.333333,0.333333}%
\pgfsetstrokecolor{textcolor}%
\pgfsetfillcolor{textcolor}%
\pgftext[x=0.165127in,y=1.330015in,,bottom,rotate=90.000000]{\color{textcolor}\rmfamily\fontsize{10.000000}{12.000000}\selectfont Cumulative Probability}%
\end{pgfscope}%
\begin{pgfscope}%
\pgfpathrectangle{\pgfqpoint{0.461616in}{0.431659in}}{\pgfqpoint{1.796714in}{1.796714in}}%
\pgfusepath{clip}%
\pgfsetrectcap%
\pgfsetroundjoin%
\pgfsetlinewidth{1.003750pt}%
\definecolor{currentstroke}{rgb}{1.000000,0.498039,0.054902}%
\pgfsetstrokecolor{currentstroke}%
\pgfsetdash{}{0pt}%
\pgfpathmoveto{\pgfqpoint{0.543285in}{0.513327in}}%
\pgfpathlineto{\pgfqpoint{0.553823in}{0.519347in}}%
\pgfpathlineto{\pgfqpoint{0.564361in}{0.521354in}}%
\pgfpathlineto{\pgfqpoint{0.627588in}{0.521354in}}%
\pgfpathlineto{\pgfqpoint{0.638126in}{0.524029in}}%
\pgfpathlineto{\pgfqpoint{0.669740in}{0.528711in}}%
\pgfpathlineto{\pgfqpoint{0.680278in}{0.535400in}}%
\pgfpathlineto{\pgfqpoint{0.690816in}{0.538076in}}%
\pgfpathlineto{\pgfqpoint{0.701354in}{0.544095in}}%
\pgfpathlineto{\pgfqpoint{0.711892in}{0.548109in}}%
\pgfpathlineto{\pgfqpoint{0.722430in}{0.558811in}}%
\pgfpathlineto{\pgfqpoint{0.732968in}{0.577539in}}%
\pgfpathlineto{\pgfqpoint{0.743505in}{0.590916in}}%
\pgfpathlineto{\pgfqpoint{0.754043in}{0.615664in}}%
\pgfpathlineto{\pgfqpoint{0.764581in}{0.629042in}}%
\pgfpathlineto{\pgfqpoint{0.775119in}{0.676531in}}%
\pgfpathlineto{\pgfqpoint{0.817271in}{0.926688in}}%
\pgfpathlineto{\pgfqpoint{0.827809in}{1.001601in}}%
\pgfpathlineto{\pgfqpoint{0.838347in}{1.092567in}}%
\pgfpathlineto{\pgfqpoint{0.859422in}{1.334698in}}%
\pgfpathlineto{\pgfqpoint{0.869960in}{1.439710in}}%
\pgfpathlineto{\pgfqpoint{0.901574in}{1.698562in}}%
\pgfpathlineto{\pgfqpoint{0.922650in}{1.845044in}}%
\pgfpathlineto{\pgfqpoint{0.933188in}{1.886514in}}%
\pgfpathlineto{\pgfqpoint{0.954264in}{1.994870in}}%
\pgfpathlineto{\pgfqpoint{0.964802in}{2.024301in}}%
\pgfpathlineto{\pgfqpoint{0.975339in}{2.036340in}}%
\pgfpathlineto{\pgfqpoint{0.985877in}{2.060420in}}%
\pgfpathlineto{\pgfqpoint{0.996415in}{2.070453in}}%
\pgfpathlineto{\pgfqpoint{1.006953in}{2.085836in}}%
\pgfpathlineto{\pgfqpoint{1.017491in}{2.097876in}}%
\pgfpathlineto{\pgfqpoint{1.028029in}{2.103227in}}%
\pgfpathlineto{\pgfqpoint{1.038567in}{2.105903in}}%
\pgfpathlineto{\pgfqpoint{1.049105in}{2.106571in}}%
\pgfpathlineto{\pgfqpoint{1.080719in}{2.121955in}}%
\pgfpathlineto{\pgfqpoint{1.101794in}{2.128644in}}%
\pgfpathlineto{\pgfqpoint{1.133408in}{2.133995in}}%
\pgfpathlineto{\pgfqpoint{1.207174in}{2.139346in}}%
\pgfpathlineto{\pgfqpoint{1.217711in}{2.141353in}}%
\pgfpathlineto{\pgfqpoint{1.228249in}{2.142021in}}%
\pgfpathlineto{\pgfqpoint{1.249325in}{2.145366in}}%
\pgfpathlineto{\pgfqpoint{1.365242in}{2.146035in}}%
\pgfpathlineto{\pgfqpoint{1.386318in}{2.146703in}}%
\pgfpathlineto{\pgfqpoint{2.176661in}{2.146703in}}%
\pgfpathlineto{\pgfqpoint{2.176661in}{2.146703in}}%
\pgfusepath{stroke}%
\end{pgfscope}%
\begin{pgfscope}%
\pgfpathrectangle{\pgfqpoint{0.461616in}{0.431659in}}{\pgfqpoint{1.796714in}{1.796714in}}%
\pgfusepath{clip}%
\pgfsetbuttcap%
\pgfsetroundjoin%
\pgfsetlinewidth{1.003750pt}%
\definecolor{currentstroke}{rgb}{0.121569,0.466667,0.705882}%
\pgfsetstrokecolor{currentstroke}%
\pgfsetdash{{3.700000pt}{1.600000pt}}{0.000000pt}%
\pgfpathmoveto{\pgfqpoint{0.543285in}{0.513327in}}%
\pgfpathlineto{\pgfqpoint{0.848885in}{0.513327in}}%
\pgfpathlineto{\pgfqpoint{0.869960in}{0.516003in}}%
\pgfpathlineto{\pgfqpoint{0.891036in}{0.516003in}}%
\pgfpathlineto{\pgfqpoint{0.922650in}{0.520016in}}%
\pgfpathlineto{\pgfqpoint{0.943726in}{0.529380in}}%
\pgfpathlineto{\pgfqpoint{0.954264in}{0.535400in}}%
\pgfpathlineto{\pgfqpoint{0.964802in}{0.546771in}}%
\pgfpathlineto{\pgfqpoint{0.975339in}{0.560148in}}%
\pgfpathlineto{\pgfqpoint{0.996415in}{0.612320in}}%
\pgfpathlineto{\pgfqpoint{1.017491in}{0.687902in}}%
\pgfpathlineto{\pgfqpoint{1.028029in}{0.729372in}}%
\pgfpathlineto{\pgfqpoint{1.038567in}{0.782881in}}%
\pgfpathlineto{\pgfqpoint{1.070181in}{0.981535in}}%
\pgfpathlineto{\pgfqpoint{1.112332in}{1.321989in}}%
\pgfpathlineto{\pgfqpoint{1.133408in}{1.453087in}}%
\pgfpathlineto{\pgfqpoint{1.143946in}{1.541378in}}%
\pgfpathlineto{\pgfqpoint{1.154484in}{1.592212in}}%
\pgfpathlineto{\pgfqpoint{1.165022in}{1.655085in}}%
\pgfpathlineto{\pgfqpoint{1.196636in}{1.788190in}}%
\pgfpathlineto{\pgfqpoint{1.249325in}{1.986175in}}%
\pgfpathlineto{\pgfqpoint{1.280939in}{2.039685in}}%
\pgfpathlineto{\pgfqpoint{1.291477in}{2.053062in}}%
\pgfpathlineto{\pgfqpoint{1.302015in}{2.063095in}}%
\pgfpathlineto{\pgfqpoint{1.312553in}{2.075803in}}%
\pgfpathlineto{\pgfqpoint{1.323091in}{2.085836in}}%
\pgfpathlineto{\pgfqpoint{1.333628in}{2.098545in}}%
\pgfpathlineto{\pgfqpoint{1.354704in}{2.121287in}}%
\pgfpathlineto{\pgfqpoint{1.365242in}{2.127306in}}%
\pgfpathlineto{\pgfqpoint{1.375780in}{2.138008in}}%
\pgfpathlineto{\pgfqpoint{1.407394in}{2.140684in}}%
\pgfpathlineto{\pgfqpoint{1.417932in}{2.142690in}}%
\pgfpathlineto{\pgfqpoint{1.428470in}{2.142690in}}%
\pgfpathlineto{\pgfqpoint{1.439008in}{2.144028in}}%
\pgfpathlineto{\pgfqpoint{1.460083in}{2.144028in}}%
\pgfpathlineto{\pgfqpoint{1.470621in}{2.146035in}}%
\pgfpathlineto{\pgfqpoint{1.491697in}{2.146703in}}%
\pgfpathlineto{\pgfqpoint{2.176661in}{2.146703in}}%
\pgfpathlineto{\pgfqpoint{2.176661in}{2.146703in}}%
\pgfusepath{stroke}%
\end{pgfscope}%
\begin{pgfscope}%
\pgfpathrectangle{\pgfqpoint{0.461616in}{0.431659in}}{\pgfqpoint{1.796714in}{1.796714in}}%
\pgfusepath{clip}%
\pgfsetbuttcap%
\pgfsetroundjoin%
\pgfsetlinewidth{1.003750pt}%
\definecolor{currentstroke}{rgb}{0.839216,0.152941,0.156863}%
\pgfsetstrokecolor{currentstroke}%
\pgfsetdash{{6.400000pt}{1.600000pt}{1.000000pt}{1.600000pt}}{0.000000pt}%
\pgfpathmoveto{\pgfqpoint{0.543285in}{0.513327in}}%
\pgfpathlineto{\pgfqpoint{1.080719in}{0.513327in}}%
\pgfpathlineto{\pgfqpoint{1.101794in}{0.518010in}}%
\pgfpathlineto{\pgfqpoint{1.112332in}{0.518010in}}%
\pgfpathlineto{\pgfqpoint{1.122870in}{0.520685in}}%
\pgfpathlineto{\pgfqpoint{1.133408in}{0.527374in}}%
\pgfpathlineto{\pgfqpoint{1.143946in}{0.531387in}}%
\pgfpathlineto{\pgfqpoint{1.154484in}{0.537407in}}%
\pgfpathlineto{\pgfqpoint{1.165022in}{0.546102in}}%
\pgfpathlineto{\pgfqpoint{1.175560in}{0.560817in}}%
\pgfpathlineto{\pgfqpoint{1.186098in}{0.573526in}}%
\pgfpathlineto{\pgfqpoint{1.196636in}{0.584228in}}%
\pgfpathlineto{\pgfqpoint{1.207174in}{0.600949in}}%
\pgfpathlineto{\pgfqpoint{1.217711in}{0.612320in}}%
\pgfpathlineto{\pgfqpoint{1.228249in}{0.629711in}}%
\pgfpathlineto{\pgfqpoint{1.238787in}{0.643088in}}%
\pgfpathlineto{\pgfqpoint{1.249325in}{0.666498in}}%
\pgfpathlineto{\pgfqpoint{1.270401in}{0.706630in}}%
\pgfpathlineto{\pgfqpoint{1.280939in}{0.734054in}}%
\pgfpathlineto{\pgfqpoint{1.291477in}{0.765491in}}%
\pgfpathlineto{\pgfqpoint{1.302015in}{0.787563in}}%
\pgfpathlineto{\pgfqpoint{1.312553in}{0.816325in}}%
\pgfpathlineto{\pgfqpoint{1.323091in}{0.840404in}}%
\pgfpathlineto{\pgfqpoint{1.344166in}{0.905953in}}%
\pgfpathlineto{\pgfqpoint{1.375780in}{1.023674in}}%
\pgfpathlineto{\pgfqpoint{1.386318in}{1.050429in}}%
\pgfpathlineto{\pgfqpoint{1.396856in}{1.099256in}}%
\pgfpathlineto{\pgfqpoint{1.407394in}{1.166812in}}%
\pgfpathlineto{\pgfqpoint{1.428470in}{1.237712in}}%
\pgfpathlineto{\pgfqpoint{1.439008in}{1.293897in}}%
\pgfpathlineto{\pgfqpoint{1.449545in}{1.360783in}}%
\pgfpathlineto{\pgfqpoint{1.460083in}{1.453087in}}%
\pgfpathlineto{\pgfqpoint{1.470621in}{1.619635in}}%
\pgfpathlineto{\pgfqpoint{1.481159in}{1.727992in}}%
\pgfpathlineto{\pgfqpoint{1.491697in}{1.741369in}}%
\pgfpathlineto{\pgfqpoint{1.512773in}{1.760098in}}%
\pgfpathlineto{\pgfqpoint{1.523311in}{1.786852in}}%
\pgfpathlineto{\pgfqpoint{1.533849in}{1.818289in}}%
\pgfpathlineto{\pgfqpoint{1.544387in}{1.835680in}}%
\pgfpathlineto{\pgfqpoint{1.554925in}{1.874474in}}%
\pgfpathlineto{\pgfqpoint{1.565462in}{1.881832in}}%
\pgfpathlineto{\pgfqpoint{1.576000in}{1.892534in}}%
\pgfpathlineto{\pgfqpoint{1.586538in}{1.900560in}}%
\pgfpathlineto{\pgfqpoint{1.597076in}{1.914606in}}%
\pgfpathlineto{\pgfqpoint{1.628690in}{1.942030in}}%
\pgfpathlineto{\pgfqpoint{1.639228in}{1.960758in}}%
\pgfpathlineto{\pgfqpoint{1.649766in}{1.971460in}}%
\pgfpathlineto{\pgfqpoint{1.660304in}{1.976142in}}%
\pgfpathlineto{\pgfqpoint{1.670842in}{1.996877in}}%
\pgfpathlineto{\pgfqpoint{1.681379in}{2.001559in}}%
\pgfpathlineto{\pgfqpoint{1.691917in}{2.004903in}}%
\pgfpathlineto{\pgfqpoint{1.702455in}{2.010254in}}%
\pgfpathlineto{\pgfqpoint{1.723531in}{2.016274in}}%
\pgfpathlineto{\pgfqpoint{1.744607in}{2.016274in}}%
\pgfpathlineto{\pgfqpoint{1.755145in}{2.018950in}}%
\pgfpathlineto{\pgfqpoint{1.776221in}{2.026307in}}%
\pgfpathlineto{\pgfqpoint{1.786759in}{2.026976in}}%
\pgfpathlineto{\pgfqpoint{1.797296in}{2.028983in}}%
\pgfpathlineto{\pgfqpoint{1.807834in}{2.028983in}}%
\pgfpathlineto{\pgfqpoint{1.818372in}{2.030320in}}%
\pgfpathlineto{\pgfqpoint{1.828910in}{2.030320in}}%
\pgfpathlineto{\pgfqpoint{1.839448in}{2.031658in}}%
\pgfpathlineto{\pgfqpoint{1.849986in}{2.031658in}}%
\pgfpathlineto{\pgfqpoint{1.860524in}{2.033665in}}%
\pgfpathlineto{\pgfqpoint{1.871062in}{2.047711in}}%
\pgfpathlineto{\pgfqpoint{1.881600in}{2.069784in}}%
\pgfpathlineto{\pgfqpoint{1.892138in}{2.101220in}}%
\pgfpathlineto{\pgfqpoint{1.902676in}{2.126637in}}%
\pgfpathlineto{\pgfqpoint{1.913213in}{2.127975in}}%
\pgfpathlineto{\pgfqpoint{1.934289in}{2.137339in}}%
\pgfpathlineto{\pgfqpoint{1.944827in}{2.137339in}}%
\pgfpathlineto{\pgfqpoint{1.955365in}{2.142021in}}%
\pgfpathlineto{\pgfqpoint{1.965903in}{2.142021in}}%
\pgfpathlineto{\pgfqpoint{1.976441in}{2.146035in}}%
\pgfpathlineto{\pgfqpoint{2.176661in}{2.146703in}}%
\pgfpathlineto{\pgfqpoint{2.176661in}{2.146703in}}%
\pgfusepath{stroke}%
\end{pgfscope}%
\begin{pgfscope}%
\pgfpathrectangle{\pgfqpoint{0.461616in}{0.431659in}}{\pgfqpoint{1.796714in}{1.796714in}}%
\pgfusepath{clip}%
\pgfsetbuttcap%
\pgfsetroundjoin%
\pgfsetlinewidth{1.003750pt}%
\definecolor{currentstroke}{rgb}{0.172549,0.627451,0.172549}%
\pgfsetstrokecolor{currentstroke}%
\pgfsetdash{{1.000000pt}{1.650000pt}}{0.000000pt}%
\pgfpathmoveto{\pgfqpoint{0.543285in}{0.513327in}}%
\pgfpathlineto{\pgfqpoint{1.460083in}{0.513327in}}%
\pgfpathlineto{\pgfqpoint{1.470621in}{0.543427in}}%
\pgfpathlineto{\pgfqpoint{1.481159in}{0.569512in}}%
\pgfpathlineto{\pgfqpoint{1.491697in}{0.570850in}}%
\pgfpathlineto{\pgfqpoint{1.502235in}{0.579545in}}%
\pgfpathlineto{\pgfqpoint{1.512773in}{0.594261in}}%
\pgfpathlineto{\pgfqpoint{1.523311in}{0.602956in}}%
\pgfpathlineto{\pgfqpoint{1.533849in}{0.626366in}}%
\pgfpathlineto{\pgfqpoint{1.544387in}{0.641750in}}%
\pgfpathlineto{\pgfqpoint{1.565462in}{0.823013in}}%
\pgfpathlineto{\pgfqpoint{1.576000in}{0.856457in}}%
\pgfpathlineto{\pgfqpoint{1.597076in}{0.891238in}}%
\pgfpathlineto{\pgfqpoint{1.618152in}{0.934046in}}%
\pgfpathlineto{\pgfqpoint{1.628690in}{0.948761in}}%
\pgfpathlineto{\pgfqpoint{1.639228in}{0.972171in}}%
\pgfpathlineto{\pgfqpoint{1.649766in}{1.012972in}}%
\pgfpathlineto{\pgfqpoint{1.660304in}{1.063137in}}%
\pgfpathlineto{\pgfqpoint{1.670842in}{1.130024in}}%
\pgfpathlineto{\pgfqpoint{1.681379in}{1.144739in}}%
\pgfpathlineto{\pgfqpoint{1.702455in}{1.217646in}}%
\pgfpathlineto{\pgfqpoint{1.712993in}{1.241056in}}%
\pgfpathlineto{\pgfqpoint{1.723531in}{1.253096in}}%
\pgfpathlineto{\pgfqpoint{1.734069in}{1.261791in}}%
\pgfpathlineto{\pgfqpoint{1.744607in}{1.291890in}}%
\pgfpathlineto{\pgfqpoint{1.755145in}{1.305936in}}%
\pgfpathlineto{\pgfqpoint{1.776221in}{1.354095in}}%
\pgfpathlineto{\pgfqpoint{1.786759in}{1.357439in}}%
\pgfpathlineto{\pgfqpoint{1.797296in}{1.367472in}}%
\pgfpathlineto{\pgfqpoint{1.818372in}{1.383525in}}%
\pgfpathlineto{\pgfqpoint{1.828910in}{1.403591in}}%
\pgfpathlineto{\pgfqpoint{1.839448in}{1.406935in}}%
\pgfpathlineto{\pgfqpoint{1.849986in}{1.422319in}}%
\pgfpathlineto{\pgfqpoint{1.860524in}{1.515292in}}%
\pgfpathlineto{\pgfqpoint{1.871062in}{1.525994in}}%
\pgfpathlineto{\pgfqpoint{1.881600in}{1.574152in}}%
\pgfpathlineto{\pgfqpoint{1.892138in}{1.784846in}}%
\pgfpathlineto{\pgfqpoint{1.902676in}{1.966109in}}%
\pgfpathlineto{\pgfqpoint{1.913213in}{1.988851in}}%
\pgfpathlineto{\pgfqpoint{1.923751in}{1.988851in}}%
\pgfpathlineto{\pgfqpoint{1.934289in}{1.996208in}}%
\pgfpathlineto{\pgfqpoint{1.944827in}{2.030989in}}%
\pgfpathlineto{\pgfqpoint{1.955365in}{2.043698in}}%
\pgfpathlineto{\pgfqpoint{1.965903in}{2.064433in}}%
\pgfpathlineto{\pgfqpoint{1.976441in}{2.104565in}}%
\pgfpathlineto{\pgfqpoint{1.986979in}{2.109916in}}%
\pgfpathlineto{\pgfqpoint{2.039668in}{2.109916in}}%
\pgfpathlineto{\pgfqpoint{2.060744in}{2.113260in}}%
\pgfpathlineto{\pgfqpoint{2.081820in}{2.113929in}}%
\pgfpathlineto{\pgfqpoint{2.092358in}{2.117942in}}%
\pgfpathlineto{\pgfqpoint{2.123972in}{2.124631in}}%
\pgfpathlineto{\pgfqpoint{2.134510in}{2.127306in}}%
\pgfpathlineto{\pgfqpoint{2.145047in}{2.132657in}}%
\pgfpathlineto{\pgfqpoint{2.166123in}{2.146703in}}%
\pgfpathlineto{\pgfqpoint{2.176661in}{2.146703in}}%
\pgfpathlineto{\pgfqpoint{2.176661in}{2.146703in}}%
\pgfusepath{stroke}%
\end{pgfscope}%
\begin{pgfscope}%
\pgfsetrectcap%
\pgfsetmiterjoin%
\pgfsetlinewidth{1.003750pt}%
\definecolor{currentstroke}{rgb}{1.000000,1.000000,1.000000}%
\pgfsetstrokecolor{currentstroke}%
\pgfsetdash{}{0pt}%
\pgfpathmoveto{\pgfqpoint{0.461616in}{0.431659in}}%
\pgfpathlineto{\pgfqpoint{0.461616in}{2.228372in}}%
\pgfusepath{stroke}%
\end{pgfscope}%
\begin{pgfscope}%
\pgfsetrectcap%
\pgfsetmiterjoin%
\pgfsetlinewidth{1.003750pt}%
\definecolor{currentstroke}{rgb}{1.000000,1.000000,1.000000}%
\pgfsetstrokecolor{currentstroke}%
\pgfsetdash{}{0pt}%
\pgfpathmoveto{\pgfqpoint{2.258330in}{0.431659in}}%
\pgfpathlineto{\pgfqpoint{2.258330in}{2.228372in}}%
\pgfusepath{stroke}%
\end{pgfscope}%
\begin{pgfscope}%
\pgfsetrectcap%
\pgfsetmiterjoin%
\pgfsetlinewidth{1.003750pt}%
\definecolor{currentstroke}{rgb}{1.000000,1.000000,1.000000}%
\pgfsetstrokecolor{currentstroke}%
\pgfsetdash{}{0pt}%
\pgfpathmoveto{\pgfqpoint{0.461616in}{0.431659in}}%
\pgfpathlineto{\pgfqpoint{2.258330in}{0.431659in}}%
\pgfusepath{stroke}%
\end{pgfscope}%
\begin{pgfscope}%
\pgfsetrectcap%
\pgfsetmiterjoin%
\pgfsetlinewidth{1.003750pt}%
\definecolor{currentstroke}{rgb}{1.000000,1.000000,1.000000}%
\pgfsetstrokecolor{currentstroke}%
\pgfsetdash{}{0pt}%
\pgfpathmoveto{\pgfqpoint{0.461616in}{2.228372in}}%
\pgfpathlineto{\pgfqpoint{2.258330in}{2.228372in}}%
\pgfusepath{stroke}%
\end{pgfscope}%
\end{pgfpicture}%
\makeatother%
\endgroup%

%% file: figures/grid_avg_delay_validity_starlink_rate.pgf
\begingroup%
\makeatletter%
\begin{pgfpicture}%
\pgfpathrectangle{\pgfpointorigin}{\pgfqpoint{2.300000in}{2.300000in}}%
\pgfusepath{use as bounding box, clip}%
\begin{pgfscope}%
\pgfsetbuttcap%
\pgfsetmiterjoin%
\definecolor{currentfill}{rgb}{1.000000,1.000000,1.000000}%
\pgfsetfillcolor{currentfill}%
\pgfsetlinewidth{0.000000pt}%
\definecolor{currentstroke}{rgb}{0.500000,0.500000,0.500000}%
\pgfsetstrokecolor{currentstroke}%
\pgfsetdash{}{0pt}%
\pgfpathmoveto{\pgfqpoint{0.000000in}{0.000000in}}%
\pgfpathlineto{\pgfqpoint{2.300000in}{0.000000in}}%
\pgfpathlineto{\pgfqpoint{2.300000in}{2.300000in}}%
\pgfpathlineto{\pgfqpoint{0.000000in}{2.300000in}}%
\pgfpathlineto{\pgfqpoint{0.000000in}{0.000000in}}%
\pgfpathclose%
\pgfusepath{fill}%
\end{pgfscope}%
\begin{pgfscope}%
\pgfsetbuttcap%
\pgfsetmiterjoin%
\definecolor{currentfill}{rgb}{0.898039,0.898039,0.898039}%
\pgfsetfillcolor{currentfill}%
\pgfsetlinewidth{0.000000pt}%
\definecolor{currentstroke}{rgb}{0.000000,0.000000,0.000000}%
\pgfsetstrokecolor{currentstroke}%
\pgfsetstrokeopacity{0.000000}%
\pgfsetdash{}{0pt}%
\pgfpathmoveto{\pgfqpoint{0.461616in}{0.431659in}}%
\pgfpathlineto{\pgfqpoint{2.258330in}{0.431659in}}%
\pgfpathlineto{\pgfqpoint{2.258330in}{2.228372in}}%
\pgfpathlineto{\pgfqpoint{0.461616in}{2.228372in}}%
\pgfpathlineto{\pgfqpoint{0.461616in}{0.431659in}}%
\pgfpathclose%
\pgfusepath{fill}%
\end{pgfscope}%
\begin{pgfscope}%
\pgfpathrectangle{\pgfqpoint{0.461616in}{0.431659in}}{\pgfqpoint{1.796714in}{1.796714in}}%
\pgfusepath{clip}%
\pgfsetrectcap%
\pgfsetroundjoin%
\pgfsetlinewidth{0.803000pt}%
\definecolor{currentstroke}{rgb}{1.000000,1.000000,1.000000}%
\pgfsetstrokecolor{currentstroke}%
\pgfsetdash{}{0pt}%
\pgfpathmoveto{\pgfqpoint{0.548180in}{0.431659in}}%
\pgfpathlineto{\pgfqpoint{0.548180in}{2.228372in}}%
\pgfusepath{stroke}%
\end{pgfscope}%
\begin{pgfscope}%
\pgfsetbuttcap%
\pgfsetroundjoin%
\definecolor{currentfill}{rgb}{0.333333,0.333333,0.333333}%
\pgfsetfillcolor{currentfill}%
\pgfsetlinewidth{0.803000pt}%
\definecolor{currentstroke}{rgb}{0.333333,0.333333,0.333333}%
\pgfsetstrokecolor{currentstroke}%
\pgfsetdash{}{0pt}%
\pgfsys@defobject{currentmarker}{\pgfqpoint{0.000000in}{-0.048611in}}{\pgfqpoint{0.000000in}{0.000000in}}{%
\pgfpathmoveto{\pgfqpoint{0.000000in}{0.000000in}}%
\pgfpathlineto{\pgfqpoint{0.000000in}{-0.048611in}}%
\pgfusepath{stroke,fill}%
}%
\begin{pgfscope}%
\pgfsys@transformshift{0.548180in}{0.431659in}%
\pgfsys@useobject{currentmarker}{}%
\end{pgfscope}%
\end{pgfscope}%
\begin{pgfscope}%
\definecolor{textcolor}{rgb}{0.333333,0.333333,0.333333}%
\pgfsetstrokecolor{textcolor}%
\pgfsetfillcolor{textcolor}%
\pgftext[x=0.548180in,y=0.334436in,,top]{\color{textcolor}\rmfamily\fontsize{7.000000}{8.400000}\selectfont \(\displaystyle {0}\)}%
\end{pgfscope}%
\begin{pgfscope}%
\pgfpathrectangle{\pgfqpoint{0.461616in}{0.431659in}}{\pgfqpoint{1.796714in}{1.796714in}}%
\pgfusepath{clip}%
\pgfsetrectcap%
\pgfsetroundjoin%
\pgfsetlinewidth{0.803000pt}%
\definecolor{currentstroke}{rgb}{1.000000,1.000000,1.000000}%
\pgfsetstrokecolor{currentstroke}%
\pgfsetdash{}{0pt}%
\pgfpathmoveto{\pgfqpoint{0.895707in}{0.431659in}}%
\pgfpathlineto{\pgfqpoint{0.895707in}{2.228372in}}%
\pgfusepath{stroke}%
\end{pgfscope}%
\begin{pgfscope}%
\pgfsetbuttcap%
\pgfsetroundjoin%
\definecolor{currentfill}{rgb}{0.333333,0.333333,0.333333}%
\pgfsetfillcolor{currentfill}%
\pgfsetlinewidth{0.803000pt}%
\definecolor{currentstroke}{rgb}{0.333333,0.333333,0.333333}%
\pgfsetstrokecolor{currentstroke}%
\pgfsetdash{}{0pt}%
\pgfsys@defobject{currentmarker}{\pgfqpoint{0.000000in}{-0.048611in}}{\pgfqpoint{0.000000in}{0.000000in}}{%
\pgfpathmoveto{\pgfqpoint{0.000000in}{0.000000in}}%
\pgfpathlineto{\pgfqpoint{0.000000in}{-0.048611in}}%
\pgfusepath{stroke,fill}%
}%
\begin{pgfscope}%
\pgfsys@transformshift{0.895707in}{0.431659in}%
\pgfsys@useobject{currentmarker}{}%
\end{pgfscope}%
\end{pgfscope}%
\begin{pgfscope}%
\definecolor{textcolor}{rgb}{0.333333,0.333333,0.333333}%
\pgfsetstrokecolor{textcolor}%
\pgfsetfillcolor{textcolor}%
\pgftext[x=0.895707in,y=0.334436in,,top]{\color{textcolor}\rmfamily\fontsize{7.000000}{8.400000}\selectfont \(\displaystyle {20}\)}%
\end{pgfscope}%
\begin{pgfscope}%
\pgfpathrectangle{\pgfqpoint{0.461616in}{0.431659in}}{\pgfqpoint{1.796714in}{1.796714in}}%
\pgfusepath{clip}%
\pgfsetrectcap%
\pgfsetroundjoin%
\pgfsetlinewidth{0.803000pt}%
\definecolor{currentstroke}{rgb}{1.000000,1.000000,1.000000}%
\pgfsetstrokecolor{currentstroke}%
\pgfsetdash{}{0pt}%
\pgfpathmoveto{\pgfqpoint{1.243234in}{0.431659in}}%
\pgfpathlineto{\pgfqpoint{1.243234in}{2.228372in}}%
\pgfusepath{stroke}%
\end{pgfscope}%
\begin{pgfscope}%
\pgfsetbuttcap%
\pgfsetroundjoin%
\definecolor{currentfill}{rgb}{0.333333,0.333333,0.333333}%
\pgfsetfillcolor{currentfill}%
\pgfsetlinewidth{0.803000pt}%
\definecolor{currentstroke}{rgb}{0.333333,0.333333,0.333333}%
\pgfsetstrokecolor{currentstroke}%
\pgfsetdash{}{0pt}%
\pgfsys@defobject{currentmarker}{\pgfqpoint{0.000000in}{-0.048611in}}{\pgfqpoint{0.000000in}{0.000000in}}{%
\pgfpathmoveto{\pgfqpoint{0.000000in}{0.000000in}}%
\pgfpathlineto{\pgfqpoint{0.000000in}{-0.048611in}}%
\pgfusepath{stroke,fill}%
}%
\begin{pgfscope}%
\pgfsys@transformshift{1.243234in}{0.431659in}%
\pgfsys@useobject{currentmarker}{}%
\end{pgfscope}%
\end{pgfscope}%
\begin{pgfscope}%
\definecolor{textcolor}{rgb}{0.333333,0.333333,0.333333}%
\pgfsetstrokecolor{textcolor}%
\pgfsetfillcolor{textcolor}%
\pgftext[x=1.243234in,y=0.334436in,,top]{\color{textcolor}\rmfamily\fontsize{7.000000}{8.400000}\selectfont \(\displaystyle {40}\)}%
\end{pgfscope}%
\begin{pgfscope}%
\pgfpathrectangle{\pgfqpoint{0.461616in}{0.431659in}}{\pgfqpoint{1.796714in}{1.796714in}}%
\pgfusepath{clip}%
\pgfsetrectcap%
\pgfsetroundjoin%
\pgfsetlinewidth{0.803000pt}%
\definecolor{currentstroke}{rgb}{1.000000,1.000000,1.000000}%
\pgfsetstrokecolor{currentstroke}%
\pgfsetdash{}{0pt}%
\pgfpathmoveto{\pgfqpoint{1.590761in}{0.431659in}}%
\pgfpathlineto{\pgfqpoint{1.590761in}{2.228372in}}%
\pgfusepath{stroke}%
\end{pgfscope}%
\begin{pgfscope}%
\pgfsetbuttcap%
\pgfsetroundjoin%
\definecolor{currentfill}{rgb}{0.333333,0.333333,0.333333}%
\pgfsetfillcolor{currentfill}%
\pgfsetlinewidth{0.803000pt}%
\definecolor{currentstroke}{rgb}{0.333333,0.333333,0.333333}%
\pgfsetstrokecolor{currentstroke}%
\pgfsetdash{}{0pt}%
\pgfsys@defobject{currentmarker}{\pgfqpoint{0.000000in}{-0.048611in}}{\pgfqpoint{0.000000in}{0.000000in}}{%
\pgfpathmoveto{\pgfqpoint{0.000000in}{0.000000in}}%
\pgfpathlineto{\pgfqpoint{0.000000in}{-0.048611in}}%
\pgfusepath{stroke,fill}%
}%
\begin{pgfscope}%
\pgfsys@transformshift{1.590761in}{0.431659in}%
\pgfsys@useobject{currentmarker}{}%
\end{pgfscope}%
\end{pgfscope}%
\begin{pgfscope}%
\definecolor{textcolor}{rgb}{0.333333,0.333333,0.333333}%
\pgfsetstrokecolor{textcolor}%
\pgfsetfillcolor{textcolor}%
\pgftext[x=1.590761in,y=0.334436in,,top]{\color{textcolor}\rmfamily\fontsize{7.000000}{8.400000}\selectfont \(\displaystyle {60}\)}%
\end{pgfscope}%
\begin{pgfscope}%
\pgfpathrectangle{\pgfqpoint{0.461616in}{0.431659in}}{\pgfqpoint{1.796714in}{1.796714in}}%
\pgfusepath{clip}%
\pgfsetrectcap%
\pgfsetroundjoin%
\pgfsetlinewidth{0.803000pt}%
\definecolor{currentstroke}{rgb}{1.000000,1.000000,1.000000}%
\pgfsetstrokecolor{currentstroke}%
\pgfsetdash{}{0pt}%
\pgfpathmoveto{\pgfqpoint{1.938288in}{0.431659in}}%
\pgfpathlineto{\pgfqpoint{1.938288in}{2.228372in}}%
\pgfusepath{stroke}%
\end{pgfscope}%
\begin{pgfscope}%
\pgfsetbuttcap%
\pgfsetroundjoin%
\definecolor{currentfill}{rgb}{0.333333,0.333333,0.333333}%
\pgfsetfillcolor{currentfill}%
\pgfsetlinewidth{0.803000pt}%
\definecolor{currentstroke}{rgb}{0.333333,0.333333,0.333333}%
\pgfsetstrokecolor{currentstroke}%
\pgfsetdash{}{0pt}%
\pgfsys@defobject{currentmarker}{\pgfqpoint{0.000000in}{-0.048611in}}{\pgfqpoint{0.000000in}{0.000000in}}{%
\pgfpathmoveto{\pgfqpoint{0.000000in}{0.000000in}}%
\pgfpathlineto{\pgfqpoint{0.000000in}{-0.048611in}}%
\pgfusepath{stroke,fill}%
}%
\begin{pgfscope}%
\pgfsys@transformshift{1.938288in}{0.431659in}%
\pgfsys@useobject{currentmarker}{}%
\end{pgfscope}%
\end{pgfscope}%
\begin{pgfscope}%
\definecolor{textcolor}{rgb}{0.333333,0.333333,0.333333}%
\pgfsetstrokecolor{textcolor}%
\pgfsetfillcolor{textcolor}%
\pgftext[x=1.938288in,y=0.334436in,,top]{\color{textcolor}\rmfamily\fontsize{7.000000}{8.400000}\selectfont \(\displaystyle {80}\)}%
\end{pgfscope}%
\begin{pgfscope}%
\definecolor{textcolor}{rgb}{0.333333,0.333333,0.333333}%
\pgfsetstrokecolor{textcolor}%
\pgfsetfillcolor{textcolor}%
\pgftext[x=1.359973in,y=0.192461in,,top]{\color{textcolor}\rmfamily\fontsize{10.000000}{12.000000}\selectfont Avg. Data Rate [MB/s]}%
\end{pgfscope}%
\begin{pgfscope}%
\pgfpathrectangle{\pgfqpoint{0.461616in}{0.431659in}}{\pgfqpoint{1.796714in}{1.796714in}}%
\pgfusepath{clip}%
\pgfsetrectcap%
\pgfsetroundjoin%
\pgfsetlinewidth{0.803000pt}%
\definecolor{currentstroke}{rgb}{1.000000,1.000000,1.000000}%
\pgfsetstrokecolor{currentstroke}%
\pgfsetdash{}{0pt}%
\pgfpathmoveto{\pgfqpoint{0.461616in}{0.513327in}}%
\pgfpathlineto{\pgfqpoint{2.258330in}{0.513327in}}%
\pgfusepath{stroke}%
\end{pgfscope}%
\begin{pgfscope}%
\pgfsetbuttcap%
\pgfsetroundjoin%
\definecolor{currentfill}{rgb}{0.333333,0.333333,0.333333}%
\pgfsetfillcolor{currentfill}%
\pgfsetlinewidth{0.803000pt}%
\definecolor{currentstroke}{rgb}{0.333333,0.333333,0.333333}%
\pgfsetstrokecolor{currentstroke}%
\pgfsetdash{}{0pt}%
\pgfsys@defobject{currentmarker}{\pgfqpoint{-0.048611in}{0.000000in}}{\pgfqpoint{-0.000000in}{0.000000in}}{%
\pgfpathmoveto{\pgfqpoint{-0.000000in}{0.000000in}}%
\pgfpathlineto{\pgfqpoint{-0.048611in}{0.000000in}}%
\pgfusepath{stroke,fill}%
}%
\begin{pgfscope}%
\pgfsys@transformshift{0.461616in}{0.513327in}%
\pgfsys@useobject{currentmarker}{}%
\end{pgfscope}%
\end{pgfscope}%
\begin{pgfscope}%
\definecolor{textcolor}{rgb}{0.333333,0.333333,0.333333}%
\pgfsetstrokecolor{textcolor}%
\pgfsetfillcolor{textcolor}%
\pgftext[x=0.220682in, y=0.479570in, left, base]{\color{textcolor}\rmfamily\fontsize{7.000000}{8.400000}\selectfont \(\displaystyle {0.0}\)}%
\end{pgfscope}%
\begin{pgfscope}%
\pgfpathrectangle{\pgfqpoint{0.461616in}{0.431659in}}{\pgfqpoint{1.796714in}{1.796714in}}%
\pgfusepath{clip}%
\pgfsetrectcap%
\pgfsetroundjoin%
\pgfsetlinewidth{0.803000pt}%
\definecolor{currentstroke}{rgb}{1.000000,1.000000,1.000000}%
\pgfsetstrokecolor{currentstroke}%
\pgfsetdash{}{0pt}%
\pgfpathmoveto{\pgfqpoint{0.461616in}{0.840003in}}%
\pgfpathlineto{\pgfqpoint{2.258330in}{0.840003in}}%
\pgfusepath{stroke}%
\end{pgfscope}%
\begin{pgfscope}%
\pgfsetbuttcap%
\pgfsetroundjoin%
\definecolor{currentfill}{rgb}{0.333333,0.333333,0.333333}%
\pgfsetfillcolor{currentfill}%
\pgfsetlinewidth{0.803000pt}%
\definecolor{currentstroke}{rgb}{0.333333,0.333333,0.333333}%
\pgfsetstrokecolor{currentstroke}%
\pgfsetdash{}{0pt}%
\pgfsys@defobject{currentmarker}{\pgfqpoint{-0.048611in}{0.000000in}}{\pgfqpoint{-0.000000in}{0.000000in}}{%
\pgfpathmoveto{\pgfqpoint{-0.000000in}{0.000000in}}%
\pgfpathlineto{\pgfqpoint{-0.048611in}{0.000000in}}%
\pgfusepath{stroke,fill}%
}%
\begin{pgfscope}%
\pgfsys@transformshift{0.461616in}{0.840003in}%
\pgfsys@useobject{currentmarker}{}%
\end{pgfscope}%
\end{pgfscope}%
\begin{pgfscope}%
\definecolor{textcolor}{rgb}{0.333333,0.333333,0.333333}%
\pgfsetstrokecolor{textcolor}%
\pgfsetfillcolor{textcolor}%
\pgftext[x=0.220682in, y=0.806245in, left, base]{\color{textcolor}\rmfamily\fontsize{7.000000}{8.400000}\selectfont \(\displaystyle {0.2}\)}%
\end{pgfscope}%
\begin{pgfscope}%
\pgfpathrectangle{\pgfqpoint{0.461616in}{0.431659in}}{\pgfqpoint{1.796714in}{1.796714in}}%
\pgfusepath{clip}%
\pgfsetrectcap%
\pgfsetroundjoin%
\pgfsetlinewidth{0.803000pt}%
\definecolor{currentstroke}{rgb}{1.000000,1.000000,1.000000}%
\pgfsetstrokecolor{currentstroke}%
\pgfsetdash{}{0pt}%
\pgfpathmoveto{\pgfqpoint{0.461616in}{1.166678in}}%
\pgfpathlineto{\pgfqpoint{2.258330in}{1.166678in}}%
\pgfusepath{stroke}%
\end{pgfscope}%
\begin{pgfscope}%
\pgfsetbuttcap%
\pgfsetroundjoin%
\definecolor{currentfill}{rgb}{0.333333,0.333333,0.333333}%
\pgfsetfillcolor{currentfill}%
\pgfsetlinewidth{0.803000pt}%
\definecolor{currentstroke}{rgb}{0.333333,0.333333,0.333333}%
\pgfsetstrokecolor{currentstroke}%
\pgfsetdash{}{0pt}%
\pgfsys@defobject{currentmarker}{\pgfqpoint{-0.048611in}{0.000000in}}{\pgfqpoint{-0.000000in}{0.000000in}}{%
\pgfpathmoveto{\pgfqpoint{-0.000000in}{0.000000in}}%
\pgfpathlineto{\pgfqpoint{-0.048611in}{0.000000in}}%
\pgfusepath{stroke,fill}%
}%
\begin{pgfscope}%
\pgfsys@transformshift{0.461616in}{1.166678in}%
\pgfsys@useobject{currentmarker}{}%
\end{pgfscope}%
\end{pgfscope}%
\begin{pgfscope}%
\definecolor{textcolor}{rgb}{0.333333,0.333333,0.333333}%
\pgfsetstrokecolor{textcolor}%
\pgfsetfillcolor{textcolor}%
\pgftext[x=0.220682in, y=1.132920in, left, base]{\color{textcolor}\rmfamily\fontsize{7.000000}{8.400000}\selectfont \(\displaystyle {0.4}\)}%
\end{pgfscope}%
\begin{pgfscope}%
\pgfpathrectangle{\pgfqpoint{0.461616in}{0.431659in}}{\pgfqpoint{1.796714in}{1.796714in}}%
\pgfusepath{clip}%
\pgfsetrectcap%
\pgfsetroundjoin%
\pgfsetlinewidth{0.803000pt}%
\definecolor{currentstroke}{rgb}{1.000000,1.000000,1.000000}%
\pgfsetstrokecolor{currentstroke}%
\pgfsetdash{}{0pt}%
\pgfpathmoveto{\pgfqpoint{0.461616in}{1.493353in}}%
\pgfpathlineto{\pgfqpoint{2.258330in}{1.493353in}}%
\pgfusepath{stroke}%
\end{pgfscope}%
\begin{pgfscope}%
\pgfsetbuttcap%
\pgfsetroundjoin%
\definecolor{currentfill}{rgb}{0.333333,0.333333,0.333333}%
\pgfsetfillcolor{currentfill}%
\pgfsetlinewidth{0.803000pt}%
\definecolor{currentstroke}{rgb}{0.333333,0.333333,0.333333}%
\pgfsetstrokecolor{currentstroke}%
\pgfsetdash{}{0pt}%
\pgfsys@defobject{currentmarker}{\pgfqpoint{-0.048611in}{0.000000in}}{\pgfqpoint{-0.000000in}{0.000000in}}{%
\pgfpathmoveto{\pgfqpoint{-0.000000in}{0.000000in}}%
\pgfpathlineto{\pgfqpoint{-0.048611in}{0.000000in}}%
\pgfusepath{stroke,fill}%
}%
\begin{pgfscope}%
\pgfsys@transformshift{0.461616in}{1.493353in}%
\pgfsys@useobject{currentmarker}{}%
\end{pgfscope}%
\end{pgfscope}%
\begin{pgfscope}%
\definecolor{textcolor}{rgb}{0.333333,0.333333,0.333333}%
\pgfsetstrokecolor{textcolor}%
\pgfsetfillcolor{textcolor}%
\pgftext[x=0.220682in, y=1.459595in, left, base]{\color{textcolor}\rmfamily\fontsize{7.000000}{8.400000}\selectfont \(\displaystyle {0.6}\)}%
\end{pgfscope}%
\begin{pgfscope}%
\pgfpathrectangle{\pgfqpoint{0.461616in}{0.431659in}}{\pgfqpoint{1.796714in}{1.796714in}}%
\pgfusepath{clip}%
\pgfsetrectcap%
\pgfsetroundjoin%
\pgfsetlinewidth{0.803000pt}%
\definecolor{currentstroke}{rgb}{1.000000,1.000000,1.000000}%
\pgfsetstrokecolor{currentstroke}%
\pgfsetdash{}{0pt}%
\pgfpathmoveto{\pgfqpoint{0.461616in}{1.820028in}}%
\pgfpathlineto{\pgfqpoint{2.258330in}{1.820028in}}%
\pgfusepath{stroke}%
\end{pgfscope}%
\begin{pgfscope}%
\pgfsetbuttcap%
\pgfsetroundjoin%
\definecolor{currentfill}{rgb}{0.333333,0.333333,0.333333}%
\pgfsetfillcolor{currentfill}%
\pgfsetlinewidth{0.803000pt}%
\definecolor{currentstroke}{rgb}{0.333333,0.333333,0.333333}%
\pgfsetstrokecolor{currentstroke}%
\pgfsetdash{}{0pt}%
\pgfsys@defobject{currentmarker}{\pgfqpoint{-0.048611in}{0.000000in}}{\pgfqpoint{-0.000000in}{0.000000in}}{%
\pgfpathmoveto{\pgfqpoint{-0.000000in}{0.000000in}}%
\pgfpathlineto{\pgfqpoint{-0.048611in}{0.000000in}}%
\pgfusepath{stroke,fill}%
}%
\begin{pgfscope}%
\pgfsys@transformshift{0.461616in}{1.820028in}%
\pgfsys@useobject{currentmarker}{}%
\end{pgfscope}%
\end{pgfscope}%
\begin{pgfscope}%
\definecolor{textcolor}{rgb}{0.333333,0.333333,0.333333}%
\pgfsetstrokecolor{textcolor}%
\pgfsetfillcolor{textcolor}%
\pgftext[x=0.220682in, y=1.786271in, left, base]{\color{textcolor}\rmfamily\fontsize{7.000000}{8.400000}\selectfont \(\displaystyle {0.8}\)}%
\end{pgfscope}%
\begin{pgfscope}%
\pgfpathrectangle{\pgfqpoint{0.461616in}{0.431659in}}{\pgfqpoint{1.796714in}{1.796714in}}%
\pgfusepath{clip}%
\pgfsetrectcap%
\pgfsetroundjoin%
\pgfsetlinewidth{0.803000pt}%
\definecolor{currentstroke}{rgb}{1.000000,1.000000,1.000000}%
\pgfsetstrokecolor{currentstroke}%
\pgfsetdash{}{0pt}%
\pgfpathmoveto{\pgfqpoint{0.461616in}{2.146703in}}%
\pgfpathlineto{\pgfqpoint{2.258330in}{2.146703in}}%
\pgfusepath{stroke}%
\end{pgfscope}%
\begin{pgfscope}%
\pgfsetbuttcap%
\pgfsetroundjoin%
\definecolor{currentfill}{rgb}{0.333333,0.333333,0.333333}%
\pgfsetfillcolor{currentfill}%
\pgfsetlinewidth{0.803000pt}%
\definecolor{currentstroke}{rgb}{0.333333,0.333333,0.333333}%
\pgfsetstrokecolor{currentstroke}%
\pgfsetdash{}{0pt}%
\pgfsys@defobject{currentmarker}{\pgfqpoint{-0.048611in}{0.000000in}}{\pgfqpoint{-0.000000in}{0.000000in}}{%
\pgfpathmoveto{\pgfqpoint{-0.000000in}{0.000000in}}%
\pgfpathlineto{\pgfqpoint{-0.048611in}{0.000000in}}%
\pgfusepath{stroke,fill}%
}%
\begin{pgfscope}%
\pgfsys@transformshift{0.461616in}{2.146703in}%
\pgfsys@useobject{currentmarker}{}%
\end{pgfscope}%
\end{pgfscope}%
\begin{pgfscope}%
\definecolor{textcolor}{rgb}{0.333333,0.333333,0.333333}%
\pgfsetstrokecolor{textcolor}%
\pgfsetfillcolor{textcolor}%
\pgftext[x=0.220682in, y=2.112946in, left, base]{\color{textcolor}\rmfamily\fontsize{7.000000}{8.400000}\selectfont \(\displaystyle {1.0}\)}%
\end{pgfscope}%
\begin{pgfscope}%
\definecolor{textcolor}{rgb}{0.333333,0.333333,0.333333}%
\pgfsetstrokecolor{textcolor}%
\pgfsetfillcolor{textcolor}%
\pgftext[x=0.165127in,y=1.330015in,,bottom,rotate=90.000000]{\color{textcolor}\rmfamily\fontsize{10.000000}{12.000000}\selectfont Cumulative Probability}%
\end{pgfscope}%
\begin{pgfscope}%
\pgfpathrectangle{\pgfqpoint{0.461616in}{0.431659in}}{\pgfqpoint{1.796714in}{1.796714in}}%
\pgfusepath{clip}%
\pgfsetrectcap%
\pgfsetroundjoin%
\pgfsetlinewidth{1.003750pt}%
\definecolor{currentstroke}{rgb}{1.000000,0.498039,0.054902}%
\pgfsetstrokecolor{currentstroke}%
\pgfsetdash{}{0pt}%
\pgfpathmoveto{\pgfqpoint{0.543285in}{0.513327in}}%
\pgfpathlineto{\pgfqpoint{0.560662in}{0.513327in}}%
\pgfpathlineto{\pgfqpoint{0.578038in}{0.522023in}}%
\pgfpathlineto{\pgfqpoint{0.595414in}{0.590247in}}%
\pgfpathlineto{\pgfqpoint{0.612791in}{0.695928in}}%
\pgfpathlineto{\pgfqpoint{0.630167in}{0.839066in}}%
\pgfpathlineto{\pgfqpoint{0.647543in}{0.978860in}}%
\pgfpathlineto{\pgfqpoint{0.664920in}{1.083872in}}%
\pgfpathlineto{\pgfqpoint{0.682296in}{1.210957in}}%
\pgfpathlineto{\pgfqpoint{0.699672in}{1.303261in}}%
\pgfpathlineto{\pgfqpoint{0.717049in}{1.388207in}}%
\pgfpathlineto{\pgfqpoint{0.734425in}{1.455094in}}%
\pgfpathlineto{\pgfqpoint{0.751801in}{1.519305in}}%
\pgfpathlineto{\pgfqpoint{0.769178in}{1.565457in}}%
\pgfpathlineto{\pgfqpoint{0.786554in}{1.605589in}}%
\pgfpathlineto{\pgfqpoint{0.803930in}{1.644384in}}%
\pgfpathlineto{\pgfqpoint{0.821307in}{1.680502in}}%
\pgfpathlineto{\pgfqpoint{0.838683in}{1.706588in}}%
\pgfpathlineto{\pgfqpoint{0.856059in}{1.727323in}}%
\pgfpathlineto{\pgfqpoint{0.873436in}{1.742707in}}%
\pgfpathlineto{\pgfqpoint{0.890812in}{1.768793in}}%
\pgfpathlineto{\pgfqpoint{0.908188in}{1.788190in}}%
\pgfpathlineto{\pgfqpoint{0.925565in}{1.808925in}}%
\pgfpathlineto{\pgfqpoint{0.942941in}{1.821634in}}%
\pgfpathlineto{\pgfqpoint{0.960317in}{1.840362in}}%
\pgfpathlineto{\pgfqpoint{0.977694in}{1.856415in}}%
\pgfpathlineto{\pgfqpoint{0.995070in}{1.863103in}}%
\pgfpathlineto{\pgfqpoint{1.012446in}{1.872468in}}%
\pgfpathlineto{\pgfqpoint{1.029823in}{1.878487in}}%
\pgfpathlineto{\pgfqpoint{1.047199in}{1.890527in}}%
\pgfpathlineto{\pgfqpoint{1.064575in}{1.899891in}}%
\pgfpathlineto{\pgfqpoint{1.081952in}{1.914606in}}%
\pgfpathlineto{\pgfqpoint{1.099328in}{1.923970in}}%
\pgfpathlineto{\pgfqpoint{1.116704in}{1.927984in}}%
\pgfpathlineto{\pgfqpoint{1.134081in}{1.936010in}}%
\pgfpathlineto{\pgfqpoint{1.151457in}{1.942699in}}%
\pgfpathlineto{\pgfqpoint{1.168833in}{1.948050in}}%
\pgfpathlineto{\pgfqpoint{1.186210in}{1.954738in}}%
\pgfpathlineto{\pgfqpoint{1.203586in}{1.960758in}}%
\pgfpathlineto{\pgfqpoint{1.220962in}{1.965440in}}%
\pgfpathlineto{\pgfqpoint{1.238339in}{1.970122in}}%
\pgfpathlineto{\pgfqpoint{1.255715in}{1.972129in}}%
\pgfpathlineto{\pgfqpoint{1.273091in}{1.975473in}}%
\pgfpathlineto{\pgfqpoint{1.290468in}{1.978149in}}%
\pgfpathlineto{\pgfqpoint{1.307844in}{1.981493in}}%
\pgfpathlineto{\pgfqpoint{1.325221in}{1.985506in}}%
\pgfpathlineto{\pgfqpoint{1.342597in}{1.989519in}}%
\pgfpathlineto{\pgfqpoint{1.359973in}{1.992195in}}%
\pgfpathlineto{\pgfqpoint{1.377350in}{1.995539in}}%
\pgfpathlineto{\pgfqpoint{1.394726in}{1.998215in}}%
\pgfpathlineto{\pgfqpoint{1.412102in}{2.002897in}}%
\pgfpathlineto{\pgfqpoint{1.429479in}{2.007579in}}%
\pgfpathlineto{\pgfqpoint{1.446855in}{2.010923in}}%
\pgfpathlineto{\pgfqpoint{1.464231in}{2.015605in}}%
\pgfpathlineto{\pgfqpoint{1.481608in}{2.015605in}}%
\pgfpathlineto{\pgfqpoint{1.498984in}{2.015605in}}%
\pgfpathlineto{\pgfqpoint{1.516360in}{2.018950in}}%
\pgfpathlineto{\pgfqpoint{1.533737in}{2.020287in}}%
\pgfpathlineto{\pgfqpoint{1.551113in}{2.023632in}}%
\pgfpathlineto{\pgfqpoint{1.568489in}{2.027645in}}%
\pgfpathlineto{\pgfqpoint{1.585866in}{2.032996in}}%
\pgfpathlineto{\pgfqpoint{1.603242in}{2.036340in}}%
\pgfpathlineto{\pgfqpoint{1.620618in}{2.037009in}}%
\pgfpathlineto{\pgfqpoint{1.637995in}{2.041022in}}%
\pgfpathlineto{\pgfqpoint{1.655371in}{2.045036in}}%
\pgfpathlineto{\pgfqpoint{1.672747in}{2.048380in}}%
\pgfpathlineto{\pgfqpoint{1.690124in}{2.053731in}}%
\pgfpathlineto{\pgfqpoint{1.707500in}{2.057075in}}%
\pgfpathlineto{\pgfqpoint{1.724876in}{2.059751in}}%
\pgfpathlineto{\pgfqpoint{1.742253in}{2.063764in}}%
\pgfpathlineto{\pgfqpoint{1.759629in}{2.065102in}}%
\pgfpathlineto{\pgfqpoint{1.777005in}{2.068446in}}%
\pgfpathlineto{\pgfqpoint{1.794382in}{2.068446in}}%
\pgfpathlineto{\pgfqpoint{1.811758in}{2.073128in}}%
\pgfpathlineto{\pgfqpoint{1.829134in}{2.077810in}}%
\pgfpathlineto{\pgfqpoint{1.846511in}{2.083161in}}%
\pgfpathlineto{\pgfqpoint{1.863887in}{2.086505in}}%
\pgfpathlineto{\pgfqpoint{1.881263in}{2.088512in}}%
\pgfpathlineto{\pgfqpoint{1.898640in}{2.091856in}}%
\pgfpathlineto{\pgfqpoint{1.916016in}{2.093863in}}%
\pgfpathlineto{\pgfqpoint{1.933392in}{2.097876in}}%
\pgfpathlineto{\pgfqpoint{1.950769in}{2.101220in}}%
\pgfpathlineto{\pgfqpoint{1.968145in}{2.107909in}}%
\pgfpathlineto{\pgfqpoint{1.985521in}{2.115267in}}%
\pgfpathlineto{\pgfqpoint{2.002898in}{2.116604in}}%
\pgfpathlineto{\pgfqpoint{2.020274in}{2.118611in}}%
\pgfpathlineto{\pgfqpoint{2.037650in}{2.118611in}}%
\pgfpathlineto{\pgfqpoint{2.055027in}{2.119949in}}%
\pgfpathlineto{\pgfqpoint{2.072403in}{2.125969in}}%
\pgfpathlineto{\pgfqpoint{2.089779in}{2.131988in}}%
\pgfpathlineto{\pgfqpoint{2.107156in}{2.135333in}}%
\pgfpathlineto{\pgfqpoint{2.124532in}{2.136002in}}%
\pgfpathlineto{\pgfqpoint{2.141909in}{2.138008in}}%
\pgfpathlineto{\pgfqpoint{2.159285in}{2.140015in}}%
\pgfpathlineto{\pgfqpoint{2.176661in}{2.146703in}}%
\pgfusepath{stroke}%
\end{pgfscope}%
\begin{pgfscope}%
\pgfpathrectangle{\pgfqpoint{0.461616in}{0.431659in}}{\pgfqpoint{1.796714in}{1.796714in}}%
\pgfusepath{clip}%
\pgfsetbuttcap%
\pgfsetroundjoin%
\pgfsetlinewidth{1.003750pt}%
\definecolor{currentstroke}{rgb}{0.121569,0.466667,0.705882}%
\pgfsetstrokecolor{currentstroke}%
\pgfsetdash{{3.700000pt}{1.600000pt}}{0.000000pt}%
\pgfpathmoveto{\pgfqpoint{0.543285in}{0.513327in}}%
\pgfpathlineto{\pgfqpoint{0.560662in}{0.513327in}}%
\pgfpathlineto{\pgfqpoint{0.578038in}{0.513327in}}%
\pgfpathlineto{\pgfqpoint{0.595414in}{0.520685in}}%
\pgfpathlineto{\pgfqpoint{0.612791in}{0.546102in}}%
\pgfpathlineto{\pgfqpoint{0.630167in}{0.619678in}}%
\pgfpathlineto{\pgfqpoint{0.647543in}{0.722014in}}%
\pgfpathlineto{\pgfqpoint{0.664920in}{0.853112in}}%
\pgfpathlineto{\pgfqpoint{0.682296in}{0.967489in}}%
\pgfpathlineto{\pgfqpoint{0.699672in}{1.068488in}}%
\pgfpathlineto{\pgfqpoint{0.717049in}{1.176845in}}%
\pgfpathlineto{\pgfqpoint{0.734425in}{1.263798in}}%
\pgfpathlineto{\pgfqpoint{0.751801in}{1.356101in}}%
\pgfpathlineto{\pgfqpoint{0.769178in}{1.418975in}}%
\pgfpathlineto{\pgfqpoint{0.786554in}{1.489206in}}%
\pgfpathlineto{\pgfqpoint{0.803930in}{1.550742in}}%
\pgfpathlineto{\pgfqpoint{0.821307in}{1.588199in}}%
\pgfpathlineto{\pgfqpoint{0.838683in}{1.629000in}}%
\pgfpathlineto{\pgfqpoint{0.856059in}{1.669132in}}%
\pgfpathlineto{\pgfqpoint{0.873436in}{1.699900in}}%
\pgfpathlineto{\pgfqpoint{0.890812in}{1.740032in}}%
\pgfpathlineto{\pgfqpoint{0.908188in}{1.766786in}}%
\pgfpathlineto{\pgfqpoint{0.925565in}{1.791535in}}%
\pgfpathlineto{\pgfqpoint{0.942941in}{1.814276in}}%
\pgfpathlineto{\pgfqpoint{0.960317in}{1.833004in}}%
\pgfpathlineto{\pgfqpoint{0.977694in}{1.849726in}}%
\pgfpathlineto{\pgfqpoint{0.995070in}{1.863103in}}%
\pgfpathlineto{\pgfqpoint{1.012446in}{1.869792in}}%
\pgfpathlineto{\pgfqpoint{1.029823in}{1.879156in}}%
\pgfpathlineto{\pgfqpoint{1.047199in}{1.888520in}}%
\pgfpathlineto{\pgfqpoint{1.064575in}{1.900560in}}%
\pgfpathlineto{\pgfqpoint{1.081952in}{1.907918in}}%
\pgfpathlineto{\pgfqpoint{1.099328in}{1.915275in}}%
\pgfpathlineto{\pgfqpoint{1.116704in}{1.920626in}}%
\pgfpathlineto{\pgfqpoint{1.134081in}{1.927315in}}%
\pgfpathlineto{\pgfqpoint{1.151457in}{1.934672in}}%
\pgfpathlineto{\pgfqpoint{1.168833in}{1.942699in}}%
\pgfpathlineto{\pgfqpoint{1.186210in}{1.949387in}}%
\pgfpathlineto{\pgfqpoint{1.203586in}{1.953401in}}%
\pgfpathlineto{\pgfqpoint{1.220962in}{1.959420in}}%
\pgfpathlineto{\pgfqpoint{1.238339in}{1.964102in}}%
\pgfpathlineto{\pgfqpoint{1.255715in}{1.968116in}}%
\pgfpathlineto{\pgfqpoint{1.273091in}{1.969453in}}%
\pgfpathlineto{\pgfqpoint{1.290468in}{1.974804in}}%
\pgfpathlineto{\pgfqpoint{1.307844in}{1.983500in}}%
\pgfpathlineto{\pgfqpoint{1.325221in}{1.989519in}}%
\pgfpathlineto{\pgfqpoint{1.342597in}{1.992864in}}%
\pgfpathlineto{\pgfqpoint{1.359973in}{1.994202in}}%
\pgfpathlineto{\pgfqpoint{1.377350in}{1.999552in}}%
\pgfpathlineto{\pgfqpoint{1.394726in}{2.001559in}}%
\pgfpathlineto{\pgfqpoint{1.412102in}{2.006241in}}%
\pgfpathlineto{\pgfqpoint{1.429479in}{2.006910in}}%
\pgfpathlineto{\pgfqpoint{1.446855in}{2.012261in}}%
\pgfpathlineto{\pgfqpoint{1.464231in}{2.012930in}}%
\pgfpathlineto{\pgfqpoint{1.481608in}{2.014268in}}%
\pgfpathlineto{\pgfqpoint{1.498984in}{2.015605in}}%
\pgfpathlineto{\pgfqpoint{1.516360in}{2.016274in}}%
\pgfpathlineto{\pgfqpoint{1.533737in}{2.019619in}}%
\pgfpathlineto{\pgfqpoint{1.551113in}{2.024301in}}%
\pgfpathlineto{\pgfqpoint{1.568489in}{2.027645in}}%
\pgfpathlineto{\pgfqpoint{1.585866in}{2.031658in}}%
\pgfpathlineto{\pgfqpoint{1.603242in}{2.033665in}}%
\pgfpathlineto{\pgfqpoint{1.620618in}{2.037678in}}%
\pgfpathlineto{\pgfqpoint{1.637995in}{2.046373in}}%
\pgfpathlineto{\pgfqpoint{1.655371in}{2.050386in}}%
\pgfpathlineto{\pgfqpoint{1.672747in}{2.053062in}}%
\pgfpathlineto{\pgfqpoint{1.690124in}{2.058413in}}%
\pgfpathlineto{\pgfqpoint{1.707500in}{2.061757in}}%
\pgfpathlineto{\pgfqpoint{1.724876in}{2.064433in}}%
\pgfpathlineto{\pgfqpoint{1.742253in}{2.072459in}}%
\pgfpathlineto{\pgfqpoint{1.759629in}{2.077810in}}%
\pgfpathlineto{\pgfqpoint{1.777005in}{2.078479in}}%
\pgfpathlineto{\pgfqpoint{1.794382in}{2.081154in}}%
\pgfpathlineto{\pgfqpoint{1.811758in}{2.083830in}}%
\pgfpathlineto{\pgfqpoint{1.829134in}{2.086505in}}%
\pgfpathlineto{\pgfqpoint{1.846511in}{2.088512in}}%
\pgfpathlineto{\pgfqpoint{1.863887in}{2.093863in}}%
\pgfpathlineto{\pgfqpoint{1.881263in}{2.101220in}}%
\pgfpathlineto{\pgfqpoint{1.898640in}{2.105234in}}%
\pgfpathlineto{\pgfqpoint{1.916016in}{2.107909in}}%
\pgfpathlineto{\pgfqpoint{1.933392in}{2.109916in}}%
\pgfpathlineto{\pgfqpoint{1.950769in}{2.110585in}}%
\pgfpathlineto{\pgfqpoint{1.968145in}{2.111922in}}%
\pgfpathlineto{\pgfqpoint{1.985521in}{2.113929in}}%
\pgfpathlineto{\pgfqpoint{2.002898in}{2.117273in}}%
\pgfpathlineto{\pgfqpoint{2.020274in}{2.126637in}}%
\pgfpathlineto{\pgfqpoint{2.037650in}{2.126637in}}%
\pgfpathlineto{\pgfqpoint{2.055027in}{2.127975in}}%
\pgfpathlineto{\pgfqpoint{2.072403in}{2.129313in}}%
\pgfpathlineto{\pgfqpoint{2.089779in}{2.133326in}}%
\pgfpathlineto{\pgfqpoint{2.107156in}{2.136670in}}%
\pgfpathlineto{\pgfqpoint{2.124532in}{2.138008in}}%
\pgfpathlineto{\pgfqpoint{2.141909in}{2.141353in}}%
\pgfpathlineto{\pgfqpoint{2.159285in}{2.144697in}}%
\pgfpathlineto{\pgfqpoint{2.176661in}{2.146703in}}%
\pgfusepath{stroke}%
\end{pgfscope}%
\begin{pgfscope}%
\pgfpathrectangle{\pgfqpoint{0.461616in}{0.431659in}}{\pgfqpoint{1.796714in}{1.796714in}}%
\pgfusepath{clip}%
\pgfsetbuttcap%
\pgfsetroundjoin%
\pgfsetlinewidth{1.003750pt}%
\definecolor{currentstroke}{rgb}{0.839216,0.152941,0.156863}%
\pgfsetstrokecolor{currentstroke}%
\pgfsetdash{{6.400000pt}{1.600000pt}{1.000000pt}{1.600000pt}}{0.000000pt}%
\pgfpathmoveto{\pgfqpoint{0.543285in}{0.513327in}}%
\pgfpathlineto{\pgfqpoint{0.560662in}{0.513327in}}%
\pgfpathlineto{\pgfqpoint{0.578038in}{0.513327in}}%
\pgfpathlineto{\pgfqpoint{0.595414in}{0.513327in}}%
\pgfpathlineto{\pgfqpoint{0.612791in}{0.513327in}}%
\pgfpathlineto{\pgfqpoint{0.630167in}{0.530718in}}%
\pgfpathlineto{\pgfqpoint{0.647543in}{0.604962in}}%
\pgfpathlineto{\pgfqpoint{0.664920in}{0.702617in}}%
\pgfpathlineto{\pgfqpoint{0.682296in}{0.816994in}}%
\pgfpathlineto{\pgfqpoint{0.699672in}{0.927357in}}%
\pgfpathlineto{\pgfqpoint{0.717049in}{1.026349in}}%
\pgfpathlineto{\pgfqpoint{0.734425in}{1.131362in}}%
\pgfpathlineto{\pgfqpoint{0.751801in}{1.219652in}}%
\pgfpathlineto{\pgfqpoint{0.769178in}{1.303261in}}%
\pgfpathlineto{\pgfqpoint{0.786554in}{1.390882in}}%
\pgfpathlineto{\pgfqpoint{0.803930in}{1.467133in}}%
\pgfpathlineto{\pgfqpoint{0.821307in}{1.513285in}}%
\pgfpathlineto{\pgfqpoint{0.838683in}{1.562782in}}%
\pgfpathlineto{\pgfqpoint{0.856059in}{1.602914in}}%
\pgfpathlineto{\pgfqpoint{0.873436in}{1.640370in}}%
\pgfpathlineto{\pgfqpoint{0.890812in}{1.665787in}}%
\pgfpathlineto{\pgfqpoint{0.908188in}{1.697893in}}%
\pgfpathlineto{\pgfqpoint{0.925565in}{1.729999in}}%
\pgfpathlineto{\pgfqpoint{0.942941in}{1.754747in}}%
\pgfpathlineto{\pgfqpoint{0.960317in}{1.783508in}}%
\pgfpathlineto{\pgfqpoint{0.977694in}{1.803574in}}%
\pgfpathlineto{\pgfqpoint{0.995070in}{1.823640in}}%
\pgfpathlineto{\pgfqpoint{1.012446in}{1.838355in}}%
\pgfpathlineto{\pgfqpoint{1.029823in}{1.854408in}}%
\pgfpathlineto{\pgfqpoint{1.047199in}{1.863772in}}%
\pgfpathlineto{\pgfqpoint{1.064575in}{1.879156in}}%
\pgfpathlineto{\pgfqpoint{1.081952in}{1.891196in}}%
\pgfpathlineto{\pgfqpoint{1.099328in}{1.898553in}}%
\pgfpathlineto{\pgfqpoint{1.116704in}{1.907249in}}%
\pgfpathlineto{\pgfqpoint{1.134081in}{1.919957in}}%
\pgfpathlineto{\pgfqpoint{1.151457in}{1.927315in}}%
\pgfpathlineto{\pgfqpoint{1.168833in}{1.936010in}}%
\pgfpathlineto{\pgfqpoint{1.186210in}{1.946712in}}%
\pgfpathlineto{\pgfqpoint{1.203586in}{1.954738in}}%
\pgfpathlineto{\pgfqpoint{1.220962in}{1.963434in}}%
\pgfpathlineto{\pgfqpoint{1.238339in}{1.971460in}}%
\pgfpathlineto{\pgfqpoint{1.255715in}{1.976811in}}%
\pgfpathlineto{\pgfqpoint{1.273091in}{1.982162in}}%
\pgfpathlineto{\pgfqpoint{1.290468in}{1.985506in}}%
\pgfpathlineto{\pgfqpoint{1.307844in}{1.989519in}}%
\pgfpathlineto{\pgfqpoint{1.325221in}{1.993533in}}%
\pgfpathlineto{\pgfqpoint{1.342597in}{1.996877in}}%
\pgfpathlineto{\pgfqpoint{1.359973in}{2.003566in}}%
\pgfpathlineto{\pgfqpoint{1.377350in}{2.006910in}}%
\pgfpathlineto{\pgfqpoint{1.394726in}{2.012261in}}%
\pgfpathlineto{\pgfqpoint{1.412102in}{2.015605in}}%
\pgfpathlineto{\pgfqpoint{1.429479in}{2.017612in}}%
\pgfpathlineto{\pgfqpoint{1.446855in}{2.020956in}}%
\pgfpathlineto{\pgfqpoint{1.464231in}{2.023632in}}%
\pgfpathlineto{\pgfqpoint{1.481608in}{2.025638in}}%
\pgfpathlineto{\pgfqpoint{1.498984in}{2.030320in}}%
\pgfpathlineto{\pgfqpoint{1.516360in}{2.032327in}}%
\pgfpathlineto{\pgfqpoint{1.533737in}{2.036340in}}%
\pgfpathlineto{\pgfqpoint{1.551113in}{2.041691in}}%
\pgfpathlineto{\pgfqpoint{1.568489in}{2.045704in}}%
\pgfpathlineto{\pgfqpoint{1.585866in}{2.053062in}}%
\pgfpathlineto{\pgfqpoint{1.603242in}{2.059082in}}%
\pgfpathlineto{\pgfqpoint{1.620618in}{2.065770in}}%
\pgfpathlineto{\pgfqpoint{1.637995in}{2.070453in}}%
\pgfpathlineto{\pgfqpoint{1.655371in}{2.073128in}}%
\pgfpathlineto{\pgfqpoint{1.672747in}{2.077810in}}%
\pgfpathlineto{\pgfqpoint{1.690124in}{2.080486in}}%
\pgfpathlineto{\pgfqpoint{1.707500in}{2.081823in}}%
\pgfpathlineto{\pgfqpoint{1.724876in}{2.087174in}}%
\pgfpathlineto{\pgfqpoint{1.742253in}{2.091187in}}%
\pgfpathlineto{\pgfqpoint{1.759629in}{2.095201in}}%
\pgfpathlineto{\pgfqpoint{1.777005in}{2.097207in}}%
\pgfpathlineto{\pgfqpoint{1.794382in}{2.105234in}}%
\pgfpathlineto{\pgfqpoint{1.811758in}{2.107240in}}%
\pgfpathlineto{\pgfqpoint{1.829134in}{2.109247in}}%
\pgfpathlineto{\pgfqpoint{1.846511in}{2.109916in}}%
\pgfpathlineto{\pgfqpoint{1.863887in}{2.112591in}}%
\pgfpathlineto{\pgfqpoint{1.881263in}{2.113260in}}%
\pgfpathlineto{\pgfqpoint{1.898640in}{2.115267in}}%
\pgfpathlineto{\pgfqpoint{1.916016in}{2.118611in}}%
\pgfpathlineto{\pgfqpoint{1.933392in}{2.119949in}}%
\pgfpathlineto{\pgfqpoint{1.950769in}{2.121955in}}%
\pgfpathlineto{\pgfqpoint{1.968145in}{2.123962in}}%
\pgfpathlineto{\pgfqpoint{1.985521in}{2.126637in}}%
\pgfpathlineto{\pgfqpoint{2.002898in}{2.127306in}}%
\pgfpathlineto{\pgfqpoint{2.020274in}{2.131320in}}%
\pgfpathlineto{\pgfqpoint{2.037650in}{2.132657in}}%
\pgfpathlineto{\pgfqpoint{2.055027in}{2.135333in}}%
\pgfpathlineto{\pgfqpoint{2.072403in}{2.138008in}}%
\pgfpathlineto{\pgfqpoint{2.089779in}{2.138677in}}%
\pgfpathlineto{\pgfqpoint{2.107156in}{2.138677in}}%
\pgfpathlineto{\pgfqpoint{2.124532in}{2.140015in}}%
\pgfpathlineto{\pgfqpoint{2.141909in}{2.142021in}}%
\pgfpathlineto{\pgfqpoint{2.159285in}{2.146703in}}%
\pgfpathlineto{\pgfqpoint{2.176661in}{2.146703in}}%
\pgfusepath{stroke}%
\end{pgfscope}%
\begin{pgfscope}%
\pgfpathrectangle{\pgfqpoint{0.461616in}{0.431659in}}{\pgfqpoint{1.796714in}{1.796714in}}%
\pgfusepath{clip}%
\pgfsetbuttcap%
\pgfsetroundjoin%
\pgfsetlinewidth{1.003750pt}%
\definecolor{currentstroke}{rgb}{0.172549,0.627451,0.172549}%
\pgfsetstrokecolor{currentstroke}%
\pgfsetdash{{1.000000pt}{1.650000pt}}{0.000000pt}%
\pgfpathmoveto{\pgfqpoint{0.543285in}{0.513327in}}%
\pgfpathlineto{\pgfqpoint{0.560662in}{0.513327in}}%
\pgfpathlineto{\pgfqpoint{0.578038in}{0.513327in}}%
\pgfpathlineto{\pgfqpoint{0.595414in}{0.513327in}}%
\pgfpathlineto{\pgfqpoint{0.612791in}{0.514665in}}%
\pgfpathlineto{\pgfqpoint{0.630167in}{0.516672in}}%
\pgfpathlineto{\pgfqpoint{0.647543in}{0.534062in}}%
\pgfpathlineto{\pgfqpoint{0.664920in}{0.595598in}}%
\pgfpathlineto{\pgfqpoint{0.682296in}{0.736061in}}%
\pgfpathlineto{\pgfqpoint{0.699672in}{0.906622in}}%
\pgfpathlineto{\pgfqpoint{0.717049in}{1.077852in}}%
\pgfpathlineto{\pgfqpoint{0.734425in}{1.225003in}}%
\pgfpathlineto{\pgfqpoint{0.751801in}{1.340049in}}%
\pgfpathlineto{\pgfqpoint{0.769178in}{1.422988in}}%
\pgfpathlineto{\pgfqpoint{0.786554in}{1.500577in}}%
\pgfpathlineto{\pgfqpoint{0.803930in}{1.576159in}}%
\pgfpathlineto{\pgfqpoint{0.821307in}{1.637026in}}%
\pgfpathlineto{\pgfqpoint{0.838683in}{1.702575in}}%
\pgfpathlineto{\pgfqpoint{0.856059in}{1.749396in}}%
\pgfpathlineto{\pgfqpoint{0.873436in}{1.794879in}}%
\pgfpathlineto{\pgfqpoint{0.890812in}{1.832335in}}%
\pgfpathlineto{\pgfqpoint{0.908188in}{1.867785in}}%
\pgfpathlineto{\pgfqpoint{0.925565in}{1.895209in}}%
\pgfpathlineto{\pgfqpoint{0.942941in}{1.919288in}}%
\pgfpathlineto{\pgfqpoint{0.960317in}{1.934003in}}%
\pgfpathlineto{\pgfqpoint{0.977694in}{1.956076in}}%
\pgfpathlineto{\pgfqpoint{0.995070in}{1.972129in}}%
\pgfpathlineto{\pgfqpoint{1.012446in}{1.982831in}}%
\pgfpathlineto{\pgfqpoint{1.029823in}{1.994870in}}%
\pgfpathlineto{\pgfqpoint{1.047199in}{2.002897in}}%
\pgfpathlineto{\pgfqpoint{1.064575in}{2.012261in}}%
\pgfpathlineto{\pgfqpoint{1.081952in}{2.020287in}}%
\pgfpathlineto{\pgfqpoint{1.099328in}{2.026976in}}%
\pgfpathlineto{\pgfqpoint{1.116704in}{2.035003in}}%
\pgfpathlineto{\pgfqpoint{1.134081in}{2.040353in}}%
\pgfpathlineto{\pgfqpoint{1.151457in}{2.047042in}}%
\pgfpathlineto{\pgfqpoint{1.168833in}{2.055069in}}%
\pgfpathlineto{\pgfqpoint{1.186210in}{2.061757in}}%
\pgfpathlineto{\pgfqpoint{1.203586in}{2.069115in}}%
\pgfpathlineto{\pgfqpoint{1.220962in}{2.075803in}}%
\pgfpathlineto{\pgfqpoint{1.238339in}{2.079148in}}%
\pgfpathlineto{\pgfqpoint{1.255715in}{2.081154in}}%
\pgfpathlineto{\pgfqpoint{1.273091in}{2.087843in}}%
\pgfpathlineto{\pgfqpoint{1.290468in}{2.092525in}}%
\pgfpathlineto{\pgfqpoint{1.307844in}{2.096538in}}%
\pgfpathlineto{\pgfqpoint{1.325221in}{2.097876in}}%
\pgfpathlineto{\pgfqpoint{1.342597in}{2.102558in}}%
\pgfpathlineto{\pgfqpoint{1.359973in}{2.107240in}}%
\pgfpathlineto{\pgfqpoint{1.377350in}{2.109916in}}%
\pgfpathlineto{\pgfqpoint{1.394726in}{2.113929in}}%
\pgfpathlineto{\pgfqpoint{1.412102in}{2.117942in}}%
\pgfpathlineto{\pgfqpoint{1.429479in}{2.118611in}}%
\pgfpathlineto{\pgfqpoint{1.446855in}{2.119949in}}%
\pgfpathlineto{\pgfqpoint{1.464231in}{2.121287in}}%
\pgfpathlineto{\pgfqpoint{1.481608in}{2.122624in}}%
\pgfpathlineto{\pgfqpoint{1.498984in}{2.123293in}}%
\pgfpathlineto{\pgfqpoint{1.516360in}{2.124631in}}%
\pgfpathlineto{\pgfqpoint{1.533737in}{2.125969in}}%
\pgfpathlineto{\pgfqpoint{1.551113in}{2.128644in}}%
\pgfpathlineto{\pgfqpoint{1.568489in}{2.129313in}}%
\pgfpathlineto{\pgfqpoint{1.585866in}{2.130651in}}%
\pgfpathlineto{\pgfqpoint{1.603242in}{2.131320in}}%
\pgfpathlineto{\pgfqpoint{1.620618in}{2.131988in}}%
\pgfpathlineto{\pgfqpoint{1.637995in}{2.131988in}}%
\pgfpathlineto{\pgfqpoint{1.655371in}{2.131988in}}%
\pgfpathlineto{\pgfqpoint{1.672747in}{2.132657in}}%
\pgfpathlineto{\pgfqpoint{1.690124in}{2.132657in}}%
\pgfpathlineto{\pgfqpoint{1.707500in}{2.132657in}}%
\pgfpathlineto{\pgfqpoint{1.724876in}{2.133326in}}%
\pgfpathlineto{\pgfqpoint{1.742253in}{2.133326in}}%
\pgfpathlineto{\pgfqpoint{1.759629in}{2.133995in}}%
\pgfpathlineto{\pgfqpoint{1.777005in}{2.133995in}}%
\pgfpathlineto{\pgfqpoint{1.794382in}{2.133995in}}%
\pgfpathlineto{\pgfqpoint{1.811758in}{2.134664in}}%
\pgfpathlineto{\pgfqpoint{1.829134in}{2.134664in}}%
\pgfpathlineto{\pgfqpoint{1.846511in}{2.135333in}}%
\pgfpathlineto{\pgfqpoint{1.863887in}{2.136002in}}%
\pgfpathlineto{\pgfqpoint{1.881263in}{2.136670in}}%
\pgfpathlineto{\pgfqpoint{1.898640in}{2.136670in}}%
\pgfpathlineto{\pgfqpoint{1.916016in}{2.138008in}}%
\pgfpathlineto{\pgfqpoint{1.933392in}{2.138008in}}%
\pgfpathlineto{\pgfqpoint{1.950769in}{2.138008in}}%
\pgfpathlineto{\pgfqpoint{1.968145in}{2.138008in}}%
\pgfpathlineto{\pgfqpoint{1.985521in}{2.140015in}}%
\pgfpathlineto{\pgfqpoint{2.002898in}{2.140015in}}%
\pgfpathlineto{\pgfqpoint{2.020274in}{2.140015in}}%
\pgfpathlineto{\pgfqpoint{2.037650in}{2.141353in}}%
\pgfpathlineto{\pgfqpoint{2.055027in}{2.143359in}}%
\pgfpathlineto{\pgfqpoint{2.072403in}{2.144697in}}%
\pgfpathlineto{\pgfqpoint{2.089779in}{2.145366in}}%
\pgfpathlineto{\pgfqpoint{2.107156in}{2.145366in}}%
\pgfpathlineto{\pgfqpoint{2.124532in}{2.146703in}}%
\pgfpathlineto{\pgfqpoint{2.141909in}{2.146703in}}%
\pgfpathlineto{\pgfqpoint{2.159285in}{2.146703in}}%
\pgfpathlineto{\pgfqpoint{2.176661in}{2.146703in}}%
\pgfusepath{stroke}%
\end{pgfscope}%
\begin{pgfscope}%
\pgfsetrectcap%
\pgfsetmiterjoin%
\pgfsetlinewidth{1.003750pt}%
\definecolor{currentstroke}{rgb}{1.000000,1.000000,1.000000}%
\pgfsetstrokecolor{currentstroke}%
\pgfsetdash{}{0pt}%
\pgfpathmoveto{\pgfqpoint{0.461616in}{0.431659in}}%
\pgfpathlineto{\pgfqpoint{0.461616in}{2.228372in}}%
\pgfusepath{stroke}%
\end{pgfscope}%
\begin{pgfscope}%
\pgfsetrectcap%
\pgfsetmiterjoin%
\pgfsetlinewidth{1.003750pt}%
\definecolor{currentstroke}{rgb}{1.000000,1.000000,1.000000}%
\pgfsetstrokecolor{currentstroke}%
\pgfsetdash{}{0pt}%
\pgfpathmoveto{\pgfqpoint{2.258330in}{0.431659in}}%
\pgfpathlineto{\pgfqpoint{2.258330in}{2.228372in}}%
\pgfusepath{stroke}%
\end{pgfscope}%
\begin{pgfscope}%
\pgfsetrectcap%
\pgfsetmiterjoin%
\pgfsetlinewidth{1.003750pt}%
\definecolor{currentstroke}{rgb}{1.000000,1.000000,1.000000}%
\pgfsetstrokecolor{currentstroke}%
\pgfsetdash{}{0pt}%
\pgfpathmoveto{\pgfqpoint{0.461616in}{0.431659in}}%
\pgfpathlineto{\pgfqpoint{2.258330in}{0.431659in}}%
\pgfusepath{stroke}%
\end{pgfscope}%
\begin{pgfscope}%
\pgfsetrectcap%
\pgfsetmiterjoin%
\pgfsetlinewidth{1.003750pt}%
\definecolor{currentstroke}{rgb}{1.000000,1.000000,1.000000}%
\pgfsetstrokecolor{currentstroke}%
\pgfsetdash{}{0pt}%
\pgfpathmoveto{\pgfqpoint{0.461616in}{2.228372in}}%
\pgfpathlineto{\pgfqpoint{2.258330in}{2.228372in}}%
\pgfusepath{stroke}%
\end{pgfscope}%
\begin{pgfscope}%
\pgfsetbuttcap%
\pgfsetmiterjoin%
\definecolor{currentfill}{rgb}{0.898039,0.898039,0.898039}%
\pgfsetfillcolor{currentfill}%
\pgfsetfillopacity{0.800000}%
\pgfsetlinewidth{0.501875pt}%
\definecolor{currentstroke}{rgb}{0.800000,0.800000,0.800000}%
\pgfsetstrokecolor{currentstroke}%
\pgfsetstrokeopacity{0.800000}%
\pgfsetdash{}{0pt}%
\pgfpathmoveto{\pgfqpoint{1.252962in}{0.487214in}}%
\pgfpathlineto{\pgfqpoint{2.180552in}{0.487214in}}%
\pgfpathquadraticcurveto{\pgfqpoint{2.202774in}{0.487214in}}{\pgfqpoint{2.202774in}{0.509436in}}%
\pgfpathlineto{\pgfqpoint{2.202774in}{1.118079in}}%
\pgfpathquadraticcurveto{\pgfqpoint{2.202774in}{1.140301in}}{\pgfqpoint{2.180552in}{1.140301in}}%
\pgfpathlineto{\pgfqpoint{1.252962in}{1.140301in}}%
\pgfpathquadraticcurveto{\pgfqpoint{1.230740in}{1.140301in}}{\pgfqpoint{1.230740in}{1.118079in}}%
\pgfpathlineto{\pgfqpoint{1.230740in}{0.509436in}}%
\pgfpathquadraticcurveto{\pgfqpoint{1.230740in}{0.487214in}}{\pgfqpoint{1.252962in}{0.487214in}}%
\pgfpathlineto{\pgfqpoint{1.252962in}{0.487214in}}%
\pgfpathclose%
\pgfusepath{stroke,fill}%
\end{pgfscope}%
\begin{pgfscope}%
\pgfsetrectcap%
\pgfsetroundjoin%
\pgfsetlinewidth{1.003750pt}%
\definecolor{currentstroke}{rgb}{1.000000,0.498039,0.054902}%
\pgfsetstrokecolor{currentstroke}%
\pgfsetdash{}{0pt}%
\pgfpathmoveto{\pgfqpoint{1.275185in}{1.056968in}}%
\pgfpathlineto{\pgfqpoint{1.386296in}{1.056968in}}%
\pgfpathlineto{\pgfqpoint{1.497407in}{1.056968in}}%
\pgfusepath{stroke}%
\end{pgfscope}%
\begin{pgfscope}%
\definecolor{textcolor}{rgb}{0.000000,0.000000,0.000000}%
\pgfsetstrokecolor{textcolor}%
\pgfsetfillcolor{textcolor}%
\pgftext[x=1.586296in,y=1.018079in,left,base]{\color{textcolor}\rmfamily\fontsize{8.000000}{9.600000}\selectfont \textsc{Dijkstra}}%
\end{pgfscope}%
\begin{pgfscope}%
\pgfsetbuttcap%
\pgfsetroundjoin%
\pgfsetlinewidth{1.003750pt}%
\definecolor{currentstroke}{rgb}{0.121569,0.466667,0.705882}%
\pgfsetstrokecolor{currentstroke}%
\pgfsetdash{{3.700000pt}{1.600000pt}}{0.000000pt}%
\pgfpathmoveto{\pgfqpoint{1.275185in}{0.902029in}}%
\pgfpathlineto{\pgfqpoint{1.386296in}{0.902029in}}%
\pgfpathlineto{\pgfqpoint{1.497407in}{0.902029in}}%
\pgfusepath{stroke}%
\end{pgfscope}%
\begin{pgfscope}%
\definecolor{textcolor}{rgb}{0.000000,0.000000,0.000000}%
\pgfsetstrokecolor{textcolor}%
\pgfsetfillcolor{textcolor}%
\pgftext[x=1.586296in,y=0.863140in,left,base]{\color{textcolor}\rmfamily\fontsize{8.000000}{9.600000}\selectfont \textsc{Stubborn}}%
\end{pgfscope}%
\begin{pgfscope}%
\pgfsetbuttcap%
\pgfsetroundjoin%
\pgfsetlinewidth{1.003750pt}%
\definecolor{currentstroke}{rgb}{0.839216,0.152941,0.156863}%
\pgfsetstrokecolor{currentstroke}%
\pgfsetdash{{6.400000pt}{1.600000pt}{1.000000pt}{1.600000pt}}{0.000000pt}%
\pgfpathmoveto{\pgfqpoint{1.275185in}{0.747091in}}%
\pgfpathlineto{\pgfqpoint{1.386296in}{0.747091in}}%
\pgfpathlineto{\pgfqpoint{1.497407in}{0.747091in}}%
\pgfusepath{stroke}%
\end{pgfscope}%
\begin{pgfscope}%
\definecolor{textcolor}{rgb}{0.000000,0.000000,0.000000}%
\pgfsetstrokecolor{textcolor}%
\pgfsetfillcolor{textcolor}%
\pgftext[x=1.586296in,y=0.708202in,left,base]{\color{textcolor}\rmfamily\fontsize{8.000000}{9.600000}\selectfont \textsc{Tenacious}}%
\end{pgfscope}%
\begin{pgfscope}%
\pgfsetbuttcap%
\pgfsetroundjoin%
\pgfsetlinewidth{1.003750pt}%
\definecolor{currentstroke}{rgb}{0.172549,0.627451,0.172549}%
\pgfsetstrokecolor{currentstroke}%
\pgfsetdash{{1.000000pt}{1.650000pt}}{0.000000pt}%
\pgfpathmoveto{\pgfqpoint{1.275185in}{0.592153in}}%
\pgfpathlineto{\pgfqpoint{1.386296in}{0.592153in}}%
\pgfpathlineto{\pgfqpoint{1.497407in}{0.592153in}}%
\pgfusepath{stroke}%
\end{pgfscope}%
\begin{pgfscope}%
\definecolor{textcolor}{rgb}{0.000000,0.000000,0.000000}%
\pgfsetstrokecolor{textcolor}%
\pgfsetfillcolor{textcolor}%
\pgftext[x=1.586296in,y=0.553264in,left,base]{\color{textcolor}\rmfamily\fontsize{8.000000}{9.600000}\selectfont \textsc{SetCover}}%
\end{pgfscope}%
\end{pgfpicture}%
\makeatother%
\endgroup%